\documentclass[10pt]{article}
\makeatletter\def\input@path{{./}{arxiv/source/}}\makeatother
\usepackage{amsmath,amssymb,amsthm,amsfonts}
\usepackage{mathtools}
\usepackage{dsfont}
\usepackage{bm}
\usepackage[sc]{mathpazo}
\usepackage[scaled=0.92]{helvet}
\usepackage[T1]{fontenc}
\usepackage[utf8]{inputenc}
\usepackage[english]{babel}
\usepackage{microtype}
\usepackage{graphicx}
\usepackage{array}
\usepackage{tabularx}
\usepackage{booktabs}
\usepackage{multirow}
\usepackage{float}
\usepackage{placeins}
\usepackage{xspace}
\usepackage{empheq}
\usepackage[letterpaper,margin=0.75in,columnsep=0.25in]{geometry}
\makeatletter
\renewcommand{\fnum@figure}{Figure~\thefigure}
\renewcommand{\fnum@table}{Table~\thetable}
\long\def\@makecaption#1#2{\vskip\abovecaptionskip{\small\def\textbf##1{{\sffamily\bfseries ##1}}\textbf{#1}\enspace #2\par}\vskip\belowcaptionskip}
\renewcommand\section{\@startsection{section}{1}{\z@}{-3ex \@plus -1ex \@minus -.2ex}{1.2ex \@plus .2ex}{\sffamily\bfseries\Large}}
\renewcommand\subsection{\@startsection{subsection}{2}{\z@}{-2.5ex \@plus -1ex \@minus -.2ex}{1ex \@plus .2ex}{\sffamily\bfseries\large}}
\renewcommand\subsubsection{\@startsection{subsubsection}{3}{\z@}{-2ex \@plus -1ex \@minus -.2ex}{0.6ex \@plus .2ex}{\sffamily\bfseries\itshape\normalsize}}
\renewcommand\paragraph{\@startsection{paragraph}{4}{\z@}{1.5ex \@plus .5ex \@minus .2ex}{-1em}{\sffamily\bfseries\small}}
\makeatother
\usepackage[round,semicolon]{natbib}
\setcitestyle{aysep={},yysep={,}}
\usepackage{url}
\allowdisplaybreaks[4]
\usepackage{xcolor}
\usepackage[colorlinks=true,allcolors={rgb:red,1;green,3;blue,8}]{hyperref}

\newcommand{\cy}[1]{#1}

\newcommand{\addtabletext}[1]{{\setlength{\leftskip}{9pt}\footnotesize#1}}
\renewcommand{\arraystretch}{1.5}

\renewcommand{\arraystretch}{1.5}

\newcommand{\nht}{\tilde{n}_H}
\newcommand{\ntt}{\tilde{n}_T}
\newcommand{\dnHt}{\dot{\tilde{n}}_H}
\newcommand{\dnTt}{\dot{\tilde{n}}_T}
\newcommand{\nt}{\tilde{n}}

\newcommand{\taut}{\tilde{\tau}}
\newcommand{\rt}{\tilde{r}}
\newcommand{\Ft}{\tilde{F}}

\newcommand{\taus}{\tau^*}
\newcommand{\tauts}{\taut^*}

\newcommand{\thres}{\theta_{\text{res}}}
\newcommand{\thmut}{\theta_{\text{mut}}}
\newcommand{\nz}{n(0)}
\newcommand{\ntz}{\nt(0)}
\newcommand{\rs}{r^*}
\newcommand{\ns}{n^*}
\newcommand{\Ns}{N^*}

\newcommand{\nth}{n_{\text{crit}}}
\newcommand{\nc}{n_c}

\newcommand{\ropt}{r_{\text{opt}}}
\newcommand{\RES}{\mathcal{R}_{\mathrm{ES}}}
\newcommand{\rmin}{r_{\min}}
\newcommand{\rmax}{r_{\max}}

\newcommand{\tauopt}{\tau_{\text{opt}}}

\newcommand{\rp}{r_{+}}
\newcommand{\rmn}{r_{-}}
\newcommand{\rpm}{r_{\pm}}

\newcommand{\mpl}{m_{+}}
\newcommand{\mmn}{m_{-}}
\newcommand{\mt}{\tilde{m}}
\newcommand{\mts}{\tilde{m}^\ast}

\DeclareMathOperator*{\argmax}{argmax\,}

\newcommand{\tT}{t_{T}}
\newcommand{\tTt}{\tilde{t}_{T}}
\newcommand{\tDel}{t_{\Delta}}
\newcommand{\tinv}{t_{\text{inv}}}

\newcommand{\Fs}{F^{\ast}}

\newcommand{\rts}{\rt^{\ast}}

\newcommand{\p}{\mathrm{p}}

\newcommand{\Fa}{F_{\alpha}}
\newcommand{\Fta}{\tilde{F}_{\alpha}}
\newcommand{\Pt}{\tilde{P}}

\newcommand{\E}{\mathbb{E}}

\newcommand{\taueff}{\tau_{\text{eff}}}
\newcommand{\neff}{n_{\text{eff}}}

\newcommand{\xt}{\tilde{x}}

\newcommand{\pnl}[1]{\textbf{#1}}

\def\papertitle{When to retire: the cultural evolution of post-reproductive oblique teaching}

\begin{document}

\twocolumn[{%
\begin{flushright}
{\sffamily\bfseries\huge \papertitle\par}
\vspace{12pt}
{\sffamily\bfseries\small Hirotaka Goto$^{1,2}$, Joshua B. Plotkin$^{2,3,\ast}$\par}
\vspace{3pt}
{\sffamily\scriptsize $^{1}$Graduate School of Advanced Mathematical Sciences, Meiji University, Nakano, Tokyo 164-8525, Japan, $^{2}$Department of Biology, University of Pennsylvania, Philadelphia, PA 19104, USA, $^{3}$Meiji Institute for Advanced Study of Mathematical Sciences, Tokyo 164-8525, Japan\par
$^\ast$Corresponding author. Email: jplotkin@sas.upenn.edu\par
\vspace{3pt}
October 21, 2026\par}
\end{flushright}
\vspace{22pt}
{\sffamily\small\noindent\textbf{ABSTRACT}\enspace Teaching is ubiquitous across human societies. Instruction from parents to offspring is the most fundamental form of teaching, explained primarily through kin selection and inclusive fitness. Instruction to non-kin, known as oblique teaching, also plays a dominant role in modern societies, where teaching is often institutionalized.
Nonetheless, it remains puzzling how or when oblique teaching can evolve. Here we address the evolution of oblique teaching by developing a mathematical model of stage-structured populations. The model assumes that younger individuals engage in a useful activity, called hunting, and they reproduce. At some point in life, they cease contributing directly to hunting and reproduction, and instead serve exclusively as teachers to help increase the hunting expertise of the young. We refer to this transition in life stage as \emph{retirement}, and we analyze the evolution of the retirement rate, and the resulting frequency of post-reproductive teachers, in a population subject to a global carrying capacity. We find that retirement and teaching can evolve without kinship or inclusive fitness, but only in populations above a threshold size.  We analyze the retirement rates that are evolutionarily stable, which depend on the absolute population size of a mutant type upon introduction, unlike in classical adaptive dynamics.
In many regimes, evolution produces fewer teachers than would maximize the population's per-capita hunting returns or intrinsic birth rate.
A mutant that differs only slightly from the resident type generally cannot invade from a minority, so teaching cannot emerge through gradual evolution. The mechanism we describe is therefore available to cultural evolution, where a distinct norm for retirement and elder care may be adopted by a sub-population simultaneously. Our account of post-reproductive oblique teaching is thus tied closely to theories of cultural evolution, including the ``information grandmother hypothesis'' proposed by Henrich and Henrich.\par}
\vspace{24pt}
}]

Humans continue to live for a long time after they stop reproducing. Among hunter--gatherers who survive childhood, a substantial part of adult life falls after the end of reproduction \citep{Gurven2007, Kaplan2000}. Elders continue to contribute to their groups, through what they know. Among Tsimane forager--horticulturalists, the skills an adult needs are acquired largely from older members of the parental generation, and not from parents alone \citep{Schniter2023}. Among the Ache, hunting returns continue to rise for two decades after physical strength has peaked \citep{Walker2002, Gurven2006, Koster2020}. Expertise that takes so long to acquire is worth sustaining, and a society that depends on it has reason to support individuals well past the age at which they cease to hunt and reproduce. Individuals who no longer contribute directly to hunting or to reproduction can nonetheless increase the hunting returns of those who do, through teaching.

Elders who teach nonetheless pose a puzzle for evolutionary theory. A teacher consumes what others hunt and produces no offspring. Why natural selection should establish a stage of life devoted purely to instruction, and why a society should provision the individuals in it, is not obvious.

Two lines of theory explain why humans live on after reproduction ends. Under the first, a post-reproductive individual who provisions her descendants gains fitness indirectly, through kin; the grandmother hypothesis in this form is supported by measured effects of grandmothers on the survival and reproduction of grandoffspring \citep{Hawkes1998, Hawkes2004, Lahdenpera2004, Sear2008}. Under the second, a long juvenile period and a long post-reproductive period are joint consequences of a skill-intensive foraging niche: expertise takes decades to acquire, so selection favors both a long period of learning and a long life over which the resulting knowledge can be used and transmitted \citep{Kaplan2000, Kaplan2002, Kaplan2010, Gurven2012}. Henrich and Henrich combined the two arguments into the \emph{information grandmother hypothesis}: once high-fidelity cultural transmission is in place, a grandmother can invest in her grandchildren by transmitting adaptive information rather than calories, and the hypothesis predicts, across species, that sophisticated cultural capacities should accompany a long post-reproductive lifespan \citep{Henrich2010}. The hypothesis is most compelling when knowledge is costly to acquire and cheap to pass on. Food taboos observed by Fijian women during pregnancy and breastfeeding, for example, single out precisely those marine species that carry dangerous toxins; and this knowledge is acquired from kin and from prestigious members of the community rather than rediscovered, at hazard, by each generation \citep{Henrich2010}. Yet the information grandmother hypothesis has never been given a mathematical model, and whether or when natural selection will establish such a stage of life has not been determined.

But teaching in human societies is not confined to kin. Cultural evolution distinguishes three modes of knowledge transmission: vertical, from parents to their own offspring; horizontal, between individuals of the same generation; and oblique, from members of the parental generation to unrelated members of the offspring generation \citep{CavalliSforza1982, Hewlett2024}.
Oblique teaching is broadly documented where cultural transmission has been measured, and it accounts for much of what individuals learn after early childhood \citep{Hewlett2011, Hewlett2013, Hewlett2024, Jang2025}. Contemporary societies have institutionalized oblique teaching, assigning some individuals to instruction as a full-time occupation. Whereas relatedness explains why a grandmother might instruct her grandchildren, it does not explain why an elder should instruct unrelated children, still less why a society should organize itself to support this norm. Formal models of evolution by natural selection confirm this puzzle: when oblique transmission among non-kin is added to vertical transmission from parents, teaching becomes harder to evolve because the benefit is delivered to individuals who do not carry the teacher's genes \citep{Aoki2017, Mullon2017, Fogarty2011}. A recent analysis finds that teaching is strongly disfavored as the probability of oblique transmission increases, so that some additional mechanism is required to explain why teaching is widespread \citep{Shimodaira2026}. Oblique teaching, although empirically common, thus remains a theoretical mystery.

A separate line of work asks not how teaching evolves, but how much teaching a population should do. Treating the fraction of a population devoted to teaching as a quantity to be allocated, we previously showed that a population must exceed a critical size before it is beneficial to assign anyone to teach at all \citep{Goto2026}. But that analysis of optimal population allocation to teachers specified no individual-level process by which the fraction of teachers could evolve. Three critical questions remain.
Does a life history in which individuals first hunt and reproduce, and then cease both in order to teach, evolve at all in the absence of relatedness? If it does, which rates of transition into the teaching stage, and what fractions of teachers, does evolution produce? And does the population evolve the frequency of teachers that will maximize its hunting returns and associated birth rate?

Here we answer these questions with a model of a stage-structured population in which propensity to teach evolves as a life-history trait. Individuals are born into a hunter stage, in which they hunt and reproduce, and they transition irreversibly, at a rate that is heritable and free to evolve, into a post-reproductive stage in which they teach and do not reproduce. We call this transition \emph{retirement}. Teachers raise the expertise of the hunters they instruct, and so the rate at which those hunters obtain returns, but teachers themselves neither hunt nor reproduce, and they are supported by the hunting returns of younger hunters. The retirement rate therefore determines the fraction of the population that teaches. Retiring at a higher rate raises the fraction of teachers, and with it the expertise of the hunters who remain, but it leaves fewer individuals hunting and reproducing, and it shortens each individual's own hunting and reproductive life. Because a population's intrinsic birth rate is proportional to the hunting returns obtained per member, teachers included, this tradeoff between the number of hunters and their expertise also governs reproduction.

We interpret the retirement rate as a social norm rather than a physiological limit: the age at which the members of a society are expected to stop hunting and reproducing, to instruct the young, and to be provisioned by those who still hunt. A norm of this kind is held by a group, and it is set by conventions about who works, who is fed, and who teaches. The competing types in our model are accordingly groups that differ in their norm of retirement, and hence in the fraction of teachers each maintains. Within a type, teachers instruct the young and hunting returns are shared among all members; relatedness plays no part, and the two types with different norms of retirement interact only by competing for a shared carrying capacity. We ask whether a group with one norm of retirement can displace a resident population with another, in a well-defined evolutionary model of births and deaths.

We find that a positive rate of retirement, and with it a post-reproductive stage devoted to teaching, can evolve among individuals who are entirely unrelated, so that oblique teaching evolves without inclusive fitness. But retirement evolves only in populations above a threshold size,
which we denote by \( \nth \), and which exceeds the size at which teaching first becomes beneficial to a population \citep{Goto2026}. Above this threshold, the retirement rates that resist invasion form an entire interval rather than a evolutionary stable value. The stable range of retirement rates, and the fractions of teachers they sustain, are
often smaller than the retirement rate that would maximize a population's per-capita hunting returns, and in those cases evolution fails to maximize the population's intrinsic birth rate.

We also find the process of evolution of such norms has non-classical behavior. A group whose retirement rate differs only slightly from the resident's type generally cannot invade from a minority: teaching is established not by the accumulation of small changes but by a group of sufficient size adopting a new norm that differs by a finite amount. The process we describe is therefore available to cultural evolution, in which a sub-population can adopt a new norm of retirement and elder care in a single step. Our results provide a theoretical and evolutionary account of the information grandmother hypothesis and extend it beyond kinship and sex: the hypothesis does not require that knowledge pass to descendants, or that teachers be women, only that the returns of expertise be shared among those with a common norm of teaching and elder care.

\section*{\label{sec:model}A model of retirement and oblique teaching}

\subsection*{A stage-structured population}

\begin{figure}[!htbp]
    \centering
    \includegraphics[width=\linewidth]{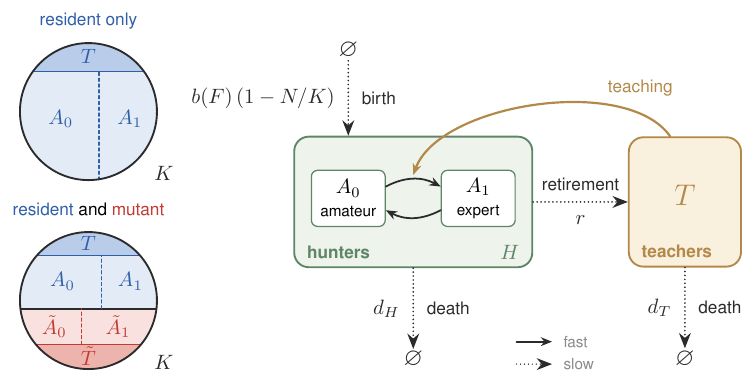}
    \caption{
    \textbf{Individuals are born into a hunter stage and retire irreversibly into a
    non-reproductive teaching stage.}
    \pnl{\emph{Right},}
    Individuals are born into the hunter stage at a rate of \( b(F)(1-N/K) \) and perform hunting activities.
    The intrinsic birth rate, $b(F)$, increases with the per-capita hunting return rate of the population, $F$, and the total birth rate is reduced as the total population size, $N$, approaches a carrying capacity, $K$.
    Hunters retire at rate \( r \), becoming teachers, or they die at rate \( d_H \).
    Teachers do not undertake any hunting work but instead teach amateur hunters (\( A_0 \)), thereby raising them to expert status (\( A_1 \)); teachers die at rate \( d_T \).
    We assume that amateur and expert hunters exchange status much faster than individuals are born, retire, or die, so that the numbers of amateur and expert hunters equilibrate instantaneously in response to the current number of teachers and the current population size
    (solid arrows).
    \pnl{\emph{Left},}
    The two diagrams on the left illustrate scenarios with only a single type with a fixed retirement rate (top) or both a resident type with retirement rate $r$ and a mutant type with rate $\tilde r$ competing for the shared carrying capacity \( K \).
    }
    \label{fig:schematic}
\end{figure}

\par
Consider first a population of size \( n \) with a fixed life-history profile governed by the rate at which individuals retire.
The population consists of two classes, a hunter (\( H \)) class and an elder-teacher (\( T \)) class, with \( n_H \) and \( n_T \) individuals respectively, so that \( n = n_H + n_T \) (Fig.~\ref{fig:schematic}).
Hunters generates returns at per-capita rate \( F \) that supports the continued growth of the population by determining the intrinsic birth rate.
All individuals are born into the hunter stage.
Hunters retire, at rate \( r \), to become teachers, or die at a constant rate.
Teachers also die at a constant rate. The model does not discriminate between genders.

\par
Hunters and teachers differ in three respects.
Hunters hunt and reproduce; teachers do neither.
Teachers instead increase the expertise of the hunters---and hence increase the mean hunting rate per hunter.
Retirement is irreversible: an individual who has entered the teacher class never returns to the hunter class.
We refer this transition as \emph{retirement}.
Because teachers neither hunt nor reproduce, but still count toward the carrying capacity of the environment, every teacher lowers the fraction of the population that hunts while raising the skill of the hunters that teachers train.
A higher rate of retirement produces a greater proportion of the population that are teachers, and also increase the proportion of hunters with greater expertise; but whether or not this increases the per-capita rate of hunting return, \( F \), and thus the birth rate of the population, hinges on the tradeoff between the number of active hunters versus their level of expertise \citep{Goto2026}.

\par
Our exposition proceeds in three steps.
First, we describe how amateur hunters acquire expertise from teachers, which  determines the per-capita hunting return rate, \( F \), of a population with a given number of teachers.
Second, we describe how births (into the hunter class), retirements (from hunter to teacher class), and deaths change the numbers of hunters and teachers over time in a population with a fixed retirement rate.  Third, we let a resident type with one retirement rate compete against a mutant type with a different retirement rate. %
The first process is assumed to be instantaneous and the remaining two slow.

\subsection*{The dynamics of amateur and expert hunters (instantaneous)}

\par
Whereas the full population is divided into hunter and teacher classes, $n=n_H+n_T$, the hunters themselves are further sub-divided by their level of expertise, either amateur or expert. The dynamics of expertise are described by
an established model of teaching and learning \citep{Goto2026}.
An amateur hunter becomes an expert hunter either by self-learning, at rate \( \lambda \), or by being taught, at a rate proportional to the number of teachers available, \( \mu n_T \); an expert loses their skill and reverts to amateur status at rate \( f \).
These population dynamics within the hunter class are described as follows:
\begin{align} \label{eq:hunter_teacher}
\begin{cases}
    \dot n_0 = f n_1 - (\lambda + \mu n_T) n_0, \\
    \dot n_1 = (\lambda + \mu n_T) n_0 - f n_1.
\end{cases}
\end{align}
where $n_0$ and $n_1$ denote the numbers of amateur and expert hunters, with $n_0+n_1=n_H$.
Measuring the three rates relative to the rate of skill loss leaves two dimensionless parameters,
the \emph{self-learning efficiency} \( \eta \coloneqq \lambda/f \) and the \emph{education efficiency} \( \nu \coloneqq \mu/f \); from this point onward, we use only the dimensionless parameters.
In a population with a given proportion of teachers, $\tau := n_T/n$ the numbers of amateur and expert hunters are assumed to reach equilibrium quickly: \( (\bar n_0, \bar n_1) = \frac{(1-\tau)n}{1 +\eta + \nu \tau n} (1, \;\eta + \nu \tau n) \).

Amateur hunters generate returns at rate \( h_0 \), whereas expert hunters generate returns at a higher rate \( h_1 > h_0 \).
We define the per-capita hunting return rate of the population as \( F = (n_0 h_0 + n_1 h_1)/n \).
The denominator is the total population size \( n \), including teachers as well as hunters: hunting returns are shared among all members of the population, so that the hunters effectively provision both themselves and the teachers.
When the number of teachers is fixed and the numbers of amateur and expert hunters have reached their equilibrium values, the per-capita hunting return rate is given by
\begin{align}
    F(\tau;n)=(1-\tau)\frac{h_0+h_1(\eta+\nu\tau n)}{1+\eta +\nu\tau n}, \label{eq:F_eq}
\end{align}
where \( \tau \) denotes the proportion of teachers (\( 0 \leq \tau \leq 1 \)).
Intuitively, Eq.~\ref{eq:F_eq} captures the basic trade-off between labor loss due to allocation of individuals to teachers (\( 1 - \tau \)) and the enhanced mean hunting rate \emph{per hunter} thanks to teaching (\( \frac{h_0+h_1(\eta+\nu\tau n)}{1+\eta +\nu\tau n} \)).

Although we refer to the hunters' activity as \emph{hunting}, we use the term only as an illustrative example of any skilled activity that benefits the group collectively and thereby raises the intrinsic birth rate of the population---gathering, fishing, cultivation, food processing and storage, the building and maintenance of shelter, or the manufacture of tools.
The term follows usage in human behavioral ecology, where return rates rise with experience \citep{Kaplan2000, Gurven2006, Walker2002, Koster2020, Gurven2012}. %
All that matters here is that expertise raises an individual's rate of return, that acquiring expertise requires instruction, and that the returns are shared among all individuals (both teachers and hunters alike). The quantity \( F \) of Eq.~\ref{eq:F_eq} is accordingly the per-capita hunting return rate, and multiplying it by the expected duration of the hunter stage, introduced below, gives an expected lifetime per-capita hunting return.

\subsection*{The dynamics of hunters and teachers}

\begin{table*}[tp]
\centering
\caption{
Symbols used in the model and the analysis.
}
\setlength{\tabcolsep}{3pt}
\renewcommand{\arraystretch}{1.2}
\begin{tabular*}{\textwidth}{
    @{\extracolsep{\fill}}
    c
    >{\raggedright\arraybackslash}m{0.35\linewidth}
    c
    >{\raggedright\arraybackslash}m{0.25\linewidth}
}
\toprule
\multicolumn{2}{c}{\textbf{Variables}} &
\multicolumn{2}{c}{\textbf{Parameters}} \\
\cmidrule(r){1-2}
\cmidrule(l){3-4}
Symbol and definition &
\multicolumn{1}{c}{Description} &
Symbol  &
\multicolumn{1}{c}{Description} \\
\midrule
\(n,\;\nt\) & Resident and mutant population sizes &
\(K\) & Shared carrying capacity \\

\(n_H,\;\nht\) & Numbers of resident and mutant hunters &
\(r,\;\rt\) & Resident and mutant retirement rates (\( \geq 0 \)) \\

\(n_T,\;\ntt\) & Numbers of resident and mutant teachers &
\( b_0 \) & Birth rate constant \\

\(n_0,\;n_1\) & Numbers of resident amateur and expert hunters &
\( d_H \) & Hunter mortality rate \\

\( N \coloneqq n + \nt \) & Total population size ( and ) &
\(d_T\) & Teacher mortality rate \\

\(\tau \coloneqq n_T/n,\;\taut \coloneqq \ntt/\nt\) & Resident and mutant teacher proportions &
\(h_0 \) & Amateur hunting rate \\

\(\rho \coloneqq n/N\) & Resident frequency &
\(h_1\) & Expert hunting rate (\( > h_0 \))\\

\(F(\tau;n)=(1-\tau)\frac{h_0+h_1(\eta+\nu\tau n)}{1+\eta +\nu\tau n}\) & Equilibrium per-capita hunting return rate &
\(h \coloneqq h_1/h_0\) & Advantage of expertise (\( > 1 \)) \\

\(b(F)\coloneqq b_0F/h_0\) & Return-dependent birth rate &
\(\eta\) & Self-learning efficiency \\

\( \Theta(\tau,r;n) \coloneqq F(\tau;n)/(r + d_H) \) & Expected per-capita lifetime hunting return  &
\(\nu\) & Education efficiency \\

\( \theta(r; n) \coloneqq \Theta(\taus(r), r; n) \) &
Expected per-capita lifetime hunting return at the quasi-stationary teacher proportion &
\( \alpha \) & Degree of sharing of hunting returns between types (\( 0 \leq \alpha \leq 1/2 \)) \\

\bottomrule
\end{tabular*}
\addtabletext{
\begin{minipage}{\textwidth}
Resident quantities are written without an accent and the corresponding mutant quantities with a tilde. Steady-state values carry a superscript asterisk.
Quasi-steady state values carry a superscript dagger; see \hyperref[sec:matmethods]{\emph{Materials and Methods}}\xspace for more details.
All parameters are strictly positive unless otherwise noted.
Note that \( n = n_H + n_T \), \( \nt = \nht + \ntt \), \( n_H = n_0 + n_1 \) by definition.
\end{minipage}
}
\label{tab:symbols}
\end{table*}

Now we turn to the dynamics of hunters and teachers in a population fixed with a fixed retirement rate.
Here the frequency of teachers,  \( \tau = n_T/n\), is no longer set exogenously as in \citep{Goto2026} but intrinsically dynamic, arising from births, deaths, and the rate of retirement from hunting and reproduction to post-reproductive teaching.
Birth, death, and retirement processes are all captured by the dynamical system
\begin{align} \label{eq:one_type}
    \begin{cases}
        \dot n_H = b(F) n_H \left( 1 - \frac{n}{K} \right) - (r+d_H) n_H, \\
        \dot n_T = r n_H - d_T n_T,
    \end{cases}
\end{align}
where \( n(t) = n_H(t) + n_T(t) \) denotes the total number of individuals in the population that may vary over time.
Hunters are produced by birth at an intrinsic rate \( b(F) \) that is modulated by the logistic factor \( 1 - n/K \), where $K$ is a fixed carrying capacity.
Hunters are lost by retirement, at rate \( r \), and by death, at rate \( d_H \).
The teacher class gains individuals only by retirement and loses them only by death, at rate \( d_T \).
Only hunters give birth, and the intrinsic birth rate of a hunter, \( b(F) \), is proportional to the hunting return per individual (not the hunting return per hunter). By contrast, teachers make no direct demographic contribution. A summary of model parameters is given in Table~\ref{tab:symbols}.

Hunting returns and reproduction are directly linked in our model: the intrinsic birth rate of a population grows proportionally to its per-capita hunting returns.
More specifically, we define \( b(F) \coloneqq b_0 F/h_0 \), where the birth rate constant \( b_0 > 0 \) denotes the intrinsic birth rate of a population consisting entirely of amateur hunters, for which \( F = h_0 \) and hence \( b = b_0 \) (Table~\ref{tab:symbols}).

We recall that the numbers of amateur and expert hunters, \( n_0 \) and \( n_1 \), are assumed to change instantaneously (Eq.~\ref{eq:hunter_teacher}), relative to the timescale of the changes in the numbers of hunters and teachers (Eq.~\ref{eq:one_type}).
This assumption justifies substituting the equillibrium expression \( F(\tau(t);n(t)) \) from Eq.~\ref{eq:F_eq} directly into the intrinsic birth rate \( b(F) \), where \( \tau(t) = n_T(t)/n(t) \).
In other words, the per-capita hunting return rate \( F(\tau;n) \) enters Eq.~\ref{eq:one_type} as a function of the current teacher proportion and population size alone, and the intrinsic birth rate
\( b(F(\tau;n)) \) changes instantaneously when the number of teachers changes.

\subsection*{Competition between types with different retirement rates}

\par
To model selection on the retirement rate, we consider a resident type, which retires at rate \( r \), competing against a mutant type, which retires at a different rate, \( \rt \), and therefore maintains a different proportion of teachers.
The two types compete for a single, shared carrying capacity, \( K \).
Each type is structured in the same way as the monomorphic population described above; the sizes of the resident and mutant populations are denoted by \( n \) and \( \nt \), and all mutant quantities carry a tilde accent (see Table~\ref{tab:symbols} for a full list of symbols and notation).
The hunter and teacher mortality rates, \( d_H \) and \( d_T \), are the same for both types.

The birth, death, and retirement processes of the two competing types, with retirement rates $r$ and $\rt$, are described by the dynamical system:
\begin{align} \label{eq:model}
    \begin{cases}
        \dot n_H = b(F) n_H \left( 1 - \frac{N}{K} \right) - (r+d_H) n_H, \\
        \dot n_T = r n_H - d_T n_T, \\
        \dnHt = b(\tilde F) \nht \left( 1 - \frac{N}{K} \right) - (\tilde r+d_H) \tilde n_H, \\
        \dnTt = \rt \nht - d_T \ntt.
    \end{cases}
\end{align}
Here \( N(t) = n(t) + \nt(t) \) denotes the total population size, counting resident hunters and teachers as well as mutant hunters and teachers.
The equations for both resident and mutant types are constructed in the same way as in Eq.~\ref{eq:one_type}, except that the logistic factor is now \( 1 - N/K \), with the total population size \( N \) in place of \( n \).
Every individual---hunter or teacher, resident or mutant---counts equally toward the shared carrying capacity $K$, so that two types  compete only in that they both seek to occupy the environment.

\par
As in the monomorphic model  with a single retirement rate, we assume that the teaching and learning dynamics occur much faster than birth, death, or retirement; so that the numbers of amateur and expert hunters within each type can be replaced by their equilibrium values.
Therefore, we define the resident and mutant per-capita hunting return rates by \( F \coloneqq F(\tau; n) \) and \( \Ft \coloneqq F(\taut; \nt) \), respectively, where \( \tau = n_T/n \) and \( \taut = \ntt/\nt \) denote the resident and mutant teacher proportions, and we substitute them into the birth rates \( b(F) \) and \( b(\Ft) \) of Eq.~\ref{eq:model}.
The two-type model generalizes the one-type model: Eq.~\ref{eq:model} reduces to Eq.~\ref{eq:one_type} when the mutant type is absent, \( (\nht, \ntt) = (0, 0) \).

\par
Two restrictions apply throughout our analysis of this model.
First, the birth rate constant is assumed to be large relative to the rate at which individuals leave the hunter class, \( r + d_H \ll b_0 \), so that a monomorphic population (of either retirement rate) at equilibrium sits close to its carrying capacity, i.e., \( \Ns \approx K \).
Second, because the hunter and teacher mortality rates are the same for both types, the retirement rate is the only trait that differs between them, and hence the only trait on selection.
We focus our analysis on the following question: which retirement rates, and hence which fractions of teachers, can  persist in a population that competes for a shared environment against a rival type that retires at a different rate?

\section*{\label{sec:results}Results}

\subsection*{\label{subsec:single_species}The retirement rate determines the fraction of teachers and the post-reproductive lifespan}

\par
We begin by studying a population in which every individual retires at the same rate, to establish what a given retirement rate implies before we ask how the retirement rate will
evolve.
The retirement rate \( r \) determines the equilibrium proportion of teachers in the population (\emph{SI Appendix}, section~\ref{sec:monomorphic}), which is given by
\begin{align}
    \taus(r) = \frac{r}{r+d_T}. %
\end{align}
Other equilibrium properties in a monomorphic population, together with the expected life-history characteristics such as total lifespan and the post-reproductive lifespan, are shown in \emph{SI Appendix}, Table~\ref{tab:single_species}.

The equilibrium fraction of teachers, \( \taus(r) \), has a simple interpretation.
The teacher class receives individuals at rate \( r \) per hunter and loses them at rate \( d_T \)
per teacher, so at equilibrium the ratio of teachers to hunters is \( r/d_T \), and the teacher
fraction follows.
A population therefore controls how many teachers it maintains only through the retirement rate
\( r \) and the teacher mortality rate \( d_T \).
A higher retirement rate raises the fraction of teachers, and with it the expertise of the hunters
those teachers train, while lowering the fraction of the population that is hunting and reproducing at any moment.
The retirement rate and teacher mortality rate also fix the life-history characteristics of individuals.
The expected fraction of an individual's lifetime spent in the post-reproductive stage is exactly equal to the equilibrium fraction of teachers, \( \text{PrR} = \taus \) (\emph{SI Appendix}, Table~\ref{tab:single_species}).
The identity is useful because the fraction of a population that is post-reproductive may be easier to measure in practice.

\subsection*{Competition between two retirement rates ends in the exclusion of one type}

\par
We now turn to competition between two types that retire at different rates (Eq.~\ref{eq:model}).
We study this competition first by deterministic and stochastic simulation, and then we derive an analytical prediction of its outcome.
Because the intrinsic birth rate is much greater than the retirement rate plus the hunter mortality rate, \( r + d_H \ll b_0 \) (and likewise \( \rt + d_H \ll b_0 \) for the mutant type), a monomorphic population of either type equilibrates near the carrying capacity, \( \ns \approx K \) (or \( \nt^* \approx K \)).

Before introducing a mutant type into the population, we first allow the resident population to reach demographic equilibrium ($n = \ns$). Then, we introduce mutants by replacing a subset of the resident hunters by mutant hunters at time $t=0$.
The mutant type is initially in the minority, i.e., $\nz \geq \ntz$.
Because they replace residents, the introduction of mutants holds the total population size fixed, so that the mutants and the residents begin their competition in an environment that is already at demographic equilibrium and near carrying capacity.

\begin{figure}[!htbp]
    \centering
    \includegraphics[width=\linewidth]{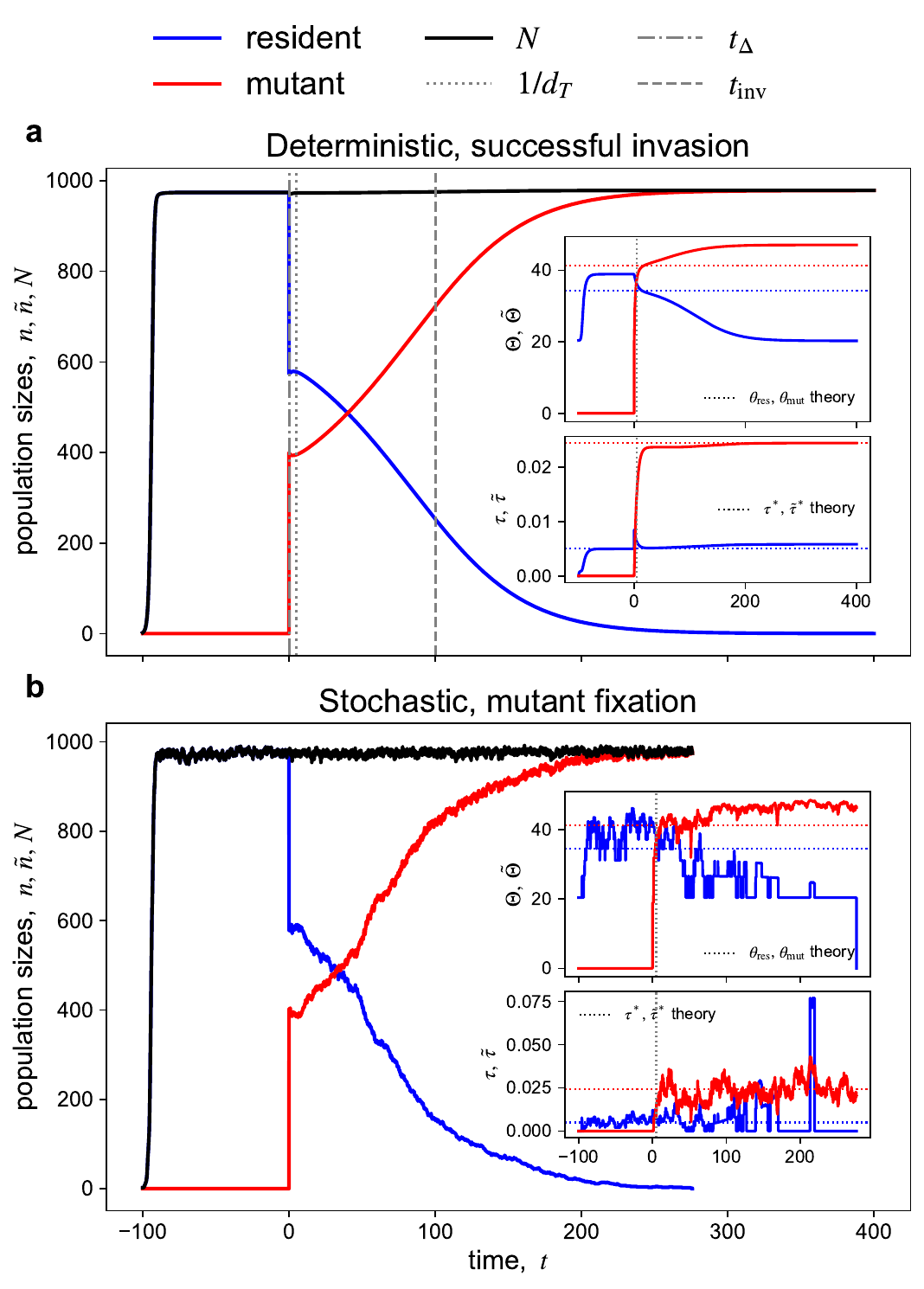}
    \caption{
    \textbf{A mutant type with a different retirement rate replaces the resident when its expected per-capita lifetime hunting return exceeds the resident's at introduction.}
    Time evolution of resident (blue), mutant (red), and total (black) population sizes following an invasion attempt at \( t = 0 \), in (\pnl{a}) deterministic and (\pnl{b}) stochastic simulations, when invasion is successful (\( \thmut > \thres \));
    see \emph{SI Appendix}, Fig.~\ref{fig:base_trajectories_full}
    for failed invasion (\( \thmut < \thres \)).
    Insets show expected per-capita lifetime hunting returns (\( \Theta \) and \( \tilde\Theta \), top) and teacher proportions (\( \tau \) and \( \taut \), bottom) over time.
    In insets, horizontal dotted lines show quasi-stationary values, \( \thres \) and \( \thmut \) (top) and \( \tau^* \) and \( \taut^* \) (bottom); the difference \( \thres - \thmut \) is used to predict the invasion outcome.
    Vertical dotted, dash-dotted, and dashed lines indicate the approximate reorganization (that is, the reorganization of the hunter-teacher ratio; \( 1/d_T \)), ecological (\( \tDel \)), and invasion (\( \tinv \)) timescales,
    representing the timescale on which the teacher proportion relaxes, the timescale on which the environment relaxes, and the timescale on which invasion occurs, respectively.
    Their expressions are given by Eq.~\ref{eq:timescales} in \hyperref[sec:matmethods]{\emph{Materials and Methods}}\xspace.
    Notice that the invasion timescale (\( \tinv \)) is much slower than the reorganization (\( 1/d_T \)) or ecological (\( \tDel \)) timescale, and in the current setting, the quantities measured in insets quickly reach quasi-stationary values over \( t \approx 1/d_T > \tDel \), which are well predicted by the horizontal dotted lines in both deterministic and stochastic simulations.
    Initial conditions at \( t = -100 \) are given by \( (n_H, n_T, \nht, \ntt) = (2,0,0,0) \) in both panels, allowing the system to quickly equilibrate in the absence of mutant individuals.
    At \( t = 0 \), 400 of the resident hunters become mutant hunters, changing the population sizes from \( (n_H, n_T, \nht, \ntt) \) to \( (n_H - \ntz, n_T, \ntz, 0) \), where \( \ntz = 400 \).
    Parameters (\pnl{a,b}):
    \( K = 1000 \), \( d_H = 0.05 \), \( d_T = 0.2 \), \( b_0 = 1 \), \( \eta = 0.02 \), \( \nu = 0.2 \), \( h_0 = 1 \), \( h_1 = 3 \), \( r = 0.001 \),
    \( \rt = 0.005 \).
    }
    \label{fig:base_trajectories}
\end{figure}

We assess invasion outcomes based on whether the mutant replaces the resident and fixes in the population, or the mutant goes extinct.
For stochastic simulations, we formulate a continuous-time discrete-state Markov process corresponding to Eq.~\ref{eq:model}, and simulate using the Gillespie algorithm. These simulations terminate upon fixation of the resident or the mutant type. For illustrative purposes, we first consider an example of such an invasion for a parameter set in which the mutant type retires at a higher rate than the resident (\( \rt > r \)).
For the parameters of this example, the mutant successfully replaces the resident, in both the deterministic and the stochastic dynamics (Fig.~\ref{fig:base_trajectories}).

Three timescales are visible in the trajectories of this example, and one is noticeably slower than the other two.
The proportion of teachers within each type relaxes very quickly, over a period of order \( 1/d_T \) for small \( r \) and \( \rt \),
and thereafter each type maintains its own quasi-stationary teacher proportion, \( \taus \) or \( \tauts \).
The total population size also relaxes quickly, over the ecological timescale \( \tDel \), and it thereafter remains close to the carrying capacity.
Only then do the two population sizes of resident and mutant, $n$ and $\tilde n$ change appreciably, over the invasion timescale \( \tinv \).
The eventual outcome for the example in Figure Fig.~\ref{fig:base_trajectories} is mutant fixation; see \emph{SI Appendix}, Fig.~\ref{fig:base_trajectories_full} for an example parameter set where the mutant goes extinct.

We emphasize our analysis must include cases when the state of the mutant is non-trivial, both in its initial population size and its phenotypic distance from the resident.
Unlike in the standard setting of adaptive dynamics \citep{Geritz1998, Brannstrom2013}, where a mutant is assumed both rare and phenotypically close to the resident, the invasion outcome will depend on the initial population sizes of the two types, through the numbers of teachers each  type maintains and thus the hunting returns and reproduction of its members. Hence a prediction of the outcome must refer to the initial population sizes of the two types, as well as to their respective rates of retirement.

\subsection*{Invasion is decided by expected per-capita lifetime hunting return at the time of introduction}

The outcome of competition between two types with different retirement rates can be predicted analytically from the state of the system at the time the mutant is introduced. We will analyze this process assuming that, when the mutant is introduced, it is no more abundant than the resident, and that the total population is at demographic equilibrium, near carrying capacity.

\par

Our analysis will assume in addition that invasion process is slow relative to the reorganization of the hunter--teacher ratios and to the relaxation of their total population sizes, namely \( \tinv \gg 1/d_T,\, \tDel \)
as seen in the example of Fig.~\ref{fig:base_trajectories}; the precise condition is given by Eq.~\ref{eq:timescale_separation} in \hyperref[sec:matmethods]{\emph{Materials and Methods}}\xspace. Under this separation of timescales, the teacher proportions and the total population size relax quickly and then remain nearly constant while the invasion proceeds, so the outcome is determined by quantities evaluated at the time of invasion, $t=0$. Competition always ends with the exclusion of one type or the other: the frequency of the resident obeys a logistic equation whose growth rate is determined by the state at the time of introduction and does not change sign thereafter
(see \emph{SI Appendix}, Eq.~\ref{eq:invasion}).

In \hyperref[sec:matmethods]{\emph{Materials and Methods}}\xspace and \emph{SI Appendix}, section~\ref{sec:predictor_derivation}, we show that the invasion outcome under our model (Eq.~\ref{eq:model}) can be predicted from the sign of the following expression:
\begin{align}
    \thres - \thmut, \label{eq:prediction}
\end{align}
where we define
\begin{align}
    \thres \coloneqq \theta(r; \nz), \quad
    \thmut \coloneqq \theta(\rt; \ntz),
\end{align}
and
\begin{align}
    \theta(r;n)
    = \frac{F(\taus(r);n)}{r + d_H}. \label{eq:theta_def}
\end{align}
The quantity \( \theta \) has an intuitive biological meaning: the expected per-capita lifetime hunting return in an equilibrium population of size \( n \) with retirement rate \( r \), evaluated at the quasi-steady state once teacher-proportion has stabilized (Table~\ref{tab:symbols}).
A positive sign predicts the resident will drive the mutant extinct (because it has a higher expected per-capital lifetime hunter return), whereas a negative sign predicts the mutant will successfully invade.,
Note that, just as in the definition of return $F$, so too is $\theta$ defined as lifetime return per all individuals in the population, meaning that returns must be shared among teachers and hunters alike.

The sign condition in Eq.~\ref{eq:prediction} establishes the principle that governs invasion: the type that will eventually fix is the one whose expected per-capita lifetime hunting return is greater, at the time the mutant is introduced.  This principle makes intuitive sense, because a type's intrinsic birth rate is proportional to its per-capita hunting return rate, and so the type with the greater expected lifetime return per individual also has the greater expected lifetime intrinsic reproductive output.

The expected per-capita lifetime hunting return, \( \theta \), is the product of two factors (Eq.~\ref{eq:theta_def}): the per-capita hunting return rate, \( F(\taus(r);n) \), evaluated once the teacher proportion has relaxed to \( \taus(r) \), and the expected duration of the hunter stage, \( 1/(r+d_H) \).
A higher retirement rate \( r \) affects both factors: it raises the fraction of teachers, and thus the skill of the hunters who remain, while leaving a smaller fraction of the population hunting (the tradeoff of Eq.~\ref{eq:F_eq}); and it shortens the hunter stage.
The population size \( n \) enters only the first factor, through the number of teachers, \( \taus(r)n \), and so the same retirement rate yields different expected lifetime returns in two populations that differ only in population size.

Note that the expected returns of the mutant and resident types are evaluated at their respective initial population sizes, so that the mutant's expected per-capita lifetime hunting return, \( \thmut = \theta(\rt;\ntz) \), depends on the size of the founding mutant sub-population, \( \ntz \), and not only on its frequency within the population as a whole. We analyze the consequences of this below, when we characterize the evolutionarily stable range of retirement rates.

\subsection*{An evolutionarily stable range of retirement rates}

The sign condition of Eq.~\ref{eq:prediction} determines the invasion outcome for any pair of resident and mutant retirement rates, given their founding population sizes.
We can therefore ask, for any resident retirement rate, which mutant retirement rates would replace it and which would fail to invade.
A resident type that resists invasion by every mutant is a stable evolutionary outcome: a population that adopts such a rate of retirement and its associated frequency of teachers cannot be selectively displaced by any competing type.

\begin{figure}[!htbp]
    \centering
    \includegraphics[width=\linewidth]{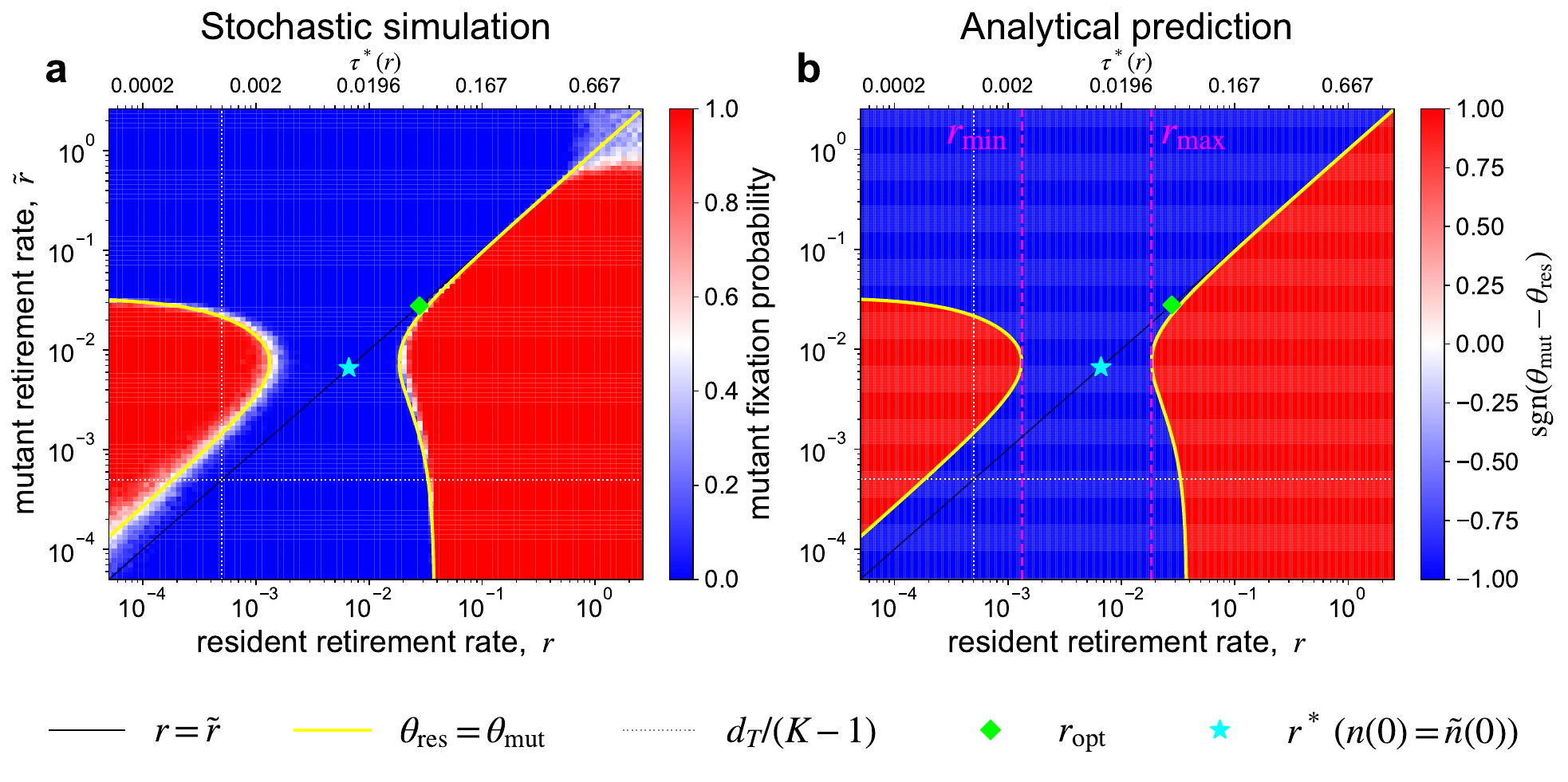}
    \caption{
    \textbf{An evolutionarily stable range of retirement rates.}
    Pairwise invasibility plots (PIPs) obtained from (\pnl{a}) stochastic simulations and (\pnl{b}) ODE analysis.
    (\pnl{a}) color gives mutant fixation probability over 100 runs per pair of \( (r, \rt) \);
    the blue regions imply resident fixation, whereas the red regions imply mutant fixation.
    (\pnl{b}) color gives the sign of the mutant relative fitness, \( \thmut - \thres \).
    In \pnl{a}, the yellow contour \( \thres = \thmut \) predicts the boundary between resident and mutant fixation and delimits an evolutionarily stable range of retirement rates, \( \RES \), that no mutant can invade.
    The separation between these regions is well predicted by the contour \( \thres = \thmut \).
    Vertical dashed lines shown in magenta delimit the evolutionarily stable range of retirement rates, \( \RES = [\rmin, \rmax] \) , implying that \( r \in \RES \) cannot be invaded by any \( \rt \).
    The cyan star marks the unique evolutionarily stable retirement rate, \( \rs \), for equal resident and mutant founding sizes \( \ntz = \nz \), and the green diamond
    the retirement rate \( \ropt \) that maximizes the per-capita hunting return rate \( F(\taus(r);K) \) of a monomorphic population at carrying capacity, which is not evolutionarily stable here.
    White dotted lines mark one expected teacher in quasi-steady states.
    The secondary axis at the top of each panel shows the quasi-stationary fraction of teachers within the resident population.
    Parameters (\pnl{a,b}):
    \( K = 1000 \), \( d_H = 0.025 \), \( d_T = 0.5 \), \( b_0 = 0.5 \), \( \eta = 0.02 \), \( \nu = 0.2 \), \( h_0 = 1 \), \( h_1 = 3 \), \( \nz = 700 \), \( \ntz = 300 \).
    Stochastic simulations are initialized as \( (n_H, n_T, \nht, \ntt) = (\nz, 0, \ntz, 0) \).
    }
    \label{fig:pip_stochastic_vs_prediction}
\end{figure}

\par
The invasion outcomes for all pairs of resident and mutant retirement rates are displayed in a pairwise invasibility plot (PIP), computed both from stochastic simulations and from the analytical prediction (Fig.~\ref{fig:pip_stochastic_vs_prediction}).
In general, a resident that retires too rarely is replaced by mutants with moderately higher retirement rates (red region at the left of each panel), and a resident that retires too often is replaced by mutants with lower retirement rates (red region at the right). Between these two regions lies a band of intermediate resident retirement rates at which the resident persists and the mutant goes extinct, whatever the mutant's rate of retirement (blue at every \( \rt \)).

The analytical prediction accounts for the pattern seen in the stochastic simulatiosn: the contour along which \( \thres = \thmut \) (yellow lines in both panels) separates the region in which the mutant fixes from the region in which the resident persists [compare Fig.~\ref{fig:pip_stochastic_vs_prediction}\pnl{a} with Fig.~\ref{fig:pip_stochastic_vs_prediction}\pnl{b}].
Agreement degrades where either retirement rate is so small that fewer than one teacher is expected at the quasi-stationary state (below the white dotted lines), and where either rate is so large that neither type reliably fixes within the time-frame simulated.

We denote by \( \RES \) the set of resident retirement rates that no mutant retirement rate can invade, and we call it the evolutionarily stable range, or set, of retirement rates.
In classical adaptive dynamics this set contains a single trait value; but here it is generally an interval.  The minimum and maximum evolutionarily stable resident retirement rates are denoted by \( \rmin \) and \( \rmax \), respectively, i.e., \( \RES \equiv [\rmin, \rmax] \) (see Fig.~\ref{fig:pip_stochastic_vs_prediction}\pnl{b}).

\begin{table}[tbp]
\footnotesize
    \centering
    \caption{
    Four regimes for the evolution of retirement and post-reproductive teaching.
    }

    \setlength{\tabcolsep}{3pt}
    \renewcommand{\arraystretch}{1.15}

    \begin{tabular*}{\columnwidth}{
        @{\extracolsep{\fill}}
        >{\centering\arraybackslash}m{0.11\columnwidth}
        >{\centering\arraybackslash}m{0.24\columnwidth}
        >{\centering\arraybackslash}m{0.24\columnwidth}
        >{\raggedright\arraybackslash}m{0.31\columnwidth}
        @{}
    }
        \toprule
        Regime
        & Condition(s)
        & ES retirement rate(s)
        & \multicolumn{1}{c}{What does it mean?} \\
        \midrule

        I
        & \( \nth > \nz \)
        & \( \RES = \{0\} \)
        & No-teaching is the only stable outcome, and so teaching never evolves. \\

        II
        & \( \nz > \nth > \ntz \)
        & \( \RES = [0,\rmax] \)
        & No-teaching is a stable outcome, and so too is some teaching. \\

        III
        & \( \ntz > \nth \)
        & \( \RES = [\rmin,\rmax] \)
        & All stable outcomes include teaching, and so teaching must evolve. \\

        IV
        & \( \nz = \ntz > \nth \)
        & \( \RES = \{\rs\} \)
        & Teaching can evolve gradually from no teaching. \\
        \bottomrule
    \end{tabular*}

    \addtabletext{
    \begin{minipage}{\columnwidth}
    In regime II, teaching cannot invade a population that lacks teaching;
    however, it can persist once established, for example, due to the immigration of a population that practices teaching.
    In cases II and III, a positive rate of teaching evolves following a \emph{discrete} change in the retirement rate, whereas in case IV teaching can evolve through continuous, gradual increases in the retirement rate (see Fig.~\ref{fig:pip_panels}).
    Here we assume $\nz \geq \ntz$ and \( \rmax > \rmin > 0 \) whenever \( \rmax \) or \( \rmin \) appears.
    \end{minipage}
    }

    \label{tab:regimes}
\end{table}

\begin{figure*}[!tp]
    \centering
    \includegraphics[width=\linewidth]{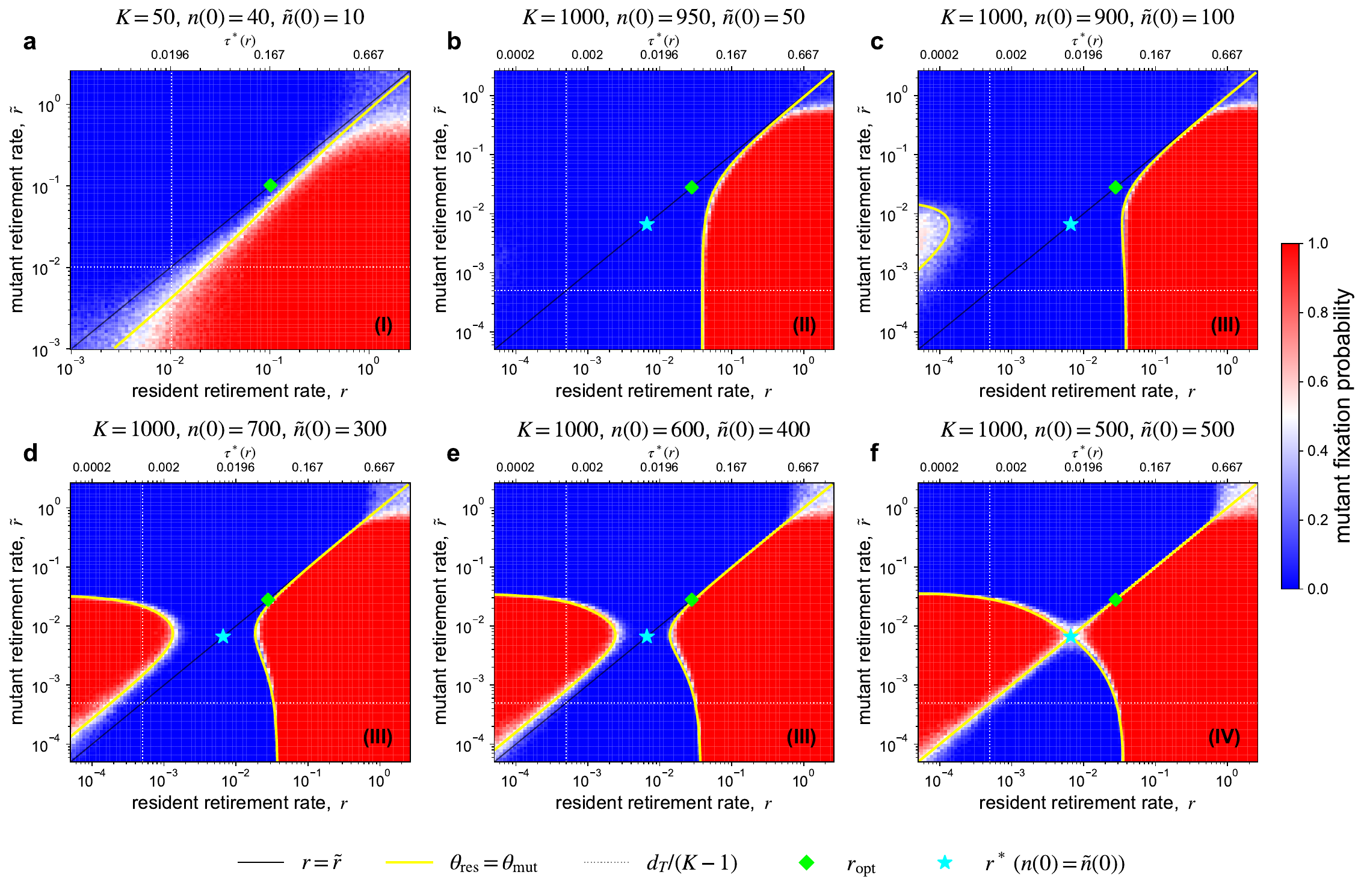}
    \caption{
    \textbf{The stable range of retirement rates contracts to a single convergence stable rate as the founding mutant population approaches the resident in size.}
    \pnl{a--f} Pairwise invasibility plots (PIPs) for various pairs of \( (\nz, \ntz) \) obtained from stochastic simulations.
    The color scale represents the mutant fixation probability over 100 realizations per pair of \( (r, \rt) \).
    As the ratio \( \ntz/\nz \) increases from near zero to one [\pnl{b--f}], the evolutionarily stable range of retirement rates contracts and collapses at the predicted value \( \rs \).
    \pnl{f} The unique evolutionarily stable retirement rate \( \rs \) is convergence stable, as the PIP becomes fully symmetric about the diagonal line \( r = \rt \).
    When the founding mutant population \( \ntz \) is very small [as in \pnl{b}], \( \rmin \) becomes zero; that is, the yellow curve on the left disappears.
    When both \( \nz \) and \( \ntz \) are very small [as in \pnl{a}], the evolutionarily stable range contracts to zero, i.e., \( \RES = \{ 0 \} \).
    Initial population sizes \( (\nz, \ntz) \) are
    \pnl{a} \( (40, 10) \) with \( K = 50 \);
    \pnl{b} \( (950, 50) \), \pnl{c} \( (900, 100) \), \pnl{d} \( (700, 300) \),
    \pnl{e} \( (600, 400) \), and \pnl{f} \( (500, 500) \), each with \( K = 1000 \).
    Parameters:
    \cy{\( d_H = 0.025 \)}, \( d_T = 0.5 \), \( b_0 = 0.5 \), \( \eta = 0.02 \), \( \nu = 0.2 \), \( h_0 = 1 \), \( h_1 = 3 \).
    All other implementations in stochastic simulations are the same as in Fig.~\ref{fig:pip_stochastic_vs_prediction}.
    The Roman numeral at the lower right corner of each panel represents the associated regime for the evolutionarily stable range (see Table~\ref{tab:regimes}).
    }
    \label{fig:pip_panels}
\end{figure*}

The evolutionarily stable range \( \RES \) depends on the initial population sizes of the resident and the mutant, as predicted by condition in Eq.~\ref{eq:prediction}.
To illustrate this dependence, we show pairwise invasibility plots as the founding size of the mutant grows from a small minority to one half of the total population (Fig.~\ref{fig:pip_panels}\pnl{b--f}), and, separately, for a much smaller total population (Fig.~\ref{fig:pip_panels}\pnl{a}).
When the founding mutant population is small relative to the resident's, the minimum ES resident retirement rate \( \rmin \) is zero (Fig.~\ref{fig:pip_panels}\pnl{b}).
As the founding size of the mutant increases towards the resident's, the range narrows from both ends: \( \rmin \) rises and \( \rmax \) falls (Fig.~\ref{fig:pip_panels}\pnl{b--f}).
Finally, when residents and mutants are equally abundant at the time of introduction, the range contracts to a single stable retirement rate, \( \rs = \rmin = \rmax \), that is also \emph{convergence stable} \citep{OTTO2011} (cyan stars in Figs.~\ref{fig:pip_stochastic_vs_prediction} and \ref{fig:pip_panels}\pnl{f}).
The contraction to a unique stable rate at \( \nz = \ntz \) follows from a symmetry: when the two types are founded at the same population size, the predictor \( \thres - \thmut = \theta(r;\nz) - \theta(\rt;\nz) \) compares one and the same function at two arguments, so a resident resists every mutant rate if and only if its own rate maximizes \( \theta(\,\cdot\,;\nz) \), and that maximizer is unique.
Equal founding population sizes thus recover the classical adaptive-dynamics result that a single trait value is both evolutionarily stable and convergence stable \citep{Geritz1998}.

The comparisons above vary the relative sizes of mutant and resident sub-populations when the total population size is fixed and large.
When the resident and the mutant are both founded at sufficiently small population sizes, no positive resident retirement rate is evolutionarily stable, so that \( \rmin = \rmax = 0 \) (Fig.~\ref{fig:pip_panels}\pnl{a}). In this regime, a small resident population is invaded by types that retire at sufficiently lower rates, down to the limit of no retirement at all, and teaching does not evolve.
These effects of founding population sizes follow from two properties of the predictor \( \thres - \thmut \), as we describe in \emph{SI Appendix}, section~\ref{sec:rES_existence}.

\subsection*{Teaching evolves only above a critical population size}

\begin{figure*}[!tp]
    \centering
    \includegraphics[width=\linewidth]{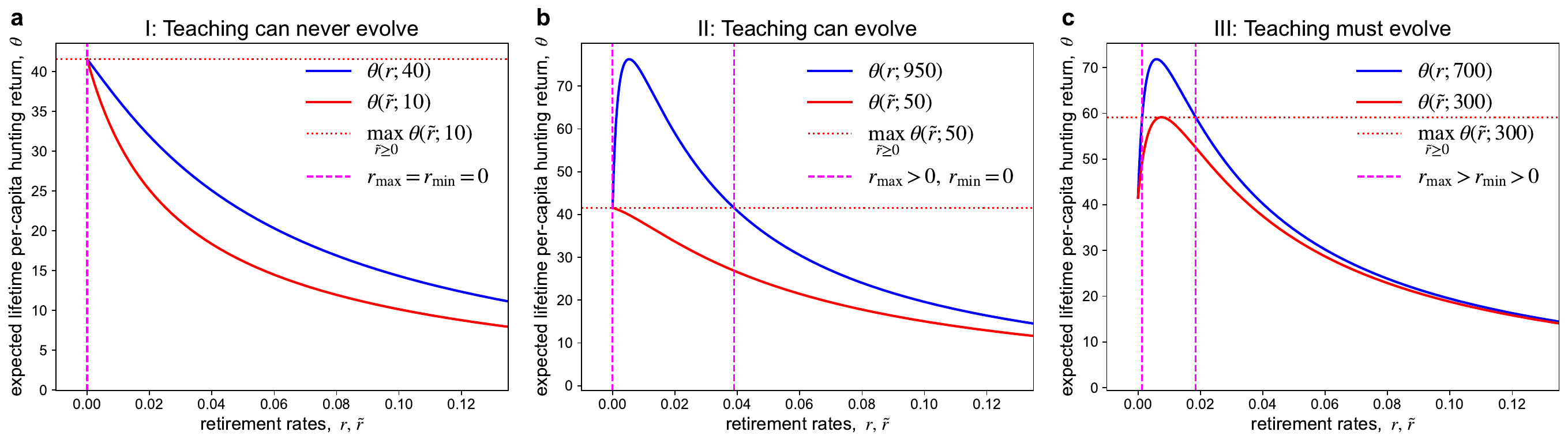}
    \caption{
    \textbf{Four regimes of the expected per-capita lifetime hunting return}
    The stable range of retirement rates \( \RES \) is governed by the shapes of the expected lifetime hunting-return curves \( \thres(r) \) and \( \thmut(\rt) \).
    The three panels illustrate qualitatively different regimes for the relationship between \( \thres(r) = \theta(r; \nz) \) (blue) and \( \thmut(\rt) = \theta(\rt; \ntz) \) (red), provided \( \nz \geq \ntz \).
    \pnl{a} Both functions decrease monotonically with \( r \) and \( \rt \), implying that \( \rmin = \rmax = 0 \).
    \pnl{b} The function \( \thres \) attains a unique interior maximum in \( r \), whereas \( \thmut \) is monotonically decreasing in \( \rt \), implying that \( \rmin = 0, \, \rmax > 0 \).
    \pnl{c} Both \( \theta(r; \nz) \) and \( \theta(\rt; \ntz) \) are uniquely interior-maximized in \( r \) and \( \rt \), respectively, implying that \( \rmax > \rmin > 0 \).
    The results shown in \pnl{a}, \pnl{b}, and \pnl{c} correspond to those in Fig.~\ref{fig:pip_panels}\pnl{a}, Fig.~\ref{fig:pip_panels}\pnl{b}, and Fig.~\ref{fig:pip_panels}\pnl{d}, respectively.
    The vertical dashed lines denote \( \rmin \) and \( \rmax \), together illustrating \( \RES \).
    This evolutionarily stable range is identified by the point(s) where the solid blue curve \( \thres \) and the dotted red horizontal line \( \theta = \max_{\rt>0} \thmut(\rt) \) intersect.
    Parameters:
    \( K = 1000 \), \( d_H = 0.025 \), \( d_T = 0.5 \), \( b_0 = 1 \), \( \eta = 0.02 \), \( \nu = 0.2 \), \( h_0 = 1 \), \( h_1 = 3 \).
    Note that in regime IV (not shown here, see Table~\ref{tab:regimes}),
    two curves are identical and $\rmin=\rmin=\rs$.
    }
    \label{fig:theta_panels}
\end{figure*}

We can obtain the stable range of retirement rates analytically from the expected per-capita lifetime hunting return \( \theta(r;n) \).
A larger initial population obtains a greater expected lifetime return than a smaller one at every positive retirement rate, and the endpoints \( \rmin \) and \( \rmax \) are the roots of the algebraic equation \( \theta(r;\nz) = \max_{\rt \geq 0} \theta(\rt;\ntz) \) (Fig.~\ref{fig:theta_panels}; see \hyperref[sec:matmethods]{\emph{Materials and Methods}}\xspace).
We also prove that \( \rmax > 0 \) only when the initial resident population exceeds a critical size, \( \nz > \nth \), where $\nth$ is a simple expression in the model parameters (see Eq.~\ref{eq:n_crit} in \hyperref[sec:matmethods]{\emph{Materials and Methods}}\xspace). If the initial resident population is smaller than this critical size, then $\rmax=0$ and teaching cannot evolve.

\subsection*{Four evolutionary regimes of teaching}

\par

\par
The evolutionarily stable range has a simple graphical construction (Fig.~\ref{fig:theta_panels}).
The resident's and the mutant's expected lifetime returns, \( \thres(r) = \theta(r;\nz) \) and \( \thmut(\rt) = \theta(\rt;\ntz) \), are the same function evaluated at the two founding population sizes, and the endpoints \( \rmin \) and \( \rmax \) are where the resident curve meets the horizontal line at the maximum of the mutant curve.
Each curve attains its single-peaked maximum at a positive retirement rate if its founding size exceeds \( \nth \) and otherwise decreases monotonically; so the two founding sizes fall into four regimes relative to \( \nth \) (Table~\ref{tab:regimes}; we do not discuss boundary cases).

When both founding sizes are below the threshold, \( \nth > \nz > \ntz \), no retirement is the only stable outcome, \( \RES = \{0\} \), so teaching cannot evolve (Fig.~\ref{fig:theta_panels}\pnl{a}; Fig.~\ref{fig:pip_panels}\pnl{a}).
When only the resident size exceeds the threshold, \( \nz > \nth > \ntz \), no retirement remains a stable outcome, and so too is every retirement rate up to a positive maximum, \( \RES = [0, \rmax] \) (Fig.~\ref{fig:theta_panels}\pnl{b}; Fig.~\ref{fig:pip_panels}\pnl{b}).
When both sizes exceed the threshold, \( \nz > \ntz > \nth \), every stable outcome includes teaching, \( \RES = [\rmin, \rmax] \) with \( \rmin > 0 \), so teaching must evolve, and it evolves only by a discrete jump from \( r = 0 \) (Fig.~\ref{fig:theta_panels}\pnl{c}; Fig.~\ref{fig:pip_panels}\pnl{c--e}).
The smaller the founding mutant population, the larger the change in retirement rate a mutant needs in order to invade a resident rate below \( \rmin \) (\emph{SI Appendix}, section~\ref{sec:rES_existence}).
When the two types are founded at equal sizes above the threshold, \( \nz = \ntz > \nth \), the two curves coincide and the stable range contracts to the single rate that maximizes them, \( \RES = \{\rs\} \); teaching then evolves gradually from a state without teaching (Fig.~\ref{fig:pip_panels}\pnl{f}).
Tables~\ref{tab:regimes} summarizes the four distinct regimes.

It is worth noting that when the founding mutant lies below \( \nth \), teaching cannot be introduced. Below the size threshold a mutant's best retirement rate is \( \rt = 0 \), so no mutant that teaches more often than the resident can ever invade, and \( \rmin = 0 \). In other words, a society cannot raise its fraction of teachers unless the group introducing the change is itself larger than \( \nth \).
If the resident size lies below the threshold as well, no positive rate of retirement is stable at all and the retirement rate collapses to zero.

\subsection*{Short-lived hunters and valuable expertise select for more teachers}

\par

\par
We now ask how the remaining model parameters shift the evolutionarily stable range of retirement, restricting attention to regime III where both founding sizes exceed \( \nth \) (Table~\ref{tab:regimes}).
The sign of the derivative of each endpoint, \( \rmin \) and \( \rmax \), with respect to each parameter is derived in \hyperref[sec:matmethods]{\emph{Materials and Methods}}\xspace and \emph{SI Appendix}, section~\ref{sec:partials} and summarized in \emph{SI Appendix}, Table~\ref{tab:partials}.

Three parameters shift the stable range (upward meaning earlier transition into the elder-teacher class and a larger fraction of teachers).
Both \( \rmin \) and \( \rmax \) increase with the hunter mortality rate \( d_H \): a higher mortality rate shortens the hunter stage whether or not an individual retires, so retirement forgoes less hunting return.
Likewise, both endpoints increase with the advantage of expertise \( h \): a greater advantage raises what a teacher adds to the hunters it trains.
And both endpoints decrease with the self-learning efficiency \( \eta \): efficient self-learning lets amateurs acquire expertise without instruction, which reduces what a teacher adds.

By contrast, a larger founding mutant size relative to the carrying capacity, \( \ntz/K \), narrows the stable range (Fig.~\ref{fig:pip_panels}).
And education efficiency \( \nu \) and the carrying capacity \( K \) act non-monotonically, while the sign for the teacher mortality rate \( d_T \) is undetermined.

\subsection*{\label{sec:ropt_condition}Evolution may fail to optimize hunting returns}

\par
Our model provides an account of which retirement rates, and associated frequencies of teachers, persist when two life histories compete for a shared carrying capacity.
The question remains whether evolution will tend to maximize a population's hunting return rate and thereby its associated intrinsic birth rate.
To answer this, we examine whether the retirement rate that maximizes the per-capita hunting return rate of a monomorphic population at carrying capacity, \( \ropt \coloneqq \argmax_{r \geq 0} F(\taus(r); K) \), is evolutionarily stable under pairwise invasion. An explicit form for $\ropt$ is given in \emph{SI Appendix}, Table~\ref{tab:single_species}.
Whether or not \( \ropt \) lies within the evolutionarily stable range \( \RES \) is equivalent to whether or not the following inequality holds:
\(
    \theta(\ropt; \nz) \geq \max_{\rt \geq 0} \theta(\rt; \ntz)
\).
We find that, in general, the return-maximizing retirement rate $\ropt$ need not be an stable evolutionary outcome, as seen in Figs.~\ref{fig:pip_stochastic_vs_prediction} and \ref{fig:pip_panels} (green diamond to the right of \( \rmax \)).
In other words, evolution may fail to achieve the frequency of post-reproductive teaching that maximizes the intrinsic birth rate of a population. The reason is that  selection favors the retirement rate that maximizes the expected lifetime hunting return of an individual, \( \theta = F/(r+d_H) \), and hence an individual's expected lifetime intrinsic reproductive output (Eq.~\ref{eq:prediction}). This rate is typically lower than the one that maximizes expected rate of hunting returns, because retiring earlier shortens the hunting and reproductive stage.

The tension between the hunting rate and evolutionarily stability is particularly acute when hunters are long-lived relative to teachers: namely, with small hunter mortality, \( d_H \to 0 \), or large teacher mortality, \( d_T \to \infty \). In both these limits, the critical population size $\nth$ grows without bound, so that that $\rmax$ approaches 0 (regime I). The return-maximizing rate of retirement, however, stays positive.
In either limit, then, the only evolutionarily stable outcome has no teachers, even though a population would maximize its birth rate by maintaining a positive proportion of teachers.

\subsection*{Evolution can maximize hunting returns when teachers are long-lived}

There are regimes, however, in which the retirement rate that maximizes the per-capita hunting return rate is itself evolutionarily stable. In the expertise-limited (\( h \gg 1 \)), resource-abundant (\( K > \ntz \gg 1 \)) regime, long-lived teachers can align return-maximization with evolutionarily stability. For a fixed proportion of initial mutants (\( 0 < \ntz/K < 1/2 \)), sufficiently low teacher mortality (\( d_T \approx d_H \)) places the return-maximizing retirement rate within the evolutionarily stable range (i.e., \( \ropt \in \RES \)) to leading order (\emph{SI Appendix}, section~\ref{sec:ropt_RES}).

This observation reveals a second way to increase the frequency of teachers.  The equilibrium frequency of teachers, \( \text{PrR} = r / (r + d_T) \), increases with the retirement rate, the trait under selection in our model, and also with lower teacher mortality, which a population may achieve through norms of elder care.
When teacher mortality equals hunter mortality (\( d_T = d_H \)), moreover, the expected total lifespan reaches its maximum, \( 1/d_H \), whatever the retirement rate (\emph{SI Appendix}, section~\ref{sec:ropt_RES}).
Reducing teacher mortality, down to the that of hunters, therefore has three consequences at once: it raises the frequency of teachers, it makes the return-maximizing retirement rate resistant to invasion, and it leaves the expected total lifespan at its maximum whichever retirement rate evolves.
(Nonetheless, we do not model the evolution of teacher mortality, treating \( d_T \) is a fixed parameter.)

Per-capita hunting returns are one measure of a population's performance.
In \emph{SI Appendix}, section~\ref{sec:alternative_ropt} we also consider the per-capita hunting return rate at the equilibrium population size, the equilibrium population size itself, and the equilibrium rate of births. Of these, only the expected lifetime reproductive output of an individual and the equilibrium population size are generally maximized within the evolutionarily stable range of retirement rates (\emph{SI Appendix}, Fig.~\ref{fig:various_optima_pips}).

\subsection*{Sharing hunting returns permits two retirement rates to coexist}
In the model so far (Eq.~\ref{eq:model}), one of the two types is excluded in every case we have observed, and each type's birth rate depends on the hunting returns of its own members alone.
In \emph{SI Appendix}, section~\ref{sec:food_sharing} we let the two types share a fraction of their hunting returns with each other.
Partial sharing permits both types to persist at a locally asymptotically stable coexistence equilibrium, in deterministic and stochastic dynamics alike (\emph{SI Appendix}, Fig.~\ref{fig:food_sharing_trajectories}, \emph{SI Appendix}, Fig.~\ref{fig:coexistence_stability}).
A society whose sub-populations share their hunting returns can therefore support more than one norm of retirement.

\section*{Discussion}

We have developed a model of a stage-structured population in which individuals are born into a hunting/reproducing stage and transition into a non-reproductive stage devoted to teaching, at a rate that is subject to selection.
The retirement rate determines what fraction of the population teaches, which improves hunter expertise, while the remaining individuals hunt and reproduce. We find that post-reproductive teaching can evolve in this setting, without either kinship or inclusive fitness. But it evolves only in populations larger than a threshold size.
Above this size threshold, the retirement rates that resist invasion form an interval rather than a point; and, often, the range of stable evolutionary outcomes sustain fewer teachers than would maximize the population's intrinsic birth rate.

The retirement rate we study is a socially set schedule for ceasing to hunt and to reproduce, and not the physiological cessation of the reproductive system.  Our results are accordingly not an account of menopause: the model makes no claim about the age at which reproduction becomes physiologically impossible, but only about when a society expects its members to stop hunting and reproducing, to begin instructing the young, and to be provisioned by those who still hunt.  A norm of this kind is adopted by a group, not by an individual, set by social conventions about the age at which one stops hunting or gardening, expectations that the able-bodied provision the old, obligations of apprenticeship, and rules about who instructs the young.  A type in our model is therefore a group that shares one such norm of social behavior. Norms and institutions of this kind can be understood as social technologies that a group adopts to sustain cooperation where individual incentives are insufficient \citep{LiePanis2024, LiePanis2026}.

The unusual form of stable evolutionary outcomes in our model, an interval of stable phenotypes, follows from the dependence of a type's fitness on the absolute size of the group that carries the trait.
In classical adaptive dynamics \citep{Geritz1998, Brannstrom2013, OTTO2011}, invasion fitness depends on the resident and mutant trait values alone; here it depends also on the founding population size of the mutant, because the rate at which amateurs acquire expertise depends on the absolute number of teachers they can reach.
A single mutant individual who retires earlier than all the residents gains nothing for its own type, because upon retirement it has nobody of the same type to instruct.
Teaching therefore cannot emerge in a society unless the group introducing a positive rate of retirement is itself above the threshold size \( \nth \); and a population that is itself below the threshold supports no positive rate of retirement, so that even teaching introduced by immigration would be lost.
Even above the threshold size, evolution cannot proceed by the substitution of mutants that differ infinitesimally from the resident, as it does in classical adaptive dynamics: the new rate of retirement must differ from the resident's by a finite amount (\hyperref[sec:matmethods]{\emph{Materials and Methods}}\xspace), and the smaller the initial population size of the mutant type, the larger the difference required for invasion.
Only when mutant and resident are founded at equal sizes, \( \ntz = \nz \), is the classical behavior with a single stable value recovered.
This size dependence, in both the initial population size and the size of the phenotypic change, confines our account to cultural rather than genetic evolution: a genetic mutation arises in one individual, whereas a group can adopt a substantially new norm for retirement and teaching in a single step \citep{Henrich2004, Boyd2011, CavalliSforza1982}.

Our results provide a theoretical and evolutionary account of the information grandmother hypothesis, under which post-reproductive individuals repay their consumption by transmitting knowledge rather than calories \citep{Henrich2010}.  That hypothesis, as originally stated, rests on relatedness: a grandmother is well placed to teach because her pupils are her descendants.  Our model contains no relatedness by parentage, and therefore extends the hypothesis to oblique teaching by elders. What makes this extension possible is that unrelated individuals can nonetheless discriminate by type, namely by the norm of retirement they share: teachers instruct individuals of the same type, kin or not, and hunting returns are shared not with offspring but with anyone of the same type.  The type with the greater expected lifetime return per individual then reproduces faster and comes to occupy more of the shared carrying capacity.  The information grandmother hypothesis therefore does not require that the information be transmitted to kin; it requires only that the returns to expertise be shared among those who share similar norms for teaching and elder care.

Our findings complement existing theory on the evolution of teaching.  Models of vertical and oblique transmission find that adding oblique transmission among non-kin makes teaching harder to evolve, because the benefit accrues to individuals who do not carry the teacher's genes \citep{Aoki2017, Mullon2017, Fogarty2011}.  Our model does not contradict that conclusion. Those models describe a teaching trait spreading within a population; ours describes whole life histories competing for one environment, so that the benefit of an oblique teacher accrues to the teacher's own type through the demography of the group rather than through descent.  This account requires the assumption that the returns to expertise are shared across the population that generates them, as the benefits of specialization are shared in models of division of labor \citep{Michod2007, Cooper2018, BellRoberts2024, Feinman2011}.
The population threshold we obtain also connects our result to the literature on demographic prerequisites for cultural complexity \citep{Powell2009, Shennan2001, Henrich2004}.

We have focused on teaching and cultural evolution, but the structure of our result has a striking parallel in the evolution of complex organisms.
In multicellular organisms, a heritable change in developmental regulation can alter the fate of many functionally coordinated cells at once, so that sterile cells need not arise independently and spread one at a time
\citep{Gavrilets2010, Michod2007}.
Theoretical work predicts that the advantage of non-reproductive somatic cells increases with the size of the collective \citep{Michod2006, Michod2007, Zhang2026}, and that somatic cells are favored above a critical collective size when they improve the survival of the collective \citep{Zhang2026}.
Retirement into teaching is a similar division within a society: a non-reproductive class whose members raise the reproductive output of the rest, and which can be established by a group of sufficient size acting at once. Comparison between model predictions and empirical patterns of germ-soma division remains a topic for future research.

Our model has several limitations.
We do not model age itself: individuals retire at a constant rate, so the duration of the hunter stage is exponentially distributed.
We expect that fixing the age at retirement, rather than the rate, would shift the threshold \( \nth \) without removing it, since the threshold arises from the comparison between hunting returns forgone and hunting returns gained rather than from the shape of the waiting-time distribution.
We also assume no relatedness whatsoever, which isolates the oblique case but understates the selection available in real populations; we speculate that adding relatedness between teacher and pupil would lower the threshold size \( \nth \) for the evolution of teaching.
We treat expertise in a single skill, whereas a population transmits many skills whose returns may interact \citep{Smolla2023, Fogarty2015, BenOren2023}.
Finally, we let only the retirement rate evolve and impose no constraint on the relationship between the total lifespan and the reproductive lifespan; holding the total lifespan fixed instead would correspond to the \emph{stopping early} hypothesis, and holding the reproductive lifespan fixed to the \emph{extended mothering} and grandmother hypotheses \citep{Watkins2021}, formulations we leave for future work.

One implication for the biological evolution of post-reproductive life follows from the cultural process we describe.
A norm under which individuals cease reproducing before reproductive senescence, and instruct the young instead, raises the value of surviving beyond the end of reproduction; and lower post-reproductive mortality in turn raises the fraction of teachers and, in the regime we analyzed, makes the return-maximizing retirement rate resistant to invasion.
Selection on genes affecting post-reproductive longevity would then act in an environment that culture has itself altered, which is the situation treated by theories of gene--culture coevolution \citep{Mullon2017}.
We speculate that culturally evolved retirement contributed in this way to the post-reproductive longevity of humans; the mortality rates \( d_H \) and \( d_T \) are fixed parameters in our model, and we do not model their evolution.

The purpose of our simple model is not to give a quantitative account of retirement and teaching in any particular society, but to identify which factors govern how those characteristics evolve.  Two factors are responsible here: hunting returns are shared among those who share cultural norms of teaching, and the benefit a teacher confers scales with the number of teachers rather than with their proportion.  Wherever those two conditions hold, a stage of life spent teaching rather than hunting can be favored by selection among unrelated individuals, provided the population exceeds a threshold size.

\section*{\label{sec:matmethods}Materials and Methods}
Here we provide a brief summary of the derivation of the predictor of invasion success, \( \thres - \thmut \), and the analysis of the fitness function \( \theta(r;n) \).
Full details are provided in \hyperref[sec:SI]{\emph{SI Appendix}}\xspace.

\subsection*{Derivation of the predictor}

We derive the predictor (selection coefficient) \( \Lambda \) using quasi-steady-state approximation.
We first derive the time evolution equations of three types of variables in the model:
teacher numbers (\( n_T, \ntt \)), i.e., second and fourth equations of Eq.~\ref{eq:model};
unused carrying capacity (carrying capacity available for further population growth, \( \Delta = K - N \));
and the frequency of residents (\( \rho = n/N \)).
Taken together, these time evolution equations describe reorganizations of stage-structured groups, ecological dynamics, and demographic turnover, which occur potentially on different timescales.
We therefore derive the corresponding timescales from these equations (namely, reorganization, ecological, and invasion timescales) and note the fact that unused carrying capacity is small by design upon mutant introduction, i.e., \( \Delta \approx 0 \) at \( t = 0 \); we then find that quasi-steady-state approximation is valid if the invasion (demographic turnover) timescale is much longer than the ecological and reorganization timescales.
This can be expressed by the following conditions:
\begin{align} \label{eq:timescale_separation}
    \tinv \gg \tT, \quad \tinv \gg \tTt, \quad \tinv \gg \tDel,
\end{align}
where the reorganization (\( \tT, \tTt \)), ecological (\( \tDel \)), and invasion (\( \tinv \)) timescales are given by
\begin{align} \label{eq:timescales}
    \tT \sim \frac{1}{r + d_T}, \
    \tTt \sim \frac{1}{\rt + d_T}, \
    \tDel \sim \frac{1}{\mathcal{B}^\dagger}, \
    \tinv \sim \frac{1}{|\Lambda|},
\end{align}
where \( \operatorname{sgn}(\Lambda) = \operatorname{sgn}(\thres - \thmut \)) determines the invasion outcome,
\( \mathcal{B} \coloneqq [(1-\tau)b(F)n + (1-\taut) b(\Ft)\nt]/K \), and \( \dagger \) indicates that the expression is evaluated at the quasi-steady state, predicted at the time of mutant introduction (\( t = 0 \)).
We refer the reader to \emph{SI Appendix}, section~\ref{sec:predictor_derivation} for a full derivation of the predictor \( \Lambda \) and to \emph{SI Appendix}, Fig.~\ref{fig:pip_panels_shading} for the timescale separation conditions (Eq.~\ref{eq:timescale_separation}) identified in the parameter space of the PIPs (Fig.~\ref{fig:pip_panels}), which shows that prediction can fail when these conditions are not satisfactorily met.

\subsection*{Analysis of the fitness function \( \theta(r;n) \)}
In the Results we applied the sign condition of the predictor (Eq.~\ref{eq:prediction}) per pair of resident and mutant retirement rates.
Here we analyze the predictor itself.
Both terms of the predictor are values of a single function \( \theta \).
The expected lifetime per-capita hunting return, \( \theta(r;n) \), of a population of size \( n \) with retirement rate \( r \) exhibits a critical population size \( \nth \) at which the shape of the function \( \theta \) changes from monotonically decreasing to having a unique interior maximizer in \( r \), as the population size \( n \) exceeds this threshold.
The critical population size is given by
\begin{align}
    \nth = \zeta \nc, \quad
    \zeta = 1 + \frac{d_T}{d_H}, \quad
    \nc = \frac{(h_0 + h_1 \eta)(1+\eta)}{(h_1-h_0)\nu}, \label{eq:n_crit}
\end{align}
and \( \nc \) is the threshold population size for teaching to be beneficial at all in terms of the rate of per-capita hunting return, derived previously in a simpler model with the teacher proportion set exogenously \citep{Goto2026};
see \emph{SI Appendix}, section~\ref{sec:rES_existence} for complete details.
See also \emph{SI Appendix}, section~\ref{sec:n_crit_alternative} for the origin of the factor \( \zeta \).

\section*{Acknowledgments}
We thank Joe Wakano and Hiraku Nishimori for insightful input during the development of this project.
H.G.~acknowledges support from the Japan Society for the Promotion of Science (JSPS) through the JSPS Research Fellowship for Young Scientists and the Grant-in-Aid for JSPS Fellows No.~JP25KJ2129.

\onecolumn
\newgeometry{margin=1in}

\renewcommand{\thefigure}{S\arabic{figure}}
\renewcommand{\thetable}{S\arabic{table}}
\renewcommand{\theequation}{S\arabic{equation}}
\renewcommand{\thesection}{S\arabic{section}}
\renewcommand{\thesubsection}{\thesection.\arabic{subsection}}
\makeatletter\renewcommand{\p@subsection}{}\makeatother
\renewcommand{\thepage}{S\arabic{page}}
\setcounter{figure}{0}
\setcounter{table}{0}
\setcounter{equation}{0}
\setcounter{section}{0}
\setcounter{page}{1}

\begin{center}
\section*{\label{sec:SI}SI Appendix for\\ \papertitle}
Hirotaka~Goto, Joshua~B.~Plotkin$^\ast$\\
\small$^\ast$Corresponding author. Email: jplotkin@sas.upenn.edu\\
\end{center}

\tableofcontents

\newpage

\section{\label{sec:monomorphic}Analysis of one-type dynamics}

We provide a detailed analysis of the one-type dynamics discussed in the main text to establish the baseline scenario. Table~\ref{tab:single_species} summarizes the analytical results.

\subsection{The equilibrium monomorphic population size}
We derive the equilibrium solution to the one-type system (see the main text and Eq.~\ref{eq:one_type} for details):
\begin{align} \label{si:eq:one_type}
    \begin{cases}
        \dot n_H = b(F) n_H \left( 1 - \frac{n}{K} \right) - (r+d_H) n_H, \\
        \dot n_T = r n_H - d_T n_T. \\
    \end{cases}
\end{align}
Stationarity of Eq.~\ref{si:eq:one_type} requires that
\begin{align}
    \ns = K\left(1 - \frac{r+d_H}{b(\Fs)}\right), \quad
    \Fs = (1-\taus) \frac{h_0 + h_1(\eta + \nu \taus \ns)}{1 + \eta + \nu \taus \ns}, \quad
    \taus = \frac{r}{r + d_T},
\end{align}
yielding the quadratic equation
\begin{align}
    a_2 (\ns)^2 + a_1 \ns + a_0 = 0,
\end{align}
where
\begin{align}
    a_0 &= K \left[h_0(1+\eta)(r+d_H) - b_0(1-\taus)(h_0 + h_1 \eta)\right], \\
    a_1 &= K h_0(r + d_H) \nu \taus - b_0(1-\taus)\left[K h_1 \nu \taus - (h_0 + h_1 \eta)\right], \\
    a_2 &= b_0(1-\taus)h_1 \nu \taus.
\end{align}
Therefore, the equilibrium monomorphic population size is given by
\begin{align}
    \ns = \frac{-a_1 \pm \sqrt{a_1^2 - 4 a_0 a_2}}{2a_2}. \label{eq:ns}
\end{align}
The solution relevant to our analysis must satisfy the condition
\begin{align}
    a_1^2 - 4 a_0 a_2 \geq 0, \quad
    0 \leq \ns \leq K.
\end{align}
Using the expression for \( \ns \), it is straightforward to find the equilibrium sizes of younger-hunter and elder-teacher classes, \( n_H^* \) and \( n_T^* \), given the equilibrium proportion of teachers, \( \taus \).

\subsection{Different measures of expected lifespan}

Let \( L \) denote the total lifespan of an individual.
Likewise, let \( L_S \) denote the duration of life stage \( S \in \{ H, T \} \), where \( H \) and \( T \) represent the hunter (reproductive) and teacher (post-reproductive) stages of life, respectively.
Given Eq.~\ref{si:eq:one_type}, the expected durations of the hunter and teacher stages are
\begin{align}
    \E[L_H] &= \E[L_H \mid \text{enter }H] = \frac{1}{r + d_H}, \\
    \E[L_T] &= \Pr(H \to T)\E[L_T \mid T] = \frac{r}{r + d_H}\frac{1}{d_T},
\end{align}
where \( \Pr(H \to T) \) denotes the probability that a hunter survives to the teacher stage.
The mean total lifespan is therefore given by
\begin{align}
    \E[L]
    &= \E[L_H] + \E[L_T] \\
    &= \frac{1}{r + d_H} + \frac{r}{r+d_H}\frac{1}{d_T} \\
    &=\frac{1}{d_T} \frac{r + d_T}{r + d_H}.
\end{align}
Similarly, the mean total lifespan given survival to the teacher stage is
\begin{align}
    \E[L \mid T]
    &= \E[L_H] + \E[L_T | T] \\
    &= \frac{1}{r + d_H} + \frac{1}{d_T} \\
    &= \frac{1}{d_T} \frac{r + d_H + d_T}{r + d_H},
\end{align}
which is strictly larger than its unconditioned counterpart \( \E[L] \), provided \( d_H > 0 \).

The post-reproductive lifespan (``PrL'' for short) is defined as the expected duration of the post-reproductive stage conditional on survival to the post-reproductive stage, i.e.,
\begin{align}
    \text{PrL} \coloneqq \E[L_T \mid T] = \frac{1}{d_T}.
\end{align}
Therefore, the conditional expected post-reproductive lifespan \emph{relative} to expected total lifespan conditional on reaching the post-reproductive stage (``rPrL'' for short) is given by
\begin{align}
    \text{rPrL} \coloneqq \frac{\E[L_T \mid T]}{\E[L \mid T]} = \frac{r + d_H}{r + d_H + d_T}.
\end{align}

\subsection{The retirement rate that maximizes per-capita hunting returns}
Here we derive an exact expression for the optimal retirement rate that maximizes the per-capita hunting return rate of a monomorphic population of size \( n \) at equilibrium.
Given the equilibrium proportion of teachers \( \taus(r) = r/(r + d_T) \), it is straightforward to compute the per-capita hunting return rate-maximizing retirement rate:
\begin{align}
    \frac{\partial F}{\partial r}
    = \left.\frac{\partial}{\partial \tau}F(\tau;n) \right|_{\tau=\taus(r)} \frac{d\taus}{d r}.
\end{align}
Since
\begin{align}
    \frac{d\taus}{dr} = \frac{d_T}{(r + d_T)^2} > 0, \quad r \geq 0,
\end{align}
we find
\begin{align}
    \frac{\partial F}{\partial r} = 0 \quad \iff \quad \frac{\partial F}{\partial \tau} = 0,
\end{align}
which yields the optimal proportion of teachers, \( \tauopt \), obtained in earlier work \citep{Goto2026}, which is given by
\begin{align}
    \tauopt(n) = \frac{1}{\nu n} \left[ - (1 + \eta) + \sqrt{\left(1 - \frac{1}{h}\right)(1 + \eta + \nu n)}\right],
\end{align}
where \( h \coloneqq h_1/h_0 > 1 \), for \( n > \nc \), where \( \nc = \frac{(h_0 + h_1 \eta)(1+\eta)}{(h_1-h_0)\nu} \) is a threshold population size;
for \( n < \nc \), \( \tauopt(n) = 0 \).
See \citep{Goto2026} for more details.
Therefore, the optimal retirement rate, defined as the per-capita hunting return rate-maximizing retirement rate of a monomorphic population of fixed size \( n \), is given by
\begin{align}
    \ropt (n)
    = \frac{\tauopt}{1 - \tauopt}d_T,
\end{align}
which is more explicitly written as
\begin{align}
    \ropt(n) = \frac{-\left(\frac{1}{h} + \eta \right)(1 + \eta + \nu n) + \nu n \sqrt{\left(1-\frac{1}{h}\right)(1 + \eta + \nu n)}}{(1 + \eta + \nu n)\left(\frac{1}{h} + \eta + \nu n\right)} d_T,
\end{align}
provided that the monomorphic population size, \( n \), is sufficiently large.

\begin{table*}
    \centering
    \caption{
    The retirement rate \( r \) determines the equilibrium composition of a monomorphic population, as well as the life-history characteristics.
    }
    \begin{tabular*}{\linewidth}{@{\extracolsep{\fill}}c >{\raggedright\arraybackslash}m{0.55\linewidth} c}
    \toprule
        Symbol & \multicolumn{1}{c}{Description} & Quantity \\
        \midrule
        \( \ns \) & Equilibrium population size & see Eq.~\ref{eq:ns} \\
        \( \taus \) & Equilibrium teacher proportion & \( \frac{r}{r + d_T} \) \\
        \( n_H^* \) & Equilibrium hunter population size & \( \frac{d_T}{r+d_T}\ns \) \\
        \( n_T^* \) & Equilibrium teacher population size & \( \frac{r}{r+d_T}\ns \) \\
        \( \ropt(n) \) & Optimal retirement rate for a monomorphic population of size \( n \) (per-capita hunting return rate-maximizing retirement rate) & \( \frac{-\left(\frac{1}{h} + \eta \right)(1 + \eta + \nu n) + \nu n \sqrt{\left(1-\frac{1}{h}\right)(1 + \eta + \nu n)}}{(1 + \eta + \nu n)\left(\frac{1}{h} + \eta + \nu n\right)} d_T \) \\
        \( \mathbb{E}[L] \) & Expected total lifespan & \( \frac{1}{d_T} \frac{r + d_T}{r + d_H} \) \\
        \( \mathbb{E}[L_H] \) & Expected hunter lifespan & \( \frac{1}{r + d_H} \) \\
        \( \mathbb{E}[L_T] \) & Expected teacher lifespan & \( \frac{1}{d_T}\frac{r}{r+d_H} \) \\
        \( \E[L \mid T] \) & Expected lifespan conditional on survival to the post-reproductive stage & \( \frac{1}{r+d_H} + \frac{1}{d_T} \) \\
        \( \text{PrL} \coloneqq \mathbb{E}[L_T \mid T] \) & Expected post-reproductive lifespan & \( \frac{1}{d_T} \) \\
        \( \text{rPrL} \coloneqq \frac{\mathbb{E}[L_T \mid T]}{\mathbb{E}[L \mid T]} \) & Ratio of conditional expected post-reproductive lifespan to conditional expected total lifespan & \( \frac{r + d_H}{r + d_H + d_T} \) \\
        \( \text{PrR} \coloneqq \frac{\mathbb{E}[L_T]}{\mathbb{E}[L]} \) & Post-reproductive representation (unconditional), i.e., the fraction of expected lifetime spent in the teacher stage & \( \frac{r}{r + d_T} \equiv \taus \) \\
    \bottomrule
    \end{tabular*}
    \addtabletext{
    \begin{minipage}{\textwidth}
    A summary of the analysis of the one-type system, including equilibrium solutions and the expected values of various types of lifespans in a monomorphic population.
    We note that \( h \coloneqq h_1 / h_0 \).
    The final row states an exact identity: the fraction of expected lifetime spent in the post-reproductive stage equals the equilibrium fraction of the population that teaches. Derivations are given in Sec.~\ref{sec:monomorphic}.
    \end{minipage}
    }
    \label{tab:single_species}
\end{table*}

\FloatBarrier

\section{\label{sec:competition}Competition between types with different retirement rates (supplementary figures)}

\begin{figure}
    \centering
    \includegraphics[width=\linewidth]{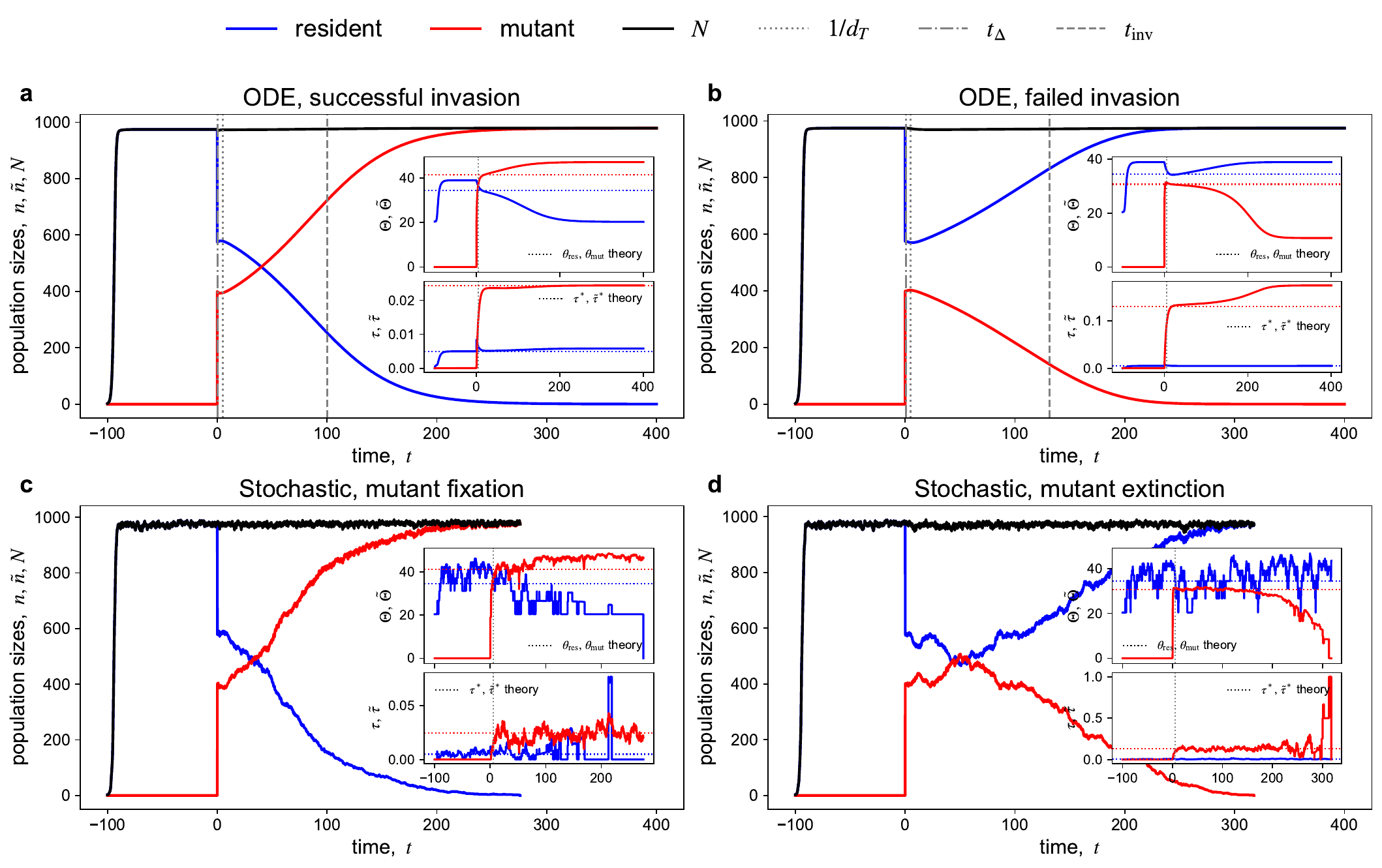}
    \caption{
    \textbf{A mutant type with a different retirement rate replaces the resident when its expected per-capita lifetime hunting return
    exceeds the resident's at the time of introduction, and is lost otherwise.}
    Time evolution of resident (blue), mutant (red), and total (black) population sizes following an invasion attempt at \( t = 0 \):
    (\pnl{a,c}) successful invasion (\( \thmut > \thres \)) and
    (\pnl{b,d}) failed invasion (\( \thmut < \thres \)).
    The top row (\pnl{a,b}) shows results from ODE simulations, and the bottom row (\pnl{c,d}) shows results from stochastic simulations.
    The approximate relaxation timescale for the proportion of teachers, \( \tau \), is indicated by the vertical dotted line (\( 1/d_T \));
    the ecological timescale by the vertical dash-dotted line (\( \tDel \));
    and the invasion timescale by the dashed line (\( \tinv \)).
    The insets show the time evolution of the expected per-capita lifetime hunting returns (\( \Theta \) and \( \tilde\Theta \), top) and the proportions of teachers (\( \tau \) and \( \taut \), bottom);  for their definitions, see Table~\ref{tab:symbols} of the main text.
    In insets, horizontal dotted lines show quasi-stationary values, \( \thres \) and \( \thmut \) (top) and \( \taus \) and \( \tauts \) (bottom); and the difference \( \thres - \thmut \) is used to predict which type goes extinct.
    The expected per-capita lifetime hunting returns of the resident and the mutant evaluated at \( t = 0 \) are given by \( \thres \coloneqq\theta(r;\nz) \) and \( \thmut \coloneqq\theta(\rt;\ntz) \), respectively.
    The prediction is valid only when \( \tinv \gg \tDel, 1/d_T \) for sufficiently small \( r \) and \( \rt \) (see \hyperref[sec:matmethods]{\emph{Materials and Methods}}\xspace of the main text); the invasion timescale is noticeably longer than the other two in both successful and failed invasion cases.
    In insets, the vertical dotted lines indicate the approximate reorganization timescale \( 1/d_T \), on which \( \Theta \) and \( \tilde\Theta \) quickly reach their quasi-stationary values (\( \thres \) and \( \thmut \)), indicated by the horizontal dotted lines.
    Initial conditions (at \( t = -100 \)) are given by \( (n_H, n_T, \nht, \ntt) = (2,0,0,0) \) across all panels, allowing the system to quickly equilibrate in the absence of mutant individuals.
    At \( t = 0 \), 400 of the resident hunters become mutant hunters, changing the population sizes from \( (n_H, n_T, \nht, \ntt) \) to \( (n_H - \ntz, n_T, \ntz, 0) \), where \( \ntz = 400 \).
    Parameters:
    \( K = 1000 \), \( d_H = 0.05 \), \( d_T = 0.2 \), \( b_0 = 1 \), \( \eta = 0.02 \), \( \nu = 0.2 \), \( h_0 = 1 \), \( h_1 = 3 \), \( r = 0.001 \);
    (\pnl{a,c}) \( \rt = 0.005 \) and (\pnl{b,d}) \( \rt = 0.03 \).
    The left panels (\pnl{a,c}) are the same as those shown in Fig.~\ref{fig:base_trajectories} of the main text.
    }
    \label{fig:base_trajectories_full}
\end{figure}

Figure~\ref{fig:base_trajectories_full} shows invasion dynamics for both successful and failed invasion.

\FloatBarrier

\section{\label{sec:predictor_derivation}Derivation of the predictor}

\subsection{Preliminaries}

We derive the predictor (selection coefficient) \( \Lambda \) using quasi-steady-state approximation.
First, the reorganizations of stage-structured groups are described by
\begin{align}
    \dot n_T = r n - (r + d_T)n_T, \quad
    \dnTt = \rt \nt - (\rt + d_T) \ntt,
\end{align}
which occur on the timescales of \( \sim \frac{1}{r + d_T}, \frac{1}{\rt + d_T} \), provided that the resident and mutant population sizes are in quasi-steady states.

Ecological dynamics are characterized by the time evolution of total population size \( N \), which is equivalently captured by remaining carrying capacity, denoted by \( \Delta \coloneqq K - N  \).
We find
\begin{align}
    \dot \Delta = \mathcal{A} - \mathcal{B} \Delta - \dot n_T - \dnTt,
\end{align}
where \( \mathcal{M} \coloneqq ab(F)n + \tilde a b(\tilde F)\tilde n \), \( a \coloneqq 1 - \tau \), \( \tilde a \coloneqq 1 - \taut \), and
\begin{align}
    \mathcal{A} \coloneqq a(r + d_H) n + \tilde a (\rt + d_H) \tilde n,
    \quad
    \mathcal{B} \coloneqq \frac{\mathcal{M}}{K}.
\end{align}

Demographic turnover, or invasion, is described by the frequency of residents, \( \rho = n/N \), which evolves according to the equation
\begin{align}
    \dot \rho = (\omega - \tilde\omega) \rho(1-\rho) + \frac{1}{N}[(1-\rho) \dot n_T - \rho \dnTt],
    \label{eq:invasion}
\end{align}
where the resident and mutant fitness, \( \omega \) and \( \tilde\omega \), are given by
\begin{align}
    \omega &= \lambda_H (1-\tau), \quad \lambda_H = b(F) \frac{\Delta}{K} - (r + d_H), \\
    \tilde \omega &= \tilde\lambda_H (1-\taut), \quad
    \tilde \lambda_H = b(\tilde F) \frac{\Delta}{K} - (\rt + d_H).
\end{align}

\subsection{Quasi-steady-state approximation}
Now suppose that reorganization and ecological dynamics occur much faster than demographic turnover.
Under this separation of timescales, the frequency of residents, \( \rho \) (and hence \( n, \nt \)), changes much more slowly than teacher proportions \( \tau \), \( \taut \), and the total population size \( N \), or equivalently, \( \Delta \).
Then, a joint quasi-steady state requires
\begin{align}
    \dot n_T \approx 0, \quad \dnTt \approx 0, \quad
    \dot \Delta \approx 0,
\end{align}
yielding
\begin{align}
    \taus &= \frac{r}{r + d_T}, \quad \tauts = \frac{\rt}{\rt + d_T}, \\
    \frac{\Delta^*}{K} &= \frac{a(r + d_H) n + \tilde a (\rt + d_H) \tilde n}{ab(F)n + \tilde a b(\Ft)\nt},
\end{align}
assuming that \( n \) and \( \nt \) are fixed at some values.
Since demographic turnover is slow, \( n \) and \( \nt \) do not change over the fast timescale, implying that \( n \approx \nz \) and \( \nt \approx \ntz \) hold over that timescale.
Therefore, over this fast timescale, we approximate \( F \) and \( \Ft \) by
\begin{align}
    F \approx F^\dagger \coloneqq F(\tau^*; n(0)), \quad \Ft \approx \Ft^\dagger \coloneqq F(\tilde\tau^*; \tilde n(0)).
\end{align}
Using these expressions, we approximate the growth rates \( \lambda_H \) and \( \tilde\lambda_H \) by
\begin{align}
    \lambda_H \approx
    \lambda_H^\dagger \coloneq \left. \frac{\tilde a \nt s(r, \rt)}{\mathcal{M}} \right|_{\dagger}, \
    \tilde\lambda_H \approx
    \tilde\lambda_H^\dagger \coloneq \left. \frac{- a n s(r, \rt)}{\mathcal{M}} \right|_{\dagger},
\end{align}
where
\(
    \left. \cdot \right|_{\dagger}
    \coloneqq
    \left. \cdot \right|_{F=F^\dagger, \Ft = \Ft^\dagger, a=a^*, \tilde a = \tilde a^*, s=s^\dagger, n=\nz, \nt=\ntz}
\)
indicates that the expression is evaluated at the quasi-steady state.
Likewise, in what follows, we use a superscript dagger to imply evaluation at the quasi-steady state.
Let \( \omega^\dagger \coloneqq \lambda_H^\dagger (1-\tau) \) and \( \tilde\omega^\dagger \coloneqq \tilde\lambda_H^\dagger (1-\taut) \).
Then we obtain
\begin{align}
    \Lambda(r,\rt) \coloneqq \omega^\dagger - \tilde\omega^\dagger
    = \left.\frac{a \tilde a N s(r, \rt)}{ab(F)n + \tilde a b(\Ft)\nt} \right|_{\dagger}, \label{eq:Lambda}
\end{align}
whose sign determines the invasion outcome.
Under the quasi-steady state the teacher-class terms in the equation for \( \dot\rho \) (Eq.~\ref{eq:invasion}) vanish, so \( \dot \rho \approx \Lambda \rho(1-\rho) \): the resident frequency follows a logistic equation with constant growth rate \( \Lambda \), and one type is excluded.

More specifically, the mutant relative fitness, defined by
\(
    s(r, \rt; n, \nt) = (\rt + d_H)b(F) - (r + d_H)b(\Ft),
\)
can be rewritten as \( s(r, \rt; n, \nt) \coloneqq (r + d_H)(\rt + d_H)\mathcal{S}(r, \rt; n, \nt) \), where
\(
    \mathcal{S}(r, \rt; n, \nt) \coloneqq \frac{b(F)}{r+d_H} - \frac{b(\Ft)}{\rt + d_H},
\)
which decomposes the effects of resident and mutant traits on the invasion outcome into two independent terms.
Since all terms in Eq.~\ref{eq:Lambda} except the mutant relative fitness \( s \) are positive, invasion outcomes are determined entirely by \( s \).
Using \( \theta(r; n) \), which denotes the expected per-capita lifetime hunting return of a population of size \( n \) with retirement rate \( r \),
we can represent the resident and mutant expected per-capita lifetime hunting returns evaluated at \( t = 0 \) by
\(
    \thres = \theta(r; \nz)%
\) and
\(
    \thmut = \theta(\rt; \ntz)
\), respectively,
yielding
\(
    \mathcal{S}(r, \rt; \nz, \ntz) \propto \thres - \thmut
\),
which completes the derivation of Eq.~\ref{eq:prediction} in the main text.

\begin{figure}
    \centering
    \includegraphics[width=\linewidth]{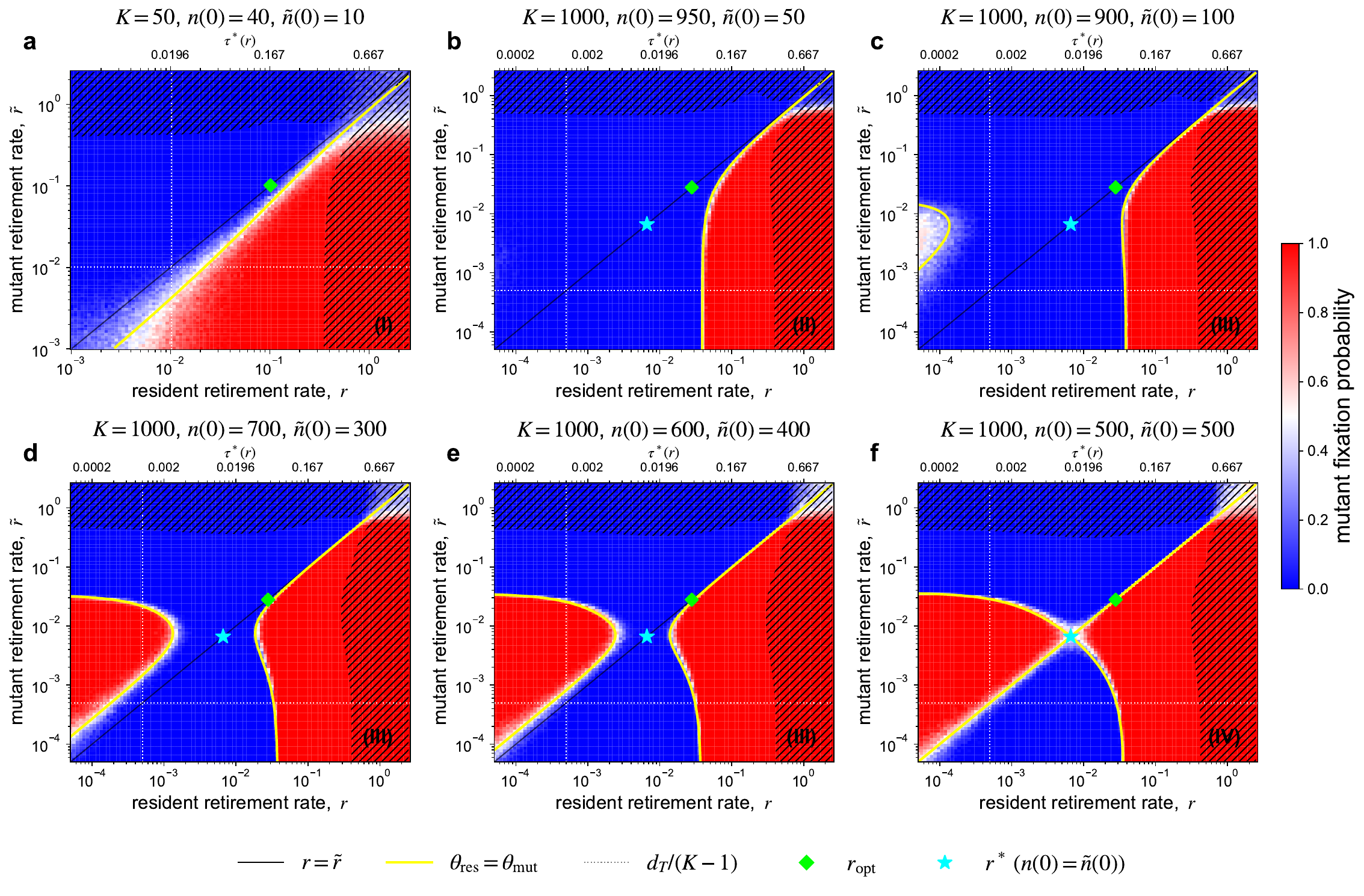}
    \caption{
    \textbf{Prediction works well when the timescale separations hold.}
    These pairwise invasibility plots are the same as those shown in Fig.~\ref{fig:pip_panels} of the main text, except for the black shading.
    The shading represents the regions where \( \max(\tDel, \tT, \tTt) > \tinv \).
    The timescale separation conditions required for quasi-steady approximation, i.e., \( \tinv \gg \tDel, \tT, \tTt \), are derived in \hyperref[sec:matmethods]{\emph{Materials and Methods}}\xspace.
    When the timescale separation fails, prediction can fail, especially when the predictor \( \Lambda \) is close to zero (see the upper right corners of the panels), i.e., selection is nearly neutral.
    Note that the yellow curves denote the zero level set \( \Lambda = 0 \).
    However, in some off-diagonal regions where at least one of the timescale separations is violated due to large \( r \) or \( \rt \), prediction still can work; we suspect this is because selection coefficient---the difference between resident fitness and mutant fitness---is so large at the time of mutant introduction that the expected lifetime per-capita hunting return can still predict the invasion outcome reliably.
    The black shaded regions are asymmetric with respect to the diagonal \( r = \rt \) when the discrepancy between mutant and resident founding population sizes is large.
    However, the shaded regions become symmetric as the ratio of the founding population sizes approaches one, yielding a unique evolutionarily stable retirement rate (panel \pnl{f}).
    Parameters:
    \cy{\( d_H = 0.025 \)}, \( d_T = 0.5 \), \( b_0 = 0.5 \), \( \eta = 0.02 \), \( \nu = 0.2 \), \( h_0 = 1 \), \( h_1 = 3 \).
    Other plotting details are the same as those in Fig.~\ref{fig:pip_panels} of the main text.
    }
    \label{fig:pip_panels_shading}
\end{figure}

The reduction above is valid if the following timescale separation holds at \( t = 0 \):
\begin{align} \label{si:eq:timescale_separation}
    \tinv \gg \tT, \quad \tinv \gg \tTt, \quad \tinv \gg \tDel,
\end{align}
where the reorganization (\( \tT, \tTt \)), ecological (\( \tDel \)), and invasion (\( \tinv \)) timescales are given by
\begin{align} \label{si:eq:timescales}
    \tT \sim \frac{1}{r + d_T}, \
    \tTt \sim \frac{1}{\rt + d_T}, \
    \tDel \sim \frac{1}{\mathcal{B}^\dagger}, \
    \tinv \sim \frac{1}{|\Lambda|},
\end{align}
which are fully determined upon introduction of mutants.
That is, the approximation is valid if \( |\Lambda| \ll \min \{ r + d_T, \rt + d_T, \mathcal{B}^{\dagger} \} \) (weak selection).
We show that prediction can fail when these timescale separations (Eq.~\ref{si:eq:timescale_separation}) are violated (see Fig.~\ref{fig:pip_panels_shading}).

\subsection{\label{subsec:theta_analysis}Analysis of the fitness function \( \theta(r;n) \)}

In the Results of the main text, we applied the sign condition of Eq.~\ref{eq:prediction} pair by pair, evaluating it for each combination of resident and mutant retirement rates.
Here we analyze the predictor itself, to explain the patterns it produces: why the stable retirement rates form an interval, and how the founding population sizes of the resident and the mutant set its endpoints, \( \rmin \) and \( \rmax \) (Fig.~\ref{fig:pip_panels}).

Both terms of the prediction \( \thres - \thmut \) are values of a single function, \( \theta(r;n) \), the expected per-capita lifetime hunting return at the quasi-steady state, evaluated at each type's own retirement rate and founding population size.
The shape of this function in its two arguments therefore determines every invasion outcome, and with it the evolutionarily stable range.
We prove the following two properties:
(i) \( \theta(\cdot \,; n) \) pointwise dominates \( \theta(\cdot \,;m) \) provided \( n > m \).
(ii) There exists a \( \nth > 0 \) such that \( \theta(r;n) \) attains its maximum at a positive retirement rate for all \( n > \nth \), in which case we say that \( \theta \) has a \emph{unique interior maximizer}.

Our first step is to show that \( \theta(\,\cdot\,; n) \) pointwise dominates \( \theta(\, \cdot \,; m) \) provided \( n > m \);
in other words, \( \theta(r;n) > \theta(r; m) \) for all \( r > 0 \) if \( n > m \).
Notice that \( W(\tau;n) \coloneqq [h_0 + h_1(\eta + \nu \tau n)]/(1+\eta+\nu\tau n)\), the weighted average of \( h_0 \) and \( h_1 > h_0 \), increases with \( n \).
Let \( V(r) \coloneqq d_T/[(r+d_H)(r+d_T)] \)
so that \( \theta(r;n) = V(r)W(\taus(r);n) \).
We establish that \( \theta(r;n)-\theta(r;m) = V(r)\left[W(\taus(r);n) - W(\taus(r);m)\right] > 0 \) for \( n > m \) at all \( r > 0 \).

Next, we prove that there exists \( \nth > 0 \) such that \( \theta(r;n) \) is uniquely interior-maximized in \( r \) for \( n > \nth \) and monotonically decreasing for \( n < \nth \).
We find that
\begin{align}
    \frac{\partial \theta}{\partial r}
    = d_T \frac{P(r)}{D(r)}, \ P(r) = Q_3 r^3 + Q_2 r^2 + Q_1 r + Q_0
\end{align}
where \( D(r) > 0 \) and \( Q_k \) (\( k = 0, \ldots, 3 \)) are constants independent of \( r \).
We find that \( Q_1 \), \( Q_2 \), and \( Q_3 \) are all negative regardless of the parameter values, but \( Q_0 \) can be positive or negative (see \ref{sec:rES_existence} for their explicit expressions).
Therefore, if \( Q_0 < 0 \), \( \theta \) monotonically decreases in \( r > 0 \), whereas if \( Q_0 > 0 \), \( \theta \) increases and then decreases in \( r \), attaining its maximum at some intermediate \( r \).
The exact expression for \( Q_0 \) is given by \( Q_0 = (h_1 - h_0) \nu n - \frac{d_H + d_T}{d_H} (h_0 + h_1 \eta)(1 + \eta) \).
Therefore,
\( \theta(r;n) \) is monotone decreasing in \( r \) if \( n < \nth \) and uniquely interior-maximized if \( n > \nth \), where the critical population size is given by
\begin{align}
    \nth = \zeta \nc, \quad
    \zeta = 1 + \frac{d_T}{d_H}, \quad
    \nc = \frac{(h_0 + h_1 \eta)(1+\eta)}{(h_1-h_0)\nu},
\end{align}
and \( \nc \) is the threshold population size for teaching to be beneficial at all, derived previously in a simpler model with the teacher proportion set exogenously \citep{Goto2026}.

Both properties of the expected per-capita lifetime hunting return \( \theta(r;n) \), one governing its
dependence on population size \( n \) and the other its dependence on the retirement rate
\( r \), have plain interpretations.
Pointwise dominance says that a larger population obtains more return per individual over a lifetime than a smaller one does, at every positive retirement rate, because the larger population contains more teachers at the same teacher proportion.
The second property says that, once a population is large enough, some intermediate retirement rate maximizes the expected per-capita lifetime hunting return: retiring too late leaves hunters unskilled, for want of teachers, and retiring too early leaves too little of life spent hunting.
Below the threshold population size \( \nth \), \( \theta(r;n) \) decreases monotonically with the retirement rate, and its maximum is attained at the boundary \( r = 0 \).

The thresholds $\nth$ and $\nc$ differ by the factor \( 1 + d_T/d_H \), which exceeds one, so a population must be strictly larger to evolve retirement than to benefit from teaching. This factor has a direct interpretation.  Assigning an individual to teach costs the population that individual's current hunting returns;  retiring an individual costs, in addition, all the hunting returns it would have contributed over the remainder of its hunting life.
The size of the additional cost is set by the ratio of the two mortality rates.
When teachers are short-lived relative to hunters, \( d_T \gg d_H \), the teaching stage that follows retirement is brief while the hunter stage it replaces is long, the factor \( 1 + d_T/d_H \) is then large, and retirement requires a much larger population than teaching does.
When teachers are instead long-lived relative to hunters, the factor is close to one and the two thresholds are close together.

\subsection{\label{subsec:define_RES}Evolutionarily stable range of retirement rates}

We have defined \( \RES \) to be the interval \( [\rmin, \rmax] \), where \( \rmax \geq \rmin \geq 0 \) are the solutions to the equation
\(
    \theta(r;n) = \max_{\rt \geq 0} \theta(\rt; \nt).
\)
By construction,
\begin{align*}
    r \in \RES \ \iff
    \ \text{\( r \) cannot be invaded by any mutant},
\end{align*}
given \( \nt \) and all other parameters.
The two properties of \( \theta \) established above guarantee that this range is nondegenerate whenever the initial resident population size $n$ exceeds the critical size $\nth$ (and provided the mutant initial size is less than the resident's).
Moreover, given such $n$ and $\nt$, we can solve analytically for the lower and upper limits of $\RES$ as the two distinct roots of $\theta(r;n) = \max_{\rt \geq 0} \theta(\rt; \nt)$ (Fig.~\ref{fig:theta_panels}\pnl{c}).
For notational convenience, we write \( \rp \) and \( \rmn \) to denote \( \rmax \) and \( \rmin \), respectively; the shorthand notation \( \rpm \) refers to both \( \rmax \) and \( \rmin \) in order.
For its graphical representation and its relevance to invasion outcomes, see Fig.~\ref{fig:theta_panels} and Table~\ref{tab:regimes}.
We note that \( \rmin \) and \( \rmax \) exist because \( \theta \) exhibits pointwise dominance at all \( r > 0 \); \( \theta \) is either interior-maximized or monotone decreasing in \( r \); and \( \theta \to \theta^* \coloneqq \frac{1}{d_H}\frac{h_0 + h_1 \eta}{1+\eta} \) as \( r \to 0 \), and \( \theta \to 0 \) as \( r \to \infty \), irrespective of \( n > 0 \).

\subsection{\label{subsec:non_CS}Evolutionarily stable retirement rates are not convergence stable}

We show that evolutionarily stable retirement rates, \( \RES \), are generically not convergence stable, unless the founding population sizes are equal (\( \nz = \ntz \)) or no retirement is the only evolutionarily stable state (that is, \( \RES = \{ 0 \} \)).

Pointwise dominance has a consequence for how the retirement rate can evolve.
Let the resident retire at a fixed rate \( r > 0 \), and let the founding population sizes satisfy
\( \nz > \ntz \).
Define
\begin{align}
    g(\rt) \coloneqq \theta(\rt; \ntz) - \theta(r; \nz),
\end{align}
so that the mutant invades if and only if \( g(\rt) > 0 \).
At \( \rt = r \), pointwise dominance gives
\( g(\rt) = \theta(r;\ntz) - \theta(r;\nz) < 0 \).
Moreover \( \theta(\,\cdot\,;\ntz) \) is continuous for \( \rt \geq 0 \), being a rational
function of \( \rt \) whose denominator
\( (\rt+d_H)(\rt+d_T)\left[1+\eta+\nu\taus(\rt)\ntz\right] \) is strictly positive there.
Hence \( g \) is continuous with \( g(r) < 0 \), and there exists \( \delta > 0 \) such that
\( g(\rt) < 0 \) for every \( \rt \in (r-\delta,\, r+\delta) \).
No mutant whose retirement rate lies within \( \delta \) of the resident's can invade, so the
retirement rate cannot evolve by the accumulation of arbitrarily small changes.

The argument above uses \( r > 0 \) and \( \nz > \ntz \), and it fails when either is violated.
At \( r = 0 \) the resident maintains no teachers, \( \taus(0) = 0 \), and
\( \theta(0;n) \) takes the same value for every \( n \), so dominance is not strict and
\( g(0) = 0 \).
When \( \ntz = \nz \) the two types share one function, \( \theta(\,\cdot\,;\nz) \), so again
\( g(r) = 0 \) rather than \( g(r) < 0 \), and continuity yields no interval on which \( g \)
is negative; a resident then resists invasion only if \( r \) maximizes
\( \theta(\,\cdot\,;\nz) \), and every other rate is invaded by rates arbitrarily close to it,
which is the classical scenario in adaptive dynamics.

\FloatBarrier

\clearpage

\section{\label{sec:rES_existence}A proof of the existence of a unique nontrivial range of evolutionarily stable retirement rates}

Here we show that the expected per-capita lifetime hunting return of a population of size \( n \) with retirement rate \( r \), given by
\begin{align}
    \theta(r;n) = \frac{F(\taus(r);n)}{r + d_H},
\end{align}
exhibits pointwise dominance with respect to \( n \) and is non-monotonic in \( r \) for sufficiently large \( n \).

\subsection{Pointwise dominance}
First we prove that \( \theta(\,\cdot \, ;n) \) dominates pointwise \( \theta(\,\cdot \,; m)\) for \( n > m > 0 \).
Let
\begin{align}
    V(r) \coloneqq \frac{d_T}{(r+d_H)(r+d_T)}, \quad W(\tau;n) \coloneqq \frac{h_0 + h_1 (\eta + \nu\tau n )}{1+\eta + \nu \tau n}.
\end{align}
Given
\begin{align}
    \taus(r) = \frac{r}{r+d_T},
\end{align}
we rewrite the fitness function \( \theta \) as
\begin{align}
    \theta(r;n) = V(r) W(\taus(r);n).
\end{align}
Since
\begin{align}
    \frac{\partial W}{\partial n} = \frac{(h_1-h_0) \nu \tau}{(1+\eta+\nu\tau n)^2} > 0,
\end{align}
\( W(\tau;n) \) is monotone increasing in \( n > 0 \).
Therefore, we find that, for \( n > m > 0 \) and for all \( r > 0 \),
\begin{align}
    W(\tau(r); n) > W(\tau(r);m),
\end{align}
establishing
\begin{align}
    \theta(r;n) - \theta(r;m) = V(r) \left[W(r;n) - W(r;m) \right] > 0,
\end{align}
for all \( r > 0 \), completing the proof.

\subsection{\( \theta \) is either unimodal or monotonically decreasing in the retirement rate \( r \)}
Next we prove that \( \theta \) either has a single interior peak or is monotone decreasing in \( r \geq 0 \).
Reorganizing \( W(\taus(r);n) \) in terms of \( r \) and \( n \) gives
\begin{align}
    W(\taus(r);n) = \frac{[h_0 + h_1(\eta + \nu n)]r + (h_0 + h_1 \eta) d_T}{(1 + \eta +\nu n)r + (1 + \eta) d_T}.
\end{align}
Let
\begin{align}
    a_1 = h_0 + h_1(\eta + \nu n), \quad a_0 = (h_0 + h_1 \eta) d_T, \quad
    b_1 = 1 + \eta +\nu n, \quad b_0 = (1 + \eta) d_T,
\end{align}
so that
\begin{align}
    W(\taus(r);n) = \frac{a_1 r + a_0}{b_1 r + b_0}.
\end{align}
Then,
\begin{align}
    \theta(r;n) = V(r)W(\taus(r);n) = \frac{d_T}{(r+d_H)(r+d_T)} \frac{a_1 r + a_0}{b_1 r + b_0}.
\end{align}
Let
\begin{align}
    A &= a_1 r + a_0, \\
    B &= (r+d_H)(r+d_T)(b_1 r + b_0).
\end{align}
Reorganizing \( B \) in powers of \( r \) gives
\begin{align}
    B = \alpha_3 r^3 + \alpha_2 r^2 + \alpha_1 r + \alpha_0,
\end{align}
where
\begin{align}
    \alpha_3 &= b_1, \\
    \alpha_2 &= (d_H + d_T) b_1 + b_0, \\
    \alpha_1 &= (d_T + d_H) b_0 + d_H d_T b_1, \\
    \alpha_0 &= d_H d_T b_0.
\end{align}
Then,
\begin{align}
    A' &= a_1 \\
    B' &= 3 \alpha_3 r^2 + 2 \alpha_2 r + \alpha_1,
\end{align}
where the prime denotes the derivative with respect to \( r \).
Using these results obtained above, we find
\begin{align}
    \frac{\partial \theta}{\partial r}
    = d_T \frac{A'B - AB'}{B^2}.
\end{align}
Therefore, the sign of \( \theta' \) is determined by
\begin{align}
    A'B - AB'
    &= a_1 (\alpha_3 r^3 + \alpha_2 r^2 + \alpha_1 r + \alpha_0) - (a_1 r + a_0)(3\alpha_3 r^2 + 2\alpha_2 r + \alpha_1)  \\
    &= - 2 a_1 \alpha_3 r^3 - (a_1 \alpha_2 + 3 a_0\alpha_3) r^2 - 2 a_0 \alpha_2 r + a_1 \alpha_0 - a_0 \alpha_1.
\end{align}
Let
\begin{align}
    Q_0 \coloneqq a_1 \alpha_0 - a_0 \alpha_1.
\end{align}
If \( Q_0 < 0 \), then \( \theta \) is monotonically decreasing for all \( r \geq 0 \).
On the other hand, if \( Q_0 > 0 \), \( \theta \) is unimodal in \( r \):
\( \theta' > 0 \) for small \( r \), \( \theta' < 0 \) for large \( r \), and \( \theta' = 0 \) for some unique intermediate \( r \) (because of the intermediate value theorem and the monotonicity of the cubic).
This guarantees that the fitness function \( \theta \) has a unique peak at some intermediate \( r \), i.e., this cubic has exactly one positive real root.
After some calculations, we find
\begin{align}
    Q_0
    &= (h_1 - h_0) \nu n - \frac{d_T + d_H}{d_H} (h_0 + h_1 \eta)(1+\eta).
\end{align}
Thus, the condition \( Q_0 > 0 \) holds if and only if \( n > \nth \), where
\begin{align}
    \nth = \frac{d_T + d_H}{d_H} \frac{(h_0 + h_1 \eta)(1+\eta)}{(h_1-h_0)\nu}.
\end{align}
Recalling the expression for \( \nc \), we find
\begin{align}
    \nth = \frac{d_T + d_H}{d_H} n_{c}, \label{eq:nc_new}
\end{align}
where \( \nc \) denotes the critical population size derived from the earlier simpler model (hunter-teacher model; see the main text or \citep{Goto2026} for more details), which identifies the minimum population size above which teaching is beneficial to the per-capita hunting return rate of a population (\( \nth > \nc \)).

Given the shape of the curve \( \theta(r;n) \) with respect to \( r \), we now see why evaluating the partial derivative \( \partial \theta /\partial r \) at \( r = 0 \) is sufficient to derive the critical population size \( \nth \), as it leaves only the zeroth-order term
\begin{align}
    \left.\frac{\partial \theta}{\partial r} \right|_{r=0} = Q_0.
\end{align}

\subsection{The evolutionarily stable range in each regime}
Let \( n = \nz \geq \nt = \ntz \) and \( M(\nt) \coloneqq \max_{\rt \geq 0} \theta(\rt; \nt) \), so that \( \RES = \{ r \geq 0 : \theta(r; n) \geq M(\nt) \} \).
Three facts established above determine the evolutionary stable set:
pointwise dominance, \( \theta(r;n) > \theta(r;\nt) \) for all \( r > 0 \) when \( n > \nt \);
\( \theta(\,\cdot\,;n) \) is interior-maximized if \( n > \nth \) and strictly decreasing otherwise;
and \( \theta(0;n) = \theta_0 \coloneqq \frac{h_0 + h_1 \eta}{(1+\eta) d_H} \) for every \( n \), because \( \taus(0) = 0 \).
In addition, \( \theta(r;n) \to 0 \) as \( r \to \infty \).
We exclude the boundary cases \( n = \nth \) and \( \nt = \nth \). The four regimes and shapes of $\RES$ follow from these properties:

\emph{Regime I}, \( \nth > n \). Both curves are strictly decreasing, so \( M(\nt) = \theta_0 = \theta(0;n) \) and \( \theta(r;n) < \theta_0 \) for every \( r > 0 \). Hence \( \RES = \{0\} \).

\emph{Regime II}, \( n > \nth > \nt \). The mutant curve is strictly decreasing, so \( M(\nt) = \theta_0 = \theta(0;n) \) and \( 0 \in \RES \). The resident curve increases from \( \theta_0 \) to its interior maximum and then decreases to zero, so \( \theta(r;n) \geq \theta_0 \) exactly on an interval \( [0, \rmax] \) with \( \rmax > 0 \). Hence \( \RES = [0, \rmax] \).

\emph{Regime III}, \( n > \nt > \nth \). Let \( \rts \) be the interior maximizer of \( \theta(\,\cdot\,;\nt) \), so \( M(\nt) = \theta(\rts;\nt) > \theta_0 \).
Pointwise dominance gives \( \theta(\rts; n) > M(\nt) \), so \( \rts \in \RES \). Since \( \theta(0;n) = \theta_0 < M(\nt) \), \( \theta(r;n) \to 0 \), and \( \theta(\,\cdot\,;n) \) is interior-maximized, the set \( \{ r : \theta(r;n) \geq M(\nt) \} \) is an interval \( [\rmin, \rmax] \) with \( 0 < \rmin < \rts < \rmax \).

\emph{Regime IV}, \( n = \nt > \nth \). The two curves coincide, so \( \RES = \{ r : \theta(r;n) \geq \max_{r} \theta(r;n) \} = \{ \rs \} \), the unique interior maximizer.

\subsection{\label{subsec:jump_size}A smaller founding mutant population requires a larger change in retirement rate}
The change in retirement rate that a mutant needs in order to invade increases as the founding size of the mutant population decreases.
Consider Regime III, \( n > \nt > \nth \), and a resident retirement rate below the lower endpoint of the evolutionarily stable range, \( 0 < r < \rmin \), so that some mutant retirement rates invade the resident.
For a mutant population founded at size \( m \leq \nt \), let
\begin{align*}
    \mathcal{I}(m) \coloneqq \{ \rt \geq 0 : \theta(\rt; m) > \theta(r;n) \}
\end{align*}
be the set of mutant retirement rates that invade the resident, and let \( \delta_{\min}(m) \coloneqq \inf \{ |\rt - r| : \rt \in \mathcal{I}(m) \} \) be the smallest change in retirement rate that permits invasion, with \( \delta_{\min}(m) = \infty \) when \( \mathcal{I}(m) \) is empty.
The quantity \( \delta_{\min}(m) \) is the largest \( \delta \) in the corollary of Section~\ref{subsec:non_CS}.
We show that, for founding mutant sizes \( m_1 < m_2 \leq \nt \), either \( \mathcal{I}(m_1) \) is empty or \( \delta_{\min}(m_1) > \delta_{\min}(m_2) \).

Every mutant retirement rate that invades a resident rate \( r < \rmin \) exceeds \( r \).
In Regime III the maximizer of the resident curve \( \theta(\,\cdot\,;n) \) lies strictly between \( \rmin \) and \( \rmax \), so the resident curve is strictly increasing on \( [0, \rmin] \).
At \( \rt = 0 \), \( \theta(0;m) = \theta_0 = \theta(0;n) < \theta(r;n) \).
For \( 0 < \rt \leq r \), pointwise dominance and the increase of the resident curve give \( \theta(\rt;m) < \theta(\rt;n) \leq \theta(r;n) \).
Hence \( \mathcal{I}(m) \subset (r, \infty) \), and \( \delta_{\min}(m) = \inf \mathcal{I}(m) - r \).

The sets of invading mutant rates are nested.
For \( m_1 < m_2 \) and every \( \rt > 0 \), pointwise dominance gives \( \theta(\rt;m_1) < \theta(\rt;m_2) \), so \( \mathcal{I}(m_1) \subseteq \mathcal{I}(m_2) \) and \( \delta_{\min}(m_1) \geq \delta_{\min}(m_2) \).

The inequality between the smallest invading changes is strict.
Suppose \( \mathcal{I}(m_1) \) is nonempty, so that \( \mathcal{I}(m_2) \) is nonempty as well, and let \( \hat{r} \coloneqq \inf \mathcal{I}(m_2) > r \).
Because \( \theta(\,\cdot\,;m_2) \) is continuous for \( \rt \geq 0 \) (Section~\ref{subsec:non_CS}), the mutant return at this infimum equals the resident return, \( \theta(\hat{r};m_2) = \theta(r;n) \).
Pointwise dominance then gives \( \theta(\hat{r};m_1) < \theta(\hat{r};m_2) = \theta(r;n) \), and continuity of \( \theta(\,\cdot\,;m_1) \) gives an \( \varepsilon > 0 \) such that \( \theta(\rt;m_1) < \theta(r;n) \) for every \( \rt \in [\hat{r}, \hat{r} + \varepsilon) \).
Since \( \mathcal{I}(m_1) \subseteq \mathcal{I}(m_2) \subset [\hat{r}, \infty) \), it follows that \( \mathcal{I}(m_1) \subset [\hat{r} + \varepsilon, \infty) \), and so \( \delta_{\min}(m_1) \geq \delta_{\min}(m_2) + \varepsilon > \delta_{\min}(m_2) \).

When the founding mutant population lies below the threshold, \( m < \nth \), no mutant invades at all.
The mutant curve \( \theta(\,\cdot\,;m) \) is then strictly decreasing, so \( \theta(\rt;m) \leq \theta_0 < \theta(r;n) \) for every \( \rt \geq 0 \), and \( \mathcal{I}(m) \) is empty.
Thus, for a fixed resident retirement rate below \( \rmin \), a smaller founding mutant population requires a strictly larger change in retirement rate in order to invade, and a founding mutant population below \( \nth \) cannot invade with any retirement rate.
The argument uses only pointwise dominance, the single-peakedness of the resident curve, and continuity; it requires \( r > 0 \), where pointwise dominance is strict.

\clearpage

\section{\label{sec:n_crit_alternative}An alternative derivation of the critical population size \( \nth \)}

\subsection{Static vs life-history views}

In Sec.~\ref{subsec:theta_analysis}, we provide an intuition for the factor \( \zeta = 1 + d_T/d_H \) that relates the two population thresholds by \( \nth = \zeta \nc \).
This intuition can be formalized more precisely.
Here we provide an alternative way to derive the factor \( \zeta \) and reveal its origin more systematically.
The earlier hunter-teacher model of optimal allocation \citep{Goto2026} is based on a \emph{static} view in which a population seeks to maximize the per-capita hunting return rate \( F \).
On the other hand, our model of retirement and oblique teaching is based on a \emph{lifetime-aware} view where the per-capita hunting return \( \theta \) accumulated over one's lifetime is to be maximized.
This discrepancy in perspective can be resolved once we describe the life-history view in an \emph{effective} static allocation view, and that is where the factor comes in.
Therefore,
instead of differentiating \( \theta \) directly as was done in Sec~\ref{sec:rES_existence}, we aim to find the critical size \( \nth \) by developing a mapping from our model of retirement to the prior model of optimal teaching.

When making this transformation, there are a few key differences that need to be accounted for.
In the static view, individuals are either hunters for life or teachers for life, whereas in the life-history view, all teachers are born as hunters and later retire from hunting and serve as teachers.
Therefore, the hunting period (as well as the teaching period) in the life-history view needs to be accounted for in the effective static allocation view that we develop below.

\subsection{Lifetime-aware teacher representation}
We first convert the \emph{snapshot} teacher proportion \( \tau \) into a \emph{lifetime-aware} teacher representation \( \taueff \), which denotes the proportion of individuals who are either currently teachers or eventually become teachers.
We find
\begin{align}
    \taueff = \tau + (1-\tau) \Pr(H \to T), \label{eq:taueff}
\end{align}
where \( \Pr(H \to T) = r/(r+d_H) \) denotes the probability that a hunter becomes a teacher at some point in life.
Substituting \( \tau = \taus(r) \), we obtain
\begin{align}
    \frac{\taueff}{\taus} = 1 + \frac{d_T}{r + d_H}.
\end{align}
Alternatively, this quantity can be found by calculating the expected total lifespan of an individual who eventually becomes a teacher relative to the period they spend teaching, i.e.,
\begin{align}
    \frac{\E[L \mid T]}{\E[L_T \mid T]} = \frac{\frac{1}{r+d_H} + \frac{1}{d_T}}{\frac{1}{d_T}} = 1 + \frac{d_T}{r + d_H}.
\end{align}
Therefore, the ratio of the inflated lifetime-aware teacher representation in an effective static allocation view to the actual teacher proportion is given by
\begin{align}
    \zeta_r \coloneqq \frac{\taueff}{\taus} \equiv \frac{\E[L \mid T]}{\E[L_T \mid T]} = 1 + \frac{d_T}{r + d_H}.
\end{align}

\subsection{Mapping the life-history view to an effective static view}

Now we transform the key variables \( (n, \taus, \theta) \) associated with the life-history view into alternative variables \( (\neff, \taueff, F) \) arising in an effective static view.
More specifically, we link \( F(\taueff, \neff) \) and \( \theta(r; n) = F(\taus(r);n)/(r + d_H) \) by enforcing \( F = \theta/L_0 \), where \( L_0 \) is a time constant to make the units consistent on both sides of the equation.
To this end,
because the mean hunting return per hunter, \( W(\tau;n) \coloneqq \frac{h_0+h_1(\eta+\nu\tau n)}{1+\eta +\nu\tau n} \), is a function of the absolute number of teachers \( n \tau \), we first impose
\begin{align}
    \neff \taueff = n \taus.
\end{align}
Then we also impose
\begin{align}
    1 - \taueff = \frac{1}{L_0} \frac{1 - \taus}{r+d_H}.
\end{align}
Note, at \( r = 0 \), we must have \( \taueff = 0 \) and \( \taus = 0 \), thereby yielding
\begin{align}
    L_0 = \frac{1}{d_H},
\end{align}
which is the baseline hunter lifespan in the absence of retirement.
Therefore, we obtain
\begin{align}
    \taueff = 1 - (1 - \taus) \frac{d_H}{r + d_H},
\end{align}
which equals \( \taus + (1-\taus) \Pr(H \to T) \) from Eq.~\ref{eq:taueff}.
In other words, we just constructed a mapping \( \taus \mapsto \taueff \) such that \( F = d_H \theta \).
Using this mapping, we can explicitly relate \( n \) to \( \neff \) via
\begin{align}
    n = \frac{\taueff}{\taus} \neff
    = \zeta_r \neff.
\end{align}
In the static allocation model, we already know that the optimal proportion of teachers becomes positive if and only if the population size exceeds a threshold \( \nc \) \citep{Goto2026}, i.e.,
\begin{align}
    \tauopt > 0 \iff \neff > \nc.
\end{align}
Therefore, taking the limit \( r \downarrow 0 \), we obtain
\begin{align}
    \nth = \zeta \nc,
\end{align}
where \( \zeta \coloneqq \lim_{r \downarrow 0} \zeta_r = 1 + d_T/d_H \) is the life-history scaling factor, naturally arising in the mapping \( \neff \mapsto n = \zeta_r \neff \) at the onset of retirement.

\clearpage
\section{\label{sec:partials}Dependence of the evolutionarily stable range of retirement rates on model parameters}

We systematically analyze how the stable range of retirement rates, \( \RES = [\rmn, \rp] \), changes with model parameters. The results are summarized in Table~\ref{tab:partials}.
We restrict our analysis to the most general case where \( \rmax > \rmin > 0 \), i.e., where \( \nz > \ntz > \nth \).

\subsection{\label{subsubsec:preliminaries}Preliminaries}

Let \( p \) denote an arbitrary model parameter.
In what follows, we write \( \theta = \theta(r; n, p) \) to emphasize the dependence of \( \theta \) on \( p \in \{ d_H, d_T, \eta, \nu, h, K \} \).
Below we identify \( n \) and \( \nz \) for notational convenience when desired; the same goes for \( \nt \) and \( \ntz \)).

\subsubsection{A general formula for the partial derivatives \( \partial \rpm/\partial p \)}

Our goal is to determine the signs of the partial derivative \( \frac{\partial \rpm}{\partial p} \) for each \( p \).
To do so, we make use of the following relationship:
\begin{align}
    \theta(\rpm; n, p) = \theta(\rt^\ast; \nt, p), \label{eq:rpm}
\end{align}
where
\begin{align}
    \rt^\ast \coloneqq \argmax_{\rt > 0} \theta(\rt; \nt, p).
\end{align}
We find
\begin{align}
    \frac{\partial}{\partial p} \theta(\rpm; n, p)
    = \frac{\partial }{\partial \rpm} \theta(\rpm; n, p) \frac{\partial \rpm}{\partial p} + \frac{\partial}{\partial p} \theta(\rpm; n, p)
    = \frac{\partial}{\partial p} M(\nt, p).
\end{align}
Provided \( \frac{\partial \theta}{\partial \rpm} \neq 0 \), we find
\begin{align}
    \frac{\partial \rpm}{\partial p} =
    \frac{\frac{\partial}{\partial p} M(\nt, p) - \frac{\partial}{\partial p} \theta(\rpm; n, p)}{\frac{\partial }{\partial \rpm} \theta(\rpm; n, p)}.
\end{align}
By applying the chain rule, we obtain
\begin{align}
    \frac{\partial}{\partial p} M(\nt, p) = \frac{\partial}{\partial p} \theta(\rt^\ast(p); \nt; p)
    = \underbrace{\frac{\partial \theta(\rt^\ast; \nt, p)}{\partial r}}_{=0 \text{ (by definition)}} \frac{\partial \rt^\ast}{\partial p} + \frac{\partial \theta}{\partial p}
    = \left. \frac{\partial}{\partial p}\theta(\rt; \nt, p) \right|_{\rt = \rt^\ast}
\end{align}
because \( \rts \) is the interior maximizer of \( \theta(\rt; \nt, p) \).
Therefore, we establish
\begin{align}
    \frac{\partial \rpm}{\partial p}
    = \frac{\theta_p(\rt^\ast; \nt^\ast, p) - \theta_p(\rpm; n, p)}{\theta_r(\rpm; n, p)},  \label{eq:drpm_dp}
\end{align}
where we used subscripts to denote partial derivatives to simplify the notation.
In addition, provided \( \rp > \rmn \geq 0 \), we find
\begin{align}
    \left. \frac{\partial \theta(r; n, p)}{\partial r} \right|_{r = \rp} < 0, \quad
    \left. \frac{\partial \theta(r; n, p)}{\partial r} \right|_{r = \rmn} > 0. \label{eq:dtheta_dr}
\end{align}

\subsubsection{Relationships between \( \mmn \), \( \mts \), and \( \mpl \)}
Since \( \RES \) is nondegenerate and unique for large founding population sizes \( \nz > \ntz > \nth \), we have, by construction,
\begin{align}
    \rmn < \rts < \rp, \label{eq:r_inequalities}
\end{align}
which implies that
\begin{align}
    V(\rmn) > V(\rt^\ast) > V(\rp).
\end{align}
Since \( \theta(\rpm; n, p) = \theta(\rts; \nt, p) \), we must have
\begin{align}
    W(\rmn; n) < W(\rt^*; \nt) < W(\rp; n).
\end{align}
Because \( W \) is monotonically decreasing in \( m = \tau n \), we find
\begin{align}
    \taus(\rmn) n < \taus(\rt^\ast) \nt < \taus(\rp) n,
\end{align}
which we henceforth denote by
\begin{align}
    \mmn < \mt^\ast < \mpl \label{eq:m_inequalities}
\end{align}
for notational convenience.

\begin{figure*}[t]
    \centering
    \includegraphics[width=\linewidth]{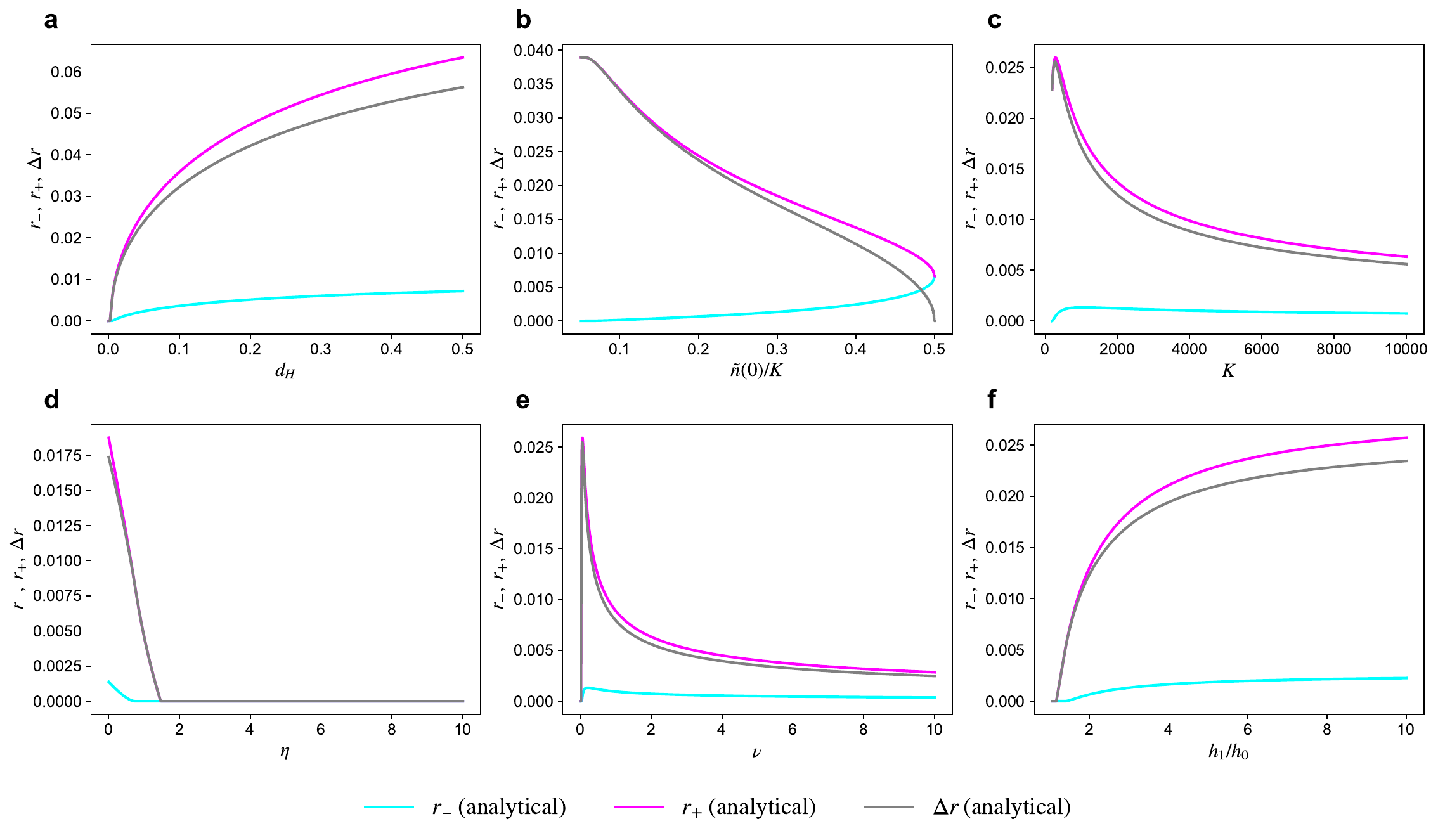}
    \caption{
    Dependence of the evolutionarily stable range of retirement rates \( \RES \) on model parameters:
    (\pnl{a}) hunter mortality rate \( d_H \);
    (\pnl{b}) \( \ntz/K \) for fixed \( K \);
    (\pnl{c}) carrying capacity \( K \), with fixed \( \ntz/K = 0.3 \);
    (\pnl{d}) self-learning efficiency \( \eta \);
    (\pnl{e}) education efficiency \( \nu \); and
    (\pnl{f}) advantage of expertise \( h_1/h_0 \).
    (\pnl{c}) \( K \) is limited to the range where \( \nz > \nth \), where \( \theta(r;n) \) is unimodal if and only if \( n > \nth  \).
    The bounds \( \rpm \) are obtained using algebraic equations and substituting coefficients and finding the roots numerically.
    The width of the stable range is defined by \( \Delta r = \rp - \rmn \).
    Analytical predictions for these behaviors are summarized in Table~\ref{tab:partials}.
    }
    \label{fig:parameter_dependence}
\end{figure*}

In what follows, we seek to identify the signs of the partial derivatives \( \frac{\partial \rpm}{\partial p} \) for various model parameters, primarily using Eqs.~\ref{eq:drpm_dp}, \ref{eq:dtheta_dr}, \ref{eq:r_inequalities}, and \ref{eq:m_inequalities}.
Specifically, we investigate the following cases: \( p \in \{ d_H, d_T, h, \eta, \nu, K \} \).
To complement our analysis, we also show results obtained from directly solving Eq.~\ref{eq:rpm} for \( \rpm \) numerically in Fig.~\ref{fig:parameter_dependence}, illustrating how the evolutionarily stable range of retirement rates changes with model parameters.
The results are also summarized in Table~\ref{tab:partials}.

\subsection{\label{subsec:}Hunter mortality rate (\( p = d_H \))}

A direct calculation shows
\begin{align}
    \theta_{d_H}= - \frac{\theta}{r + d_H},
\end{align}
which is negative by construction.
Equation~\ref{eq:r_inequalities} implies
\begin{align}
    \frac{1}{\rmn + d_H} > \frac{1}{\rt^\ast + d_H} > \frac{1}{\rp + d_H}.
\end{align}
Since \( \theta(\rpm; n, p) = \theta(\rt^*; \nt, p) \), we find
\begin{align}
    \frac{\theta(\rmn; n, p)}{\rmn + d_H} > \frac{\theta(\rt^\ast; \nt, p)}{\rt^\ast + d_H} > \frac{\theta(\rp; n, p)}{\rp + d_H}.
\end{align}
Thus, we obtain
\begin{align}
    \theta_{d_H}(\rmn; n, p) < \theta_{d_H}(\rt^*; n, p) < \theta_{d_H}(\rp; n, p).
\end{align}
Using Eqs.~\ref{eq:drpm_dp} and \ref{eq:dtheta_dr}, we obtain
\begin{align}
    \frac{\partial \rp}{\partial d_H} > 0, \quad
    \frac{\partial \rmn}{\partial d_H} > 0.
\end{align}
In short: a higher hunter mortality rate increases the evolutionarily stable retirement rates overall, shifting \( \RES \) upward.

\subsection{Elder-teacher mortality rate (\( p = d_T \))}

Similarly, we find
\begin{align}
    \theta_{d_T}
    &= \frac{1}{r + d_H} \frac{\partial F}{\partial d_T} \\
    &= - \frac{1}{r + d_H} \frac{\partial F}{\partial \tau} \frac{d\tau}{d d_T} \\
    &= - \frac{1}{r + d_H} \frac{r}{(r + d_T)^2} \frac{\partial F}{\partial \tau}.
\end{align}
Therefore, the signs of \( \partial \rpm/\partial d_T \) remain undetermined by our analytical method.

\subsection{Self-learning efficiency (\( p = \eta \))}

Given
\begin{align}
    \frac{\partial W}{\partial \eta} = \frac{h_1 - h_0}{(1 + \eta + \nu \tau n)^2},
\end{align}
we find
\begin{align}
    \theta_\eta
    &= V(r) \frac{\partial W}{\partial \eta}
    = V(r) \frac{h_1 - h_0}{(1 + \eta + \nu \taus n)^2} \\
    &= V(r) W(\taus; n) \frac{h_1 - h_0}{[h_0 + h_1 (\eta + \nu \taus n)](1 + \eta + \nu \taus n)} \\
    &= \theta(r; n, p) \frac{h_1 - h_0}{[h_0 + h_1 (\eta + \nu \taus n)](1 + \eta + \nu \taus n)} \\
    &= \theta(r; n, p) g(\taus n),
\end{align}
where we defined
\begin{align}
    g(m) = \frac{h_1 - h_0}{[h_0 + h_1 (\eta + \nu m)](1 + \eta + \nu m)},
\end{align}
and \( m = \taus n \).
Since \( g(m) \) is decreasing in \( m \), we obtain
\begin{align}
    g(\mmn) > g(\mt^\ast) > g(\mpl)
\end{align}
given Eq.~\ref{eq:m_inequalities}.
Therefore, we find
\begin{align}
    \theta_\eta(\rmn; n, p) > \theta_\eta(\rt^*; \nt, p) > \theta_\eta(\rp; n, p),
\end{align}
establishing
\begin{align}
    \frac{\partial \rp}{\partial \eta} < 0, \quad
    \frac{\partial \rmn}{\partial \eta} < 0.
\end{align}
In short: efficient self-learning decreases the evolutionarily stable retirement rates overall, shifting \( \RES \) downward.

\subsection{Education efficiency (\( p = \nu \))}

Since
\begin{align}
    \frac{\partial W}{\partial \nu} = \frac{(h_1 - h_0) \taus n}{(1 + \eta + \nu \taus n)^2},
\end{align}
we find
\begin{align}
    \theta_\nu = V(r) \frac{\partial W}{\partial \nu}
    = \theta(r;n,p) w(m),
\end{align}
where
\begin{align}
    w(m) = \frac{(h_1 - h_0)m}{(1 + \eta + \nu m)[h_0 + h_1(\eta + \nu m)]}.
\end{align}
Differentiating \( w(m) \) with respect to \( m \) yields
\begin{align}
    w'(m) = \frac{(h_1 - h_0)}{(1 + \eta + \nu m)^2 [h_0 + h_1(\eta + \nu m)]^2} \left[(1 + \eta)(h_0 + h_1 \eta) - h_1 (\nu m)^2 \right].
    \label{eq:w_derivative}
\end{align}
Let
\begin{align}
    \nu_c = \frac{1}{m} \sqrt{(1 + \eta)\left(\frac{1}{h} + \eta \right)}.
\end{align}
We then obtain
\begin{align}
    \begin{dcases}
        w'(m) > 0 \qquad \nu < \nu_c, \\
        w'(m) < 0 \qquad \nu > \nu_c,
    \end{dcases}
\end{align}
implying
\begin{align}
    \begin{dcases}
        \frac{\partial \rpm}{\partial \nu} > 0 \qquad \nu < \nu_c, \\
        \frac{\partial \rpm}{\partial \nu} < 0 \qquad \nu > \nu_c.
    \end{dcases}
\end{align}
This means that \( \partial_\nu \rpm > 0 \) for small \( \nu \), whereas \( \partial_\nu \rpm < 0 \) for large \( \nu \).
If education is highly efficient, the evolutionarily stable retirement rates decrease as it becomes even more efficient;
if education is highly inefficient, the stable retirement rates increase as education becomes more efficient.

\subsection{Advantage of expertise (\( p = h_1 \))}

Given
\begin{align}
    \frac{\partial W}{\partial h} = \frac{\eta + \nu \tau n}{1 + \eta + \nu \tau n},
\end{align}
we find
\begin{align}
    \theta_{h_1}
    &= V(r) \frac{\partial W}{\partial h_1} \\
    &= \theta(r; n, p) \frac{\eta + \nu \tau n}{h_0 + h_1(\eta + \nu \tau n)} \\
    &= \theta(r; n, p) s(m),
\end{align}
where we defined
\begin{align}
    s(m) = \frac{\eta + \nu m}{h_0 + h_1(\eta + \nu m)}.
\end{align}
Since
\begin{align}
    s'(m)
    = \frac{h_0 \nu}{[h_0 + h_1 (\eta + \nu m)]^2} > 0,
\end{align}
we obtain
\begin{align}
    s(\mmn) < s(\mt^*) < s(\mpl).
\end{align}
Thus,
\begin{align}
    \frac{\partial \rp}{\partial h_1} > 0, \quad
    \frac{\partial \rmn}{\partial h_1} > 0.
\end{align}
In short: valuable expertise increases the evolutionarily stable retirement rates overall, shifting \( \RES \) upward.

\subsection{Carrying capacity (\( p = K \))}

Let \( \xt \coloneqq \frac{\nt}{K} \in (0, 1/2] \) be fixed as \( K \) changes (i.e., \( \nt \) grows proportionally to \( K \)).
Since \( n + \nt = K \), we can rewrite
\begin{align}
    n = K(1-\xt), \quad \nt = K \xt.
\end{align}
Then,
we find
\begin{align}
    \theta_K
    &= V \frac{\partial}{\partial K} W(\taus(r); n(K))
    = V \frac{\partial W}{\partial n} \frac{dn}{dK} \\
    &= V \frac{(h_1 - h_0)\nu \taus}{(1 + \eta + \nu \taus n)^2} (1-\xt) \\
    &= VW \frac{(h_1 - h_0)\nu \taus n}{(1 + \eta + \nu \taus n)(h_0 + h_1(\eta + \nu \taus n))} \frac{1-\xt}{n}\\
    &= \frac{\nu}{K} \theta(r;n(K)) w(m),
\end{align}
where \( m = \taus n \).
Notice from Eq.~\ref{eq:w_derivative} that
\begin{align}
    \begin{dcases}
        w'(m) < 0 \qquad m > m_c, \\
        w'(m) > 0 \qquad m < m_c, \\
    \end{dcases}
\end{align}
where
\begin{align}
    m_c = \frac{1}{\nu} \sqrt{(1 + \eta)\left(\frac{1}{h} + \eta \right)}.
\end{align}
This suggests that there is no universal sign for \( w'(m) \).
Therefore, for \( m \gtrless m_c \), \( w'(m) \lessgtr 0 \).
Then, because \( \mmn < \mt^* < \mpl \), we find
\begin{align}
    w(\mmn) \gtrless w(\mt^*) \gtrless w(\mpl),
\end{align}
giving
\begin{align}
    \underbrace{\frac{\nu}{K} \theta(\rmn; n)w(\mmn)}_{\theta_K(\rmn;n)} \gtrless \underbrace{\frac{\nu}{K} \theta(\rts; \nt)w(\mt^\ast)}_{\theta_K(\rts; \nt)} \gtrless \underbrace{\frac{\nu}{K} \theta(\rp; n)w(\mpl)}_{\theta_K(\rp;n)}.
\end{align}
Therefore,
\begin{align}
    \begin{dcases}
        \frac{\partial \rpm}{\partial K} < 0 \qquad m > m_c, \\
        \frac{\partial \rpm}{\partial K} > 0 \qquad m < m_c. \\
    \end{dcases}
\end{align}
In short:
when there are few resident teachers at the quasi-steady state, increasing \( K \) moves \( \RES \) downward;
conversely, when there are many resident teachers at the quasi-steady state, increasing \( K \) moves \( \RES \) upward.

\clearpage

\begin{table*}
    \setlength{\tabcolsep}{5pt}
    \caption{
    Parameters that raise the return to instruction shift the evolutionarily stable range of retirement rates upward.
    }
    \centering
    \begin{tabular*}{\textwidth}{@{\extracolsep{\fill}} c c l}
        \toprule
        Parameter & Predicted dependence of \( \rmax \) and \( \rmin \) & Description or comment \\
        \midrule
        \( d_H \) & \( \frac{\partial \rmax}{\partial d_H} > 0, \; \frac{\partial \rmin}{\partial d_H} > 0 \) & Higher hunter mortality rate shifts \( \RES \) upward. \\
        \( d_T \) & indeterminate & \textemdash \\
        \( \ntz/K \) & \( \frac{\partial \rmax}{\partial (\ntz/K)} < 0, \; \frac{\partial \rmin}{\partial (\ntz/K)} > 0 \) & The behavior follows from the definition of \( \RES \);  \( K \) is fixed. \\
        \( K \) & nonmonotonic &  \( \ntz/K \) is fixed. \\
        \( \eta \) & \( \frac{\partial \rmax}{\partial \eta} < 0, \; \frac{\partial \rmin}{\partial \eta} < 0 \) & More efficient self-learning shifts \( \RES \) downward. \\
        \( \nu \) & nonmonotonic & \textemdash \\
        \( h_1/h_0 \) & \( \frac{\partial \rmax}{\partial (h_1/h_0)} > 0, \; \frac{\partial \rmin}{\partial (h_1/h_0)} > 0 \) & Greater returns to expertise shift \( \RES \) upward. \\
        \bottomrule
    \end{tabular*}
    \addtabletext{
    \begin{minipage}{\textwidth}
    Analytical predictions for the dependence of the range of ES retirement rates, \( \RES \), on model parameters.
    Each entry is obtained from the sign of the derivatives \( \partial \rmax/\partial \p \) and \( \partial \rmin/\partial \p \), where \( \p \) denotes a focal model parameter.
    Predictions are derived under \( \nz > \ntz > \nth \), the regime in which
    \( \rmax > \rmin > 0 \).
    Numerical results for all parameters other than \( d_T \) are shown in Fig.~\ref{fig:parameter_dependence}.
    \end{minipage}
    }
    \label{tab:partials}
\end{table*}

\FloatBarrier

\clearpage

\section{\label{sec:ropt_RES}Is optimal teaching evolutionarily stable?}

In the main text, we identify the exact condition (\hyperref[sec:ropt_condition]{\emph{Evolution may fail to optimize hunting returns}})
under which the return-maximizing retirement rate is evolutionarily stable within our analytical framework.
However, it does not provide useful intuition as the expressions remain general and are not necessarily decomposable.
Here we provide asymptotic analyses to assess whether optimal teaching is evolutionarily stable in some regimes that are biologically relevant---that is, whether the retirement rate producing the optimal proportion of teachers, \( \ropt \), in the absence of competitors at population size \( K \) is included in the associated evolutionarily stable range of retirement rates, \( \RES \).
More specifically, we examine small-\( d_H \), large-\( d_T \), and large-\( h \) (expertise-limited) regimes.
In particular, we analyze the expertise-limited regime in detail to provide insight on potential selective pressure toward low teacher mortality rate, which is different from selection on retirement rate.

\subsection{Preliminaries}

Let
\begin{align}
    \xi(p) \coloneqq \theta(\ropt(K); n, p) - \theta(\rt^*; \nt, p).
\end{align}
Expanding \( \xi \) in powers of a small parameter \( p \) reveals whether \( \ropt \) falls inside or outside of the interval \( \RES \) in the small-\( p \) limit.
To simplify the notation, we introduce
\begin{align}
    \Theta_1 &\coloneqq \theta(\ropt(K); n), \\
    \Theta_2 &\coloneqq \theta(\rts; \nt),
\end{align}
so that
\begin{align}
    \xi = \Theta_1 - \Theta_2.
\end{align}
From this point onward, we drop the dependence of \( \ropt \) on \( K \), i.e., \( \ropt = \ropt(K) \).
In addition, we occasionally write \( F(r;n) \) for \( F(\taus(r);n) \) for ease of notation.

\subsection{Small-\( d_H \) limit}

We expand \( \theta \) in powers of \( d_H \):
\begin{align}
    \theta(r;n) = \frac{F(r; n)}{r + d_H} = \frac{F(r;n)}{r} - \frac{F(r;n)}{r^2}d_H + O\left(d_H^2\right).
\end{align}
Then, as \( d_H \to 0 \), we find
\begin{align}
    \Theta_1 \to \frac{F(\ropt; n)}{\ropt}.
\end{align}
We next expand \( F \) around \( r = 0 \):
\begin{align}
    F(r;n) = F(0;n) + r \frac{\partial}{\partial r}F(0;n) + O(r^2).
\end{align}
We note that \( \rts \to 0 \) as \( d_H \to 0 \).
Thus,
\begin{align}
    \Theta_2 = \frac{F(0; \nt) + O(\rt^*)}{\rt^* + d_H} \sim \frac{F(0; \nt)}{d_H},
\end{align}
as \( d_H \to 0 \), where
\begin{align}
    F(0; \nt) = \frac{h_0 + h_1 \eta}{1 + \eta},
\end{align}
which is strictly positive.
Therefore, we obtain
\begin{align}
    \xi(d_H) \sim \frac{F(\ropt; n)}{\ropt} - \frac{F(0; \nt)}{d_H},
\end{align}
which tends to \( - \infty \) as \( d_H \to 0 \).
Thus,
\begin{align}
    \xi(d_H) < 0
\end{align}
in the small-\( d_H \) limit.
That is,
\begin{align}
    \ropt \notin \RES, \quad d_H \to 0.
\end{align}

\clearpage

\subsection{Large-\( d_T \) limit}

\subsubsection{Value Function Expansion}

We first derive a general formula for a value function \( \varphi(p) \) when the function is perturbed with respect to its argument \( p \).
Later we will take \( p \) to be \( d_T^{-1} \) and consider the limit \( d_T \to \infty \) (i.e., \( p \to 0 \)).

We define
\begin{align}
    \varphi(p) \coloneqq \max_r \theta(r; n, p) \equiv \theta(r^*; n, p).
\end{align}
Note that \( r^* \) depends on \( p \).
We then expand both \( \varphi \) and \( r^* \) in powers of \( p \):
\begin{align}
    \varphi(p) &= \varphi_0 + \varphi_1 p + \varphi_2 p^2 + \cdots, \\
    r^*(p) &= r_0 + r_1 p + r_2 p^2 + \cdots.
\end{align}
Next we compute the expansion coefficient \( \varphi_k \) up to a desired order (up to first order in the discussion below).

First, we find
\begin{align}
    \varphi_0 = \varphi(0) = \theta(r_0; n, 0).
\end{align}
Second,
by the envelope theorem (or standard calculations), we find
\begin{align}
    \frac{d\varphi}{dp} = \frac{\partial \theta(r^*; n, p)}{\partial p}
    = \theta_p(r^*; n, p) + \underbrace{\theta_r(r^*; n, p)}_{=0 \; \text{(FOC)}} \frac{dr^*}{dp}
    = \theta_p(r^*; n, p).
\end{align}
Here \( \theta_z \) denotes the derivative of \( \theta \) with respect to its \textit{independent} variable \( z \).
Therefore,
\begin{align}
    \varphi_1
    = \left. \frac{d\varphi}{dp}\right|_{p=0}
    = \theta_p(r_0; n, 0).
\end{align}
Therefore, provided that \( p \) is small, we obtain the approximate expansion
\begin{align}
    \theta(r^*; n, p) = \theta(r_0; n, 0) + \theta_p(r_0; n, 0) p + O(p^2). \label{eq:val_func_expansion}
\end{align}

\subsubsection{Preliminaries}

Reorganizing \( F \) in terms of \( r \) and \( d_T \), we find
\begin{align}
    F(r; n) = \frac{d_T}{r + d_T} \frac{(h_0 + h_1\eta) d_T + [h_0 + h_1 (\eta + \nu n)]r}{(1 + \eta)d_T + (1 + \eta + \nu n)r}.
\end{align}
Therefore,
\begin{align}
    F(r; n) = \frac{d_T}{r + d_T}\frac{A d_T + Br}{C d_T + D  r},
\end{align}
where
\begin{align}
    A = h_0 + h_1 \eta, \quad
    B(n) = h_0 + h_1(\eta + \nu n), \quad
    C = 1 + \eta, \quad
    D(n) = 1 + \eta + \nu n.
\end{align}
We note that
\begin{align}
    \ropt(n) = R(n) d_T, \quad
    R(n) = \frac{-\left(\frac{1}{h} + \eta \right)(1 + \eta + \nu n) + \nu n \sqrt{\left(1-\frac{1}{h}\right)(1 + \eta + \nu n)}}{(1 + \eta + \nu n)\left(\frac{1}{h} + \eta + \nu n\right)},
\end{align}
for \( n > \nc \).

\subsubsection{Asymptotic analysis}

Rewriting \( F \), we obtain
\begin{align}
    F(r; n) = \frac{1}{\frac{r}{d_T} + 1} \frac{A + B \frac{r}{d_T}}{C + D\frac{r}{d_T}}.
\end{align}
Setting \( p = 1/d_T \), we find
\begin{align}
    F(r; n) = \frac{1}{rp + 1}\frac{A + B rp}{C + Drp}.
\end{align}
We note
\begin{align}
    \ropt(K) = \frac{R(K)}{p}.
\end{align}
Therefore,
\begin{align}
    \Theta_1
    &= \theta(\ropt(K); n, p) \\
    &= \frac{1}{\ropt(K) + d_H} \frac{1}{\ropt(K)p + 1} \frac{A + B \ropt(K)p}{C + D\ropt(K)p} \\
    &= \frac{1}{\frac{R(K)}{p} + d_H} \frac{1}{R(K) + 1} \frac{A + B R(K)}{C + D R(K)}.
\end{align}
Note
\begin{align}
    \frac{1}{\frac{R(K)}{p} + d_H} = \frac{p}{R(K)} \frac{1}{1 + \frac{d_H}{R(K)}p}
    = \frac{p}{R(K)} \left[1 + \frac{d_H}{R(K)} p + O(p^2)\right].
\end{align}
Therefore,
\begin{align}
    \Theta_1 = \frac{1}{R(K) + 1} \frac{A + B R(K)}{C + D R(K)} \frac{1}{R(K)} p + O(p^2).
\end{align}
Next, we expand
\begin{align}
    \theta(r; n, p) = \frac{1}{r + d_H} \frac{1}{rp + 1}\frac{A + B rp}{C + Drp}
\end{align}
in powers of \( p \).
We note
\begin{align}
    F(r; n)
    &= \left(1 - rp + \cdots \right)\left[ \frac{A}{C} + \frac{r}{C} \left(B- \frac{AD}{C}\right)p + \cdots \right] \\
    &= \frac{A}{C} + \frac{r}{C} \left[\left(B- \frac{AD}{C}\right) - A\right]p + O(p^2).
\end{align}
Therefore,
\begin{align}
    \theta(r;n,p) = \frac{1}{r + d_H} \left\{ \frac{A}{C} + \frac{r}{C} \left[\left(B- \frac{AD}{C}\right) - A\right] p + O(p^2) \right\},
\end{align}
yielding
\begin{align}
    \left. \frac{\partial \theta}{\partial p} \right|_{p=0} = \frac{r}{C} \left[\left(B- \frac{AD}{C}\right) - A\right].
\end{align}
Also,
\begin{align}
    \theta(r_0; \tilde n, 0)= \frac{F(r_0; \tilde n, 0)}{r_0 + d_H} = \frac{1}{r_0 + d_H}\frac{A}{C}.
\end{align}
Therefore, we obtain
\begin{align}
    \Theta_2 = \varphi_0 + \varphi_1 p + O(p^2), \quad
    \varphi_0 = \frac{1}{r_0 + d_H}\frac{A}{C}, \quad
    \varphi_1 = \frac{r_0}{C} \left[\left(B(\nt) - \frac{AD(\nt)}{C}\right) - A\right].
\end{align}
Thus,
\begin{align}
    \xi(d_T^{-1}) = \Theta_1 - \Theta_2
    = - \frac{1}{r_0 + d_H}\frac{A}{C} + O(d_T^{-1}),
\end{align}
implying, up to leading order,
\begin{align}
    \xi(d_T^{-1}) < 0
\end{align}
in the large-\( d_T \) limit.
That is,
\begin{align}
    \ropt \notin \RES, \quad d_T \to \infty.
\end{align}

\clearpage

\subsection{\label{sec:expertise_limited_regime}Large-\( h \) limit (expertise-limited regime)}

\subsubsection{Asymptotic analysis}

Here we work with \( \taus \) instead of \( r \) as it is more convenient.
In this section, we occasionally write \( \tau \) for \( \taus \) for ease of notation.
Reorganizing \( \theta \) gives
\begin{align}
    \theta = \underbrace{\frac{(1-\tau)^2}{\tau d_T + (1-\tau) d_H}}_{\equiv V} \underbrace{\frac{h_0 + h_1(\eta + \nu \tau n)}{1 + \eta + \nu \tau n}}_{\equiv W}.
\end{align}
We seek to find \( \rts \) by imposing the first-order condition
\begin{align}
    \frac{\partial \theta}{\partial \tau} = 0.
\end{align}
We find
\begin{align}
    \frac{\partial V}{\partial \tau}
    = - (1 - \tau)\frac{(1+\tau)d_T + (1-\tau)d_H}{[\tau d_T + (1-\tau) d_H]^2}, \quad
    \frac{\partial W}{\partial \tau}
    = \frac{(h_1-h_0)\nu n}{(1+\eta +\nu \tau n)^2}.
\end{align}
Thus,
\begin{align}
    \frac{\partial \theta}{\partial \tau}
    &= \frac{\partial V}{\partial \tau} W + V \frac{\partial W}{\partial \tau} \\
    &= \frac{1-\tau}{[\tau d_T + (1-\tau) d_H](1+\eta + \nu \tau n)}
    \left[- \frac{[(1+\tau)d_T + (1-\tau)d_H][h_0 + h_1(\eta + \nu \tau n)]}{\tau d_T + (1-\tau) d_H} + (1-\tau)\frac{(h_1-h_0)\nu n}{1+\eta +\nu \tau n} \right].
\end{align}
Therefore, provided \( \tau \neq 1 \), the first-order condition \( \partial_\tau \theta = 0 \) yields
\begin{align}
    [(1+\tau)d_T + (1-\tau)d_H][1 + h(\eta + \nu \tau n)](1+\eta +\nu \tau n)
    = (1-\tau)(h-1)\nu n [\tau d_T + (1-\tau) d_H]. \label{eq:foc_algebraic}
\end{align}
Consider the regime where  \( h \gg 1 \) and \( n \gg 1 \).
First,
we note that, if the above equation is satisfied, \( \tau \to 0 \) as \( n \to \infty \).
In the large-\( n \) limit, \( \tau \) scales as
\begin{align}
    \tau \sim n^{-1/2},
\end{align}
implying that \( \tau n \gg 1 \).
This is because the dominant balance of Eq.~\ref{eq:foc_algebraic} reveals that neither \( \tau n \ll 1 \) nor \( \tau n = O(1) \) would maintain consistency of the equation, so the correct balance is \( \tau n \gg 1 \).
In particular, the dominant balance implies that
\begin{align}
    [(1+\tau) d_T + (1-\tau) d_H](\nu \tau n)^2 \sim (1-\tau) \nu n [\tau d_T + (1-\tau) d_H].
\end{align}
Since \( \tau \to 0 \) as \( n \to \infty \), the leading term behaves like
\begin{align}
    \tau \sim \sqrt{\frac{d_H}{d_H + d_T}} \frac{1}{\sqrt{\nu n}}.
\end{align}
Therefore,
\( \tauts \), which maximizes \( \theta(\rt; \nt) \) at \( \rts = d_T \tauts/(1-\tauts) \), behaves like
\begin{align}
    \tauts \sim \sqrt{\frac{d_H}{d_H + d_T}} \frac{1}{\sqrt{\nu \nt}} \quad \text{as} \quad
    \nt \to \infty \quad\text{and}\quad h \to \infty. \label{eq:asymptotic_tauts}
\end{align}
Given \( \Theta_2 = \theta(\tauts; \nt) \), we expand \( \theta \) up to leading order.
First, we find
\begin{align}
    V
    &= \frac{1}{d_H} \left(1 - 2\tau + \tau^2 \right) \left[1 - \frac{d_T - d_H}{d_H} \tau + O(\tau^2) \right] \\
    &= \frac{1}{d_H} \left[1 - \frac{d_H + d_T}{d_H} \tau + O(\tau^2) \right].
\end{align}
Likewise, we find
\begin{align}
    W
    &= \left[h_0 + h_1(\eta + \nu \tau n) \right] \frac{1}{\nu \tau n} \left( 1 - \frac{1+\eta}{\nu \tau n} + O((\nu \tau n)^{-2}) \right) \\
    &= h_1 \left[1 + \left(\frac{1}{h} - 1\right) \frac{1}{\nu \tau n} + O((\nu \tau n)^{-2})  \right].
\end{align}
Therefore,
in the limit where \( n \to \infty \) and \( h \to \infty \), we find
\begin{align}
    \theta
    &= \frac{h_1}{d_H} \left[1 - \frac{d_H + d_T}{d_H} \tau - \frac{1}{\nu \tau n} + o(n^{-1/2}) \right].
\end{align}
Thus, given the leading-order behavior of \( \tauts \) (Eq.~\ref{eq:asymptotic_tauts}),
we obtain
\begin{align}
    \Theta_2
    &\sim
    \frac{h_1}{d_H} \left[1 - \sqrt{\frac{d_H + d_T}{d_H}} \frac{1}{\sqrt{\nu \nt}} - \sqrt{\frac{d_H + d_T}{d_H}} \frac{1}{\sqrt{\nu \nt}} \right] \\
    &= \frac{h_1}{d_H} \left[1 - 2 \sqrt{\frac{d_H + d_T}{d_H}}\frac{1}{\sqrt{\nu \nt}} \right].
\end{align}
Given \( \Theta_2 \coloneqq \max_{\rt \geq 0} \theta(\rt; \nt) = \theta(\rts;\nt) \), we can alternatively derive the same leading-order behavior by noting the arithmetic--geometric mean inequality
\begin{align}
    \zeta \taut + \frac{1}{\nu \taut \nt} \geq 2 \sqrt{\frac{\zeta}{\nu \nt}},
\end{align}
with equality exactly when the two negative contributions at leading order to \( \theta \) are balanced, i.e., \( \zeta \taut = \frac{1}{\nu \taut \nt} \).

Next, we derive the leading-term of \( \Theta_1 \).
Since
\begin{align}
    \tauopt(K)
    &= \frac{1}{\nu K} \left[-(1+\eta) + \sqrt{\left(1 - \frac{1}{h}\right)(1 + \eta + \nu K)} \right] \\
    & \sim \frac{1}{\sqrt{\nu K}}
\end{align}
as \( h \to \infty \) and \( K \to \infty \),
we find
\begin{align}
    \Theta_1
    &= \theta(\ropt(K); n) \\
    &\sim \frac{h_1}{d_H} \left[1 - \frac{d_H+d_T}{d_H} \frac{1}{\sqrt{\nu K}} - \sqrt{\frac{K}{\nu}} \frac{1}{n} + o(n^{-1/2})\right].
\end{align}

Taken together, these results give rise to the asymptotic behavior
\begin{align}
    \xi
    = \Theta_1 - \Theta_2
    \sim \frac{h_1}{d_H}\frac{1}{\sqrt{\nu}}
    \left[ 2 \sqrt{\frac{\zeta}{\nt}} - \frac{1}{\sqrt{K}} \zeta - \frac{\sqrt{K}}{n} \right],
\end{align}
yielding the relation
\begin{align}
    \xi \gtrless 0 \iff
    2 \sqrt{\zeta} \gtrless \sqrt{\frac{\nt}{K}} \zeta + \frac{\sqrt{\nt K}}{n},
\end{align}
to leading order and away from the asymptotic boundary.
At the boundary, the leading-order contribution to \( \xi \) vanishes, and subleading corrections determine its sign.

Let \( \zeta_c \) denote the leading-order phase boundary, given by
\begin{align}
    2 \sqrt{\zeta_c} = \sqrt{\frac{\nt}{K}} \zeta_c + \frac{\sqrt{\nt K}}{n}.
\end{align}
After some manipulation, we find
\begin{align}
    \sqrt{\zeta_c}
    \approx \sqrt{\frac{1}{\xt}} \pm \sqrt{\frac{1}{\xt} - \frac{1}{1-\xt}},
\end{align}
where \( \xt = \nt/K \) and assume \( n + \nt \approx K \) as in the main text.
Notice that the phase boundary depends only on a subset of parameters: the teacher-to-hunter mortality ratio (\( d_T/d_H \)), and the initial mutant population size relative to carrying capacity (\( \xt \)).
By contrast, the learning and teaching parameters, including self-learning and education efficiencies, do not figure into the leading-order phase boundary.

Suppose the teacher mortality rate is no less than the hunter mortality rate, that is, \( d_T \geq d_H \) (i.e., \( \zeta \geq 2 \)), which is a natural assumption given the biological process the model describes.
Then, we find, for \( 0 < \xt \leq 1/2 \),
\begin{align}
    \sqrt{\frac{1}{\xt}} - \sqrt{\frac{1}{\xt} - \frac{1}{1-\xt}} \leq \sqrt{2},
\end{align}
where equality holds at \( \xt = 1/2 \), i.e., when the mutant population takes up exactly half the carrying capacity.
Therefore, the biologically relevant phase boundary is
\begin{align}
    \sqrt{\zeta_c} \approx \sqrt{\frac{1}{\xt}} + \sqrt{\frac{1}{\xt} - \frac{1}{1-\xt}}, \quad 0 < \xt \leq \frac{1}{2}.
\end{align}

To summarize, we find the following condition to leading order:
\begin{align}
    \zeta &> \zeta_c \iff \xi < 0 \iff \ropt \notin \RES, \\
    \zeta &< \zeta_c  \iff \xi > 0 \iff \ropt \in \RES.
\end{align}
In addition, there are four simple regimes in which whether the return-maximizing retirement rate is included in the evolutionarily stable range is clearly determined:
(i) \( \xi < 0 \) if \( \xt \) is sufficiently large;
(ii) \( \xi > 0 \) if \( \xt \) is sufficiently small;
(iii) \( \xi < 0 \) if \( \zeta \) is sufficiently large; and
(iv) \( \xi > 0 \) if \( \zeta \) is sufficiently small (\( \zeta \approx 2 \));
see Table~\ref{tab:large_h_limit} for a summary of the results.
We do not include degenerate boundary cases, such as \( \xt = 1/2 \) together with \( \zeta = 2 \), for which the leading-order contribution vanishes and higher-order corrections are required to determine the sign of \( \xi \), as these cases are not of interest here.

\begin{table}[t]
    \centering
    \caption{
    Four simple regimes for evolutionary stability of the return-maximizing retirement rate in the limit of a large advantage of expertise, a large mutant population size, and a large carrying with a fixed ratio \( \xt = \nt / K \).
    }
    \begin{tabular}{l l c c}
    \toprule
        Regime & Condition & Leading sign of \( \xi \) & Return-maximizing retirement rate, \( \ropt \) \\
    \midrule
        (i) & \( \xt \) is large, approaching \( (1/2)^- \) & \( \xi < 0 \) & evolutionarily unstable \\
        (ii) & \( \xt \) is sufficiently small & \( \xi > 0 \) & evolutionarily stable \\
        (iii) & \( \zeta \) is sufficiently large & \( \xi < 0 \) & evolutionarily unstable \\
        (iv) & \( \zeta \) is sufficiently small, approaching \( 2^+ \) & \( \xi > 0 \) & evolutionarily stable \\
    \bottomrule
    \end{tabular}
    \addtabletext{
    \begin{minipage}{\textwidth}
        The ``large'' and ``small'' are relative to the other parameters, not on absolute terms.
        The fourth regime is particularly important because it supports the evolution of a prolonged post-reproductive lifespan by favoring a smaller teacher mortality rate, independent of the retirement rate.
        In the four regimes above, we limit \( \xt \in (0,1/2) \) to avoid degeneracy at leading order, although \( \xt = 1/2 \) is admissible in the model.
    \end{minipage}
    }
    \label{tab:large_h_limit}
\end{table}

The fourth regime (iv) in Table~\ref{tab:large_h_limit} in particular provides an important insight on the relationship between return-maximization and evolutionarily stability.
Given a relative founding mutant population size \( \xt \), the return-maximizing retirement rate is evolutionarily stable if an average teacher lifespan is sufficiently large, provided \( \nt, K, h \to \infty \).
In other words, if teachers, on average, live for as long as hunters do (\( d_T = d_H \)), the return-maximizing retirement rate is evolutionarily stable, to leading order, for \( 0 < \xt=\nt/K < 1/2 \) when the environment accommodates a large population (\( \nt, K \gg 1 \)) and expertise boosts hunting returns (\( h \gg 1 \)).

\subsubsection{Long-lived teachers align return maximization and evolutionary stability in a resource-abundant environment where expertise provides a disproportionate advantage}

In the expertise-limited, large carrying capacity regime (with a fixed relative mutant founding population size \( 0 < \xt < 1/2 \)), low teacher mortality (\( d_T \approx d_H \)) implies that the per-capita return maximizing retirement rate is evolutionarily stable (\( \ropt \in \RES \)), to leading order.
This is significant and especially relevant to understanding the origin of a prolonged post-reproductive lifespan because in this asymptotic regime evolution can both achieve return-maximization (when a population fully occupies the environment) and resist all possible invasion.
Simply put, in a resource-abundant environment with a disproportionate advantage of expertise, prolonged post-reproductive lifespan enables evolution to select for the return-maximizing retirement rate.
This gives teacher longevity a distinct adaptive role beyond selection on retirement rate:
longevity after retirement (due to low teacher mortality) can help align evolutionary stability with return-maximization.

Increased post-reproductive representation (PrR) can be favored through two distinct and complementary mechanisms:
evolution can increase the retirement rate \( r \), while selection on an established monomorphic population seeking to maximize per-capita hunting return---while resisting invasion---can decrease teacher mortality rate \( d_T \) to a value nearly as small as hunter mortality rate (i.e., \( d_T \approx d_H \)).
Because PrR is given by \( \taus = r/(r+d_T) \), both of these two routes independently increase post-reproductive representation.

Extending post-reproductive lifespan after retirement can be achieved through collective effort, such as providing material benefits (e.g., nutritious food) or extra care for retirees to support their post-retirement longevity.

Low teacher mortality provides an additional adaptive advantage to the population:
it allows the population to align evolutionary stability and return-maximization, without sacrificing the expected total lifespan.
The (unconditional) expected total lifespan is given by (see Table~\ref{tab:single_species})
\begin{align}
    \mathcal{L}(r, d_T;d_H) \coloneqq \frac{1}{d_T}\frac{r + d_T}{r + d_H},
\end{align}
where we define \( d_T \geq d_H \) and fix \( d_H > 0 \).
Notice that \( \mathcal{L} \) monotonically decreases in \( r \geq 0 \).
It is also monotonically decreasing in \( d_T \) for \( r > 0 \), while being fixed at its maximal value \( \mathcal{L} = 1/d_H \) for \( r = 0 \), irrespective of the value of \( d_T \).
If evolution selects for a positive retirement rate (\( r > 0 \)), which by itself hurts the expected total lifespan as mentioned above, the smallest possible teacher mortality rate (\( d_T = d_H \)) always preserves the maximal expected total lifespan (\( \mathcal{L} = 1/d_H \)), regardless of how large the retirement rate (\( r > 0 \)) becomes due to selection.

Therefore, there is strong selective pressure toward lowering teacher mortality (relative to hunter mortality) in a resource-abundant, expertise-limited regime;
sufficiently lower teacher mortality guides evolution to select for the return-maximizing, invasion-resistant retirement rate, without sacrificing the total expected lifespan, even if the mutant founding population size is nearly as large as half the carrying capacity.

Finally we note that none of the limits considered above violates the assumption \( \ns \approx K \), i.e., the above argument holds, to leading order, even if the monomorphic population size is evaluated at \( \ns \).

\clearpage

\section{\label{sec:alternative_ropt}Comparison between the evolutionarily stable range of \( r \) and various formulations for \( \ropt \)}

We discuss alternative formulas for \( \ropt \) that maximize different equilibrium quantities of a monomorphic population.
In the previous section, we defined \( \ropt \) as the retirement rate that maximizes the per-capita hunting returns of a monomorphic population, assuming its size is fixed at carrying capacity \( K \).
Here, we systematically analyze other possible formulations that may, in their own right, be appropriate for assessing the performance of a monomorphic population.
In particular, we examine four different objectives according to which the performance of a monomorphic population may be assessed:
\begin{align}
    r_{\Fs, \text{opt}}(K) &\coloneqq \argmax_{r} F(\taus(r);K) \\
    r_{\Fs, \text{opt}}[\ns] &\coloneqq \argmax_{r} F(\taus(r);\ns(r)) \\
    r_{\ns, \text{opt}} &\coloneqq \argmax_{r} \ns(r) \\
    r_{\text{birth}, \text{opt}} &\coloneqq \argmax_{r} b(\Fs)n_H^*
\end{align}
Here, \( r_{\mathcal{Q}, \text{opt}} \) denotes the optimal value of \( r \) that maximizes the quantity \( \mathcal{Q} \in \{\Fs, \ns, \text{``birth''} \} \) in a monomorphic population, with its argument indicating the function or parameter  on which the optimizer depends.
The first formula maximizes the equilibrium per-capita hunting return per unit time of a monomorphic population evaluated at carrying capacity.
This provides a benchmark criterion given a common maximal population density (also used in Fig.~\ref{fig:pip_stochastic_vs_prediction} and Fig.~\ref{fig:pip_panels} of the main text):
it isolates the effect of the retirement rate on per-capita hunting returns from its endogenous effect on the equilibrium total population size.
The second formula maximizes the equilibrium per-capita hunting return rate of a monomorphic population evaluated at the corresponding equilibrium population size.
The first formula should approximate the second, provided that \( \ns \approx K \) (near optimizers), which is assumed throughout the analysis in the paper.
By contrast, the third formula maximizes the equilibrium monomorphic population size.
The fourth formula maximizes the intrinsic birth-event rate in a monomorphic population at equilibrium, before accounting for density-dependent regulation through the logistic factor.

\begin{figure*}
    \centering
    \includegraphics[width=\linewidth]{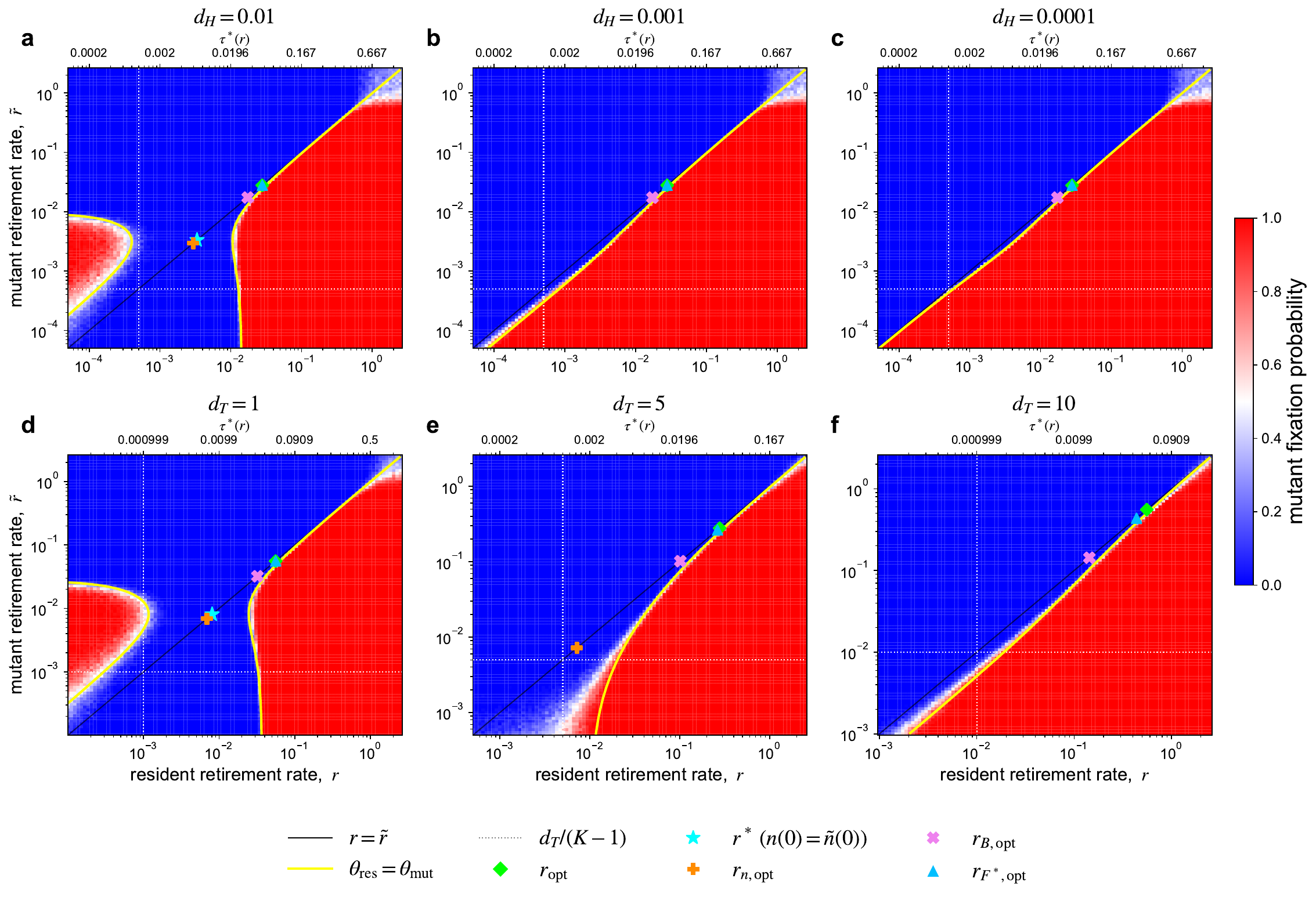}
    \caption{
    Different measures of the performance of a monomorphic population relative to the evolutionarily stable retirement rates in the small-\( d_H \) (top row) and the large-\( d_T \) (bottom row) regimes.
    As the panels go from the left to the right, the focal parameter becomes increasingly closer to its limiting value.
    In both cases, when \( d_H \) is sufficiently small (panel \pnl{c}) or \( d_T \) is sufficiently large (panel \pnl{f}), all optima except for \( r_{n, \text{opt}} \) are not evolutionarily stable.
    An optimum is not shown if its value is found to be zero, and that is why \( r_{n, \text{opt}} \) are absent in panels \pnl{b,c,f}.
    Likewise, the unique ES point \( \rs \) that arises when founding populations are at equal size also tends to zero in both regimes (\pnl{b,c,e,f}).
    }
    \label{fig:various_optima_pips}
\end{figure*}

The choice of \( \ropt \in \{ r_{\Fs, \text{opt}}(K), r_{\Fs, \text{opt}}[\ns], r_{\ns, \text{opt}}, r_{\text{birth}, \text{opt}} \} \) depends on the purpose and nature of the analysis at hand.
Here we compare them all against the evolutionarily stable range of retirement rates to assess whether the optimal retirement rate, however defined, falls inside or outside of that range in a given parameter setting.
In accordance with the analysis in Sec.~\ref{sec:ropt_RES}, we investigate the small-\( d_H \) and large-\( d_T \) regimes by changing the focal parameter gradually, with fixed founding population sizes, to illuminate the relationship between each \( \ropt \) and \( \RES \) that develops in these limits (Fig.~\ref{fig:various_optima_pips}).

Figure~\ref{fig:various_optima_pips} shows where each optimal retirement rate \( \ropt \) falls relative to the stable retirement range \( \RES \).
First, as the focal parameter approaches its limiting value, both \( \rpm \) tend to zero in these limits, in which case \( \rs \) also becomes zero and therefore are not shown in Figs.~\ref{fig:various_optima_pips}\pnl{b,c,e,f}.
Consequently, three of the candidate optimal retirement rates fall outside that stable range in these limits, namely, \( r_{\Fs, \text{opt}}(K) \), \( r_{\Fs, \text{opt}}[\ns] \), and \( r_{\text{birth}, \text{opt}} \).
By contrast, \( r_{\ns, \text{opt}} \) becomes zero for sufficiently small \( d_H \) or large \( d_T \) (and hence not shown in Figs.~\ref{fig:various_optima_pips}\pnl{b,c,f}).
This result is intuitive: \( \ns \) is maximized when \( \theta(r;\ns) \) is maximized because \( \ns = K [1 - h_0/(b_0\theta(r;\ns))] \), and since \( r \approx 0 \) maximizes \( \theta \) in these limits, \( \ns \) must also be maximized correspondingly.
In addition we note that \( r_{\Fs, \text{opt}}[\ns] \) remains near \( r_{\Fs, \text{opt}}(K) \) in these limits, consistent with the expectation from the assumption that \( \ns \approx K \).

Taken together, these results demonstrate that most candidate optimal retirement rates (expect for the one that maximizes the equilibrium monomorphic population size) may not always be evolutionarily stable, even when the founding population sizes are fixed at distinct values and the stable range is nondegenerate.
This creates a dilemma and trade-off between evolutionary stability under pairwise invasion and the performance of a monomorphic population: evolution selects for an evolutionarily stable rate of retirement with which a population may not perform optimally once it successfully occupies the environment.

We synthesize the results obtained so far.
In societies where teachers are short-lived relative to hunters, evolution tends to produce fewer teachers than would maximize population performance in the absence of evolutionary competition.
By contrast, in large environments where teachers are almost as long-lived as hunters, evolution can produce an optimal number of teachers when expertise provides a disproportionate advantage (Sec.~\ref{sec:expertise_limited_regime}).

\FloatBarrier

\clearpage

\section{\label{sec:food_sharing}Food sharing extension}

\subsection{\label{subsec:food_sharing_coexistence}Sharing hunting returns allows both types to coexist}

In the model described in the main text (Eq.~\ref{eq:model}), evolution leads to the complete exclusion of one of the two types in all observed cases, which differ in their rates of retirement and associated frequencies of teachers.
But an alternative formulation may allow the coexistence of multiple types with distinct life-history characteristics.
Our model assumes that the birth rate of each type is governed only by the hunting returns of its own members, which may not be reasonable in many societies.
Therefore, we consider cases where hunting returns are partially and preferentially shared across both types, and ask the following question:
can partial sharing of hunting returns promote the coexistence of resident and mutant populations?
To this end, we reformulate the per-capita hunting return functions as follows:
\begin{align}
    F_{\alpha} = \frac{(1-\alpha)P + \alpha\tilde P}{(1-\alpha)n + \alpha \nt }, \quad
    \tilde F_\alpha = \frac{(1-\alpha)\tilde P + \alpha P}{(1-\alpha)\nt + \alpha n},
\end{align}
where \( 0 \leq \alpha \leq 1/2 \) represents the degree of sharing and
\begin{align}
    P \coloneqq n_0 h_0 + n_1 h_1, \quad
    \tilde P \coloneqq \tilde n_0 h_0 + \tilde n_1 h_1
\end{align}
denote the total hunting returns of the resident and mutant populations, respectively.

Each type contributes a fraction \( \alpha \) of its own hunting returns, and a fraction \( \alpha \) of its own members, to a pool that the other type draws on.
This can easily be interpreted as follows:
both sub-populations throw ``parties'' and have some visitors over from the other group; visitors in turn bring food from their own stock.

The numerator of \( \Fa \) is therefore the hunting returns available to the resident population, and the denominator is the number of individuals among whom it is divided, so that \( \Fa \) remains a per-capita rate.
We refer to the exchange as \emph{food sharing} hereafter.

We note that this reformulation recovers the original hunting return functions \( \Fa \equiv F \) and \( \Fta \equiv \Ft \) when \( \alpha = 0 \); by contrast, \( \Fa = \Fta = (P + \Pt)/N \) when \( \alpha = 1/2 \), in which case food is shared equally among all individuals, without differentiating residents and mutants.

The time evolution of resident and mutant population sizes differs between the absence and the presence of food sharing (Fig.~\ref{fig:food_sharing_trajectories}).
Partial sharing prevents the complete exclusion of either type (\pnl{b,d}), which indeed would happen in the absence of food sharing given all other parameters being the same (\pnl{a,c}).
Both types persist at positive population size for the duration of the simulations, both in the deterministic dynamics (\pnl{b}) and in the stochastic dynamics (\pnl{d}) alike.

Sharing can permit coexistence of types with different rates of retirement, and hence different frequencies of teachers, because it reduces the advantage that one type otherwise holds over the other at every population size.
Without sharing, the birth rate of each type depends only on the per-capita hunting returns of its own members, so the type with the greater expected per-capita lifetime hunting return grows at the expense of the other until the other is gone.
With sharing, a fraction \( \alpha \) of each type's hunting returns is available to the other, and the amount a type receives is divided among its own members: the denominator of \( \Fa \) falls as the resident becomes rarer while the term \( \alpha \tilde P \) in its numerator does not, so the per-capita hunting return rate of a type rises as its population size falls.
A rare type is therefore supported by the common one, and the two population sizes settle at an interior equilibrium (Fig.~\ref{fig:food_sharing_trajectories}\pnl{b,d}).

Sharing the returns of hunting activity between types with different life histories therefore sustains a persistent polymorphism in the retirement rate.
Two life histories, with two different retirement rates and therefore two different fractions of teachers, are maintained together in a single environment, whereas the baseline model of Eq.~\ref{eq:model} sustains only one.
Exclusion in the baseline model is therefore a consequence of the assumption that each type lives on its own hunting returns, and not of the stage structure itself.

\subsection{\label{subsec:coexistence_stability}Local stability of the resident--mutant coexistence equilibria under food sharing}

\begin{figure*}
    \centering
    \includegraphics[width=\linewidth]{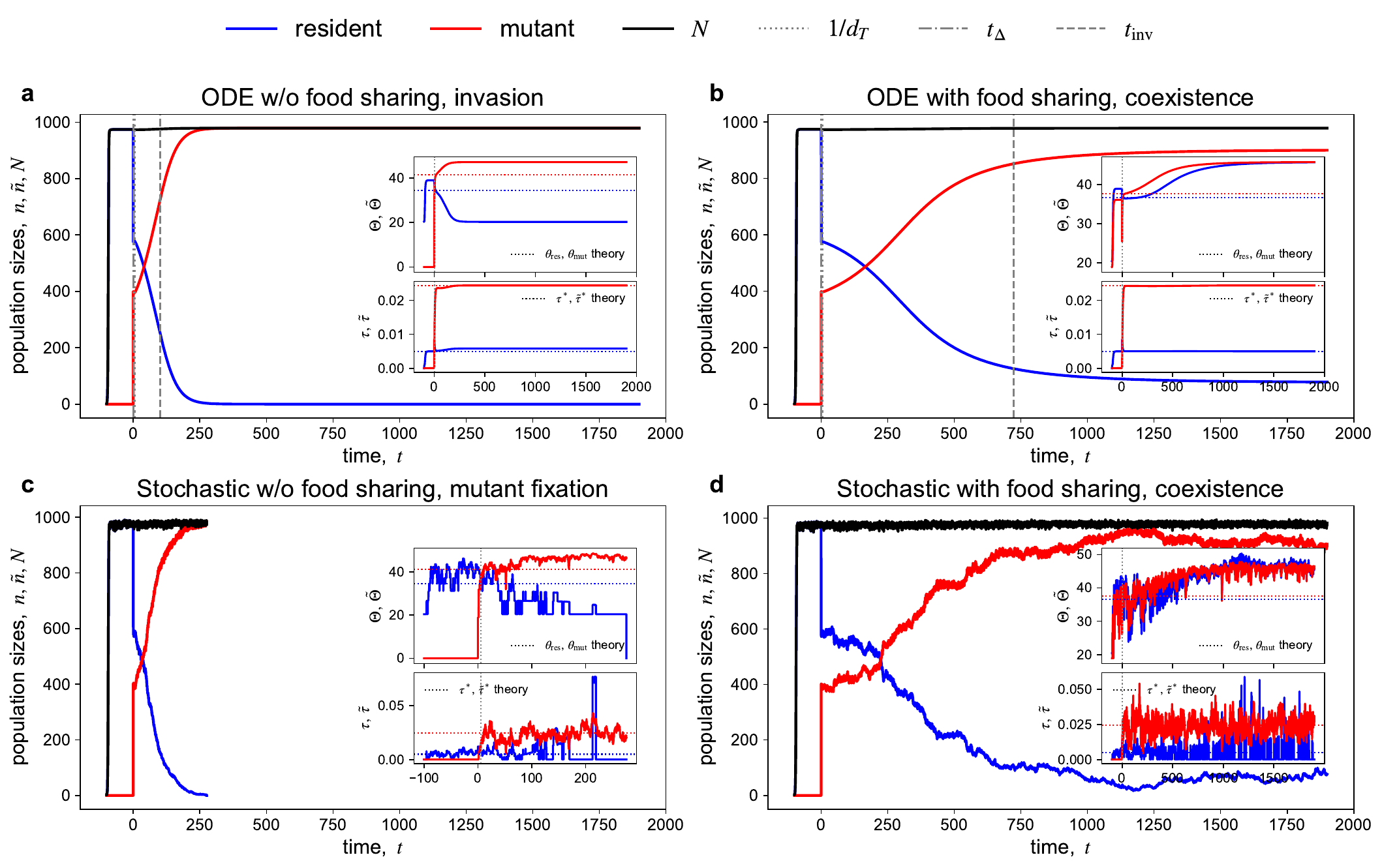}
    \caption{
    \textbf{Partial food sharing between types permits the resident and mutant to coexist,
    whereas without sharing one type always excludes the other.}
    Time evolution of resident (blue), mutant (red), and total (black) population sizes:
    (\pnl{a,c}) populations do not share their hunting returns with each other at all (same as before);
    (\pnl{b,d}) populations share some portion of their food with one another.
    The settings used in (\pnl{a,c}) are the same as those in Fig.~\ref{fig:base_trajectories}\pnl{a,c}, leading to the exclusion of residents.
    Using the same parameters but under partial food sharing (\pnl{b,d}), neither the resident nor mutant population goes extinct, allowing for both types to coexist.
    ODE (\pnl{a,b}) and stochastic (\pnl{c,d}) simulations are implemented the same way as in Fig.~\ref{fig:base_trajectories}.
    The same plotting styles (including insets and labels) are also used.
    Parameters:
    \( K = 1000 \), \( d_H = 0.05 \), \( d_T = 0.2 \), \( b_0 = 1 \), \( \eta = 0.02 \), \( \nu = 0.2 \), \( h_0 = 1 \), \( h_1 = 3 \), \( r = 0.001 \), \cy{\( \rt = 0.005 \), and \( \ntz = 400 \)}.
    The degree of sharing in \pnl{b,d} is \( \alpha = 0.3 \);
    panels \pnl{a,c} use \( \alpha = 0 \).
    Pairwise invasibility plots for food sharing cases are presented in \hyperref[sec:SI]{\emph{SI Appendix}}\xspace.
    }
    \label{fig:food_sharing_trajectories}
\end{figure*}

\begin{figure*}[b]
    \centering
    \includegraphics[width=\linewidth]{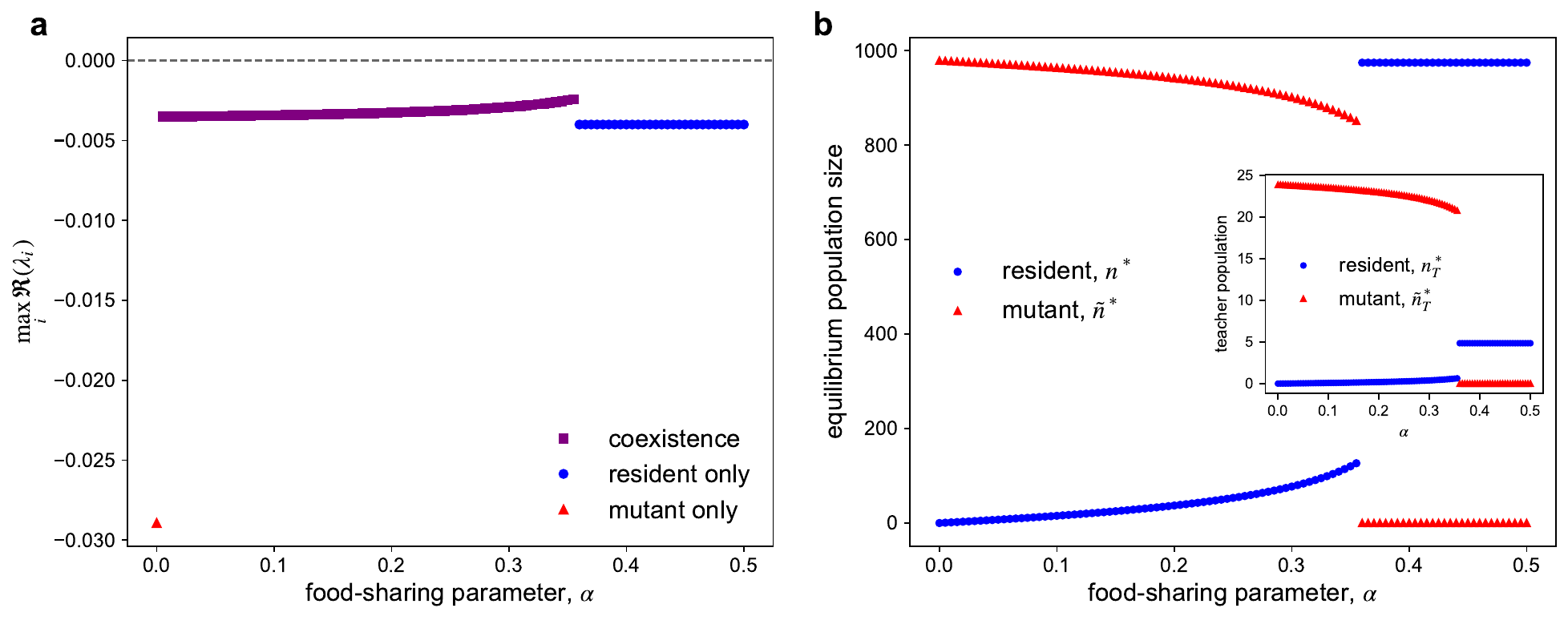}
    \caption{
    \textbf{Local stability analysis of dynamically selected steady states under food sharing.}
    \pnl{a,} The largest real part of the Jacobian eigenvalues, evaluated numerically at the corresponding steady state obtained from ODE simulations, over \( \alpha \in [0, 0.5] \).
    The coexistence, resident-only, mutant-only equilibria found in ODE simulations are all locally asymptotically stable.
    The cases where \( \alpha = 0 \) (mutant only) and \( \alpha = 0.3 \) (coexistence) correspond to Fig.~\ref{fig:food_sharing_trajectories}a and b, respectively.
    \pnl{b,} Corresponding equilibrium resident and mutant population sizes.
    Inset: equilibrium population sizes of the resident and mutant elder-teacher classes.
    The superscript asterisk indicates that the equilibrium values were obtained from ODE simulations.
    In this parameter regime,
    when there is no food sharing (\( \alpha = 0 \)), the mutant takes over and the resident goes extinct.
    When there is a positive rate of food sharing (\( \alpha_c > \alpha > 0 \)), neither type goes extinct; both the resident and mutant types remain in the population.
    When the rate of food sharing exceeds a threshold value (\( \alpha > \alpha_c \)), the resident persists and the mutant goes extinct.
    These observations are also evident in panel \pnl{b}, where the threshold value is \( \alpha_c \approx 0.36 \).
    We note that both panels show only dynamically selected equilibria and do not show all equilibria of the system, especially those that are locally unstable.
    Therefore, these figures are not intended to provide a complete bifurcation analysis but to demonstrate local stability of the dynamically selected coexistence equilibria over a wide range of food-sharing rates.
    Parameters:
    \( K = 1000 \), \( d_H = 0.05 \), \( d_T = 0.2 \), \( b_0 = 1 \), \( \eta = 0.02 \), \( \nu = 0.2 \), \( h_0 = 1 \), \( h_1 = 3 \), \( r = 0.001 \), \( \rt = 0.005 \), \( \ntz = 400 \).
    }
    \label{fig:coexistence_stability}
\end{figure*}

Here we perform numerical stability analysis on coexistence equilibria found in simulations, in which neither type goes extinct and instead a persistent polymorphism arises in the retirement rate.
We vary the degree of food sharing over \( \alpha \in [0, 0.5] \) and examine the largest real part of the Jacobian eigenvalues evaluated at the equilibrium states identified from deterministic simulations.
Figure~\ref{fig:coexistence_stability} implies three types of equilibria across varying degrees of food sharing:
resident-only, mutant-only, and resident--mutant coexistence equilibria.
As Fig.~\ref{fig:coexistence_stability}\pnl{a} shows, the coexistence equilibria found in deterministic simulations, including the one shown in Fig.~\ref{fig:food_sharing_trajectories}, are all locally asymptotically stable (purple squares), as expected.
So too are the resident-only and mutant-only equilibria reached in simulations.
The transitions between mutant-only, coexistence, and resident-only states give us useful insight into how food sharing affects the outcome of competition between types with different retirement rates:
disproportionate food sharing increasingly devalues within-type teaching as the hunting returns are shared across types, thereby disadvantaging the type with the higher retirement rate due to its resulting shorter reproductive lifespan.
This provides an intuition for why a large value of \( \alpha \) leads to the exclusion of a type with the higher retirement rate (the mutant in the present case).

\FloatBarrier

\FloatBarrier

\end{document}